\documentclass{aa}  

\usepackage{graphicx}
\usepackage{txfonts}
\usepackage{lipsum}
\usepackage{subcaption}         % necessary for continued figures
\usepackage{lscape}             % to rotate a single page table
\usepackage{placeins}           % useful with \FloatBarrier, to keep 
\usepackage{xcolor}
\usepackage{lipsum}

\usepackage{hyperref}
\hypersetup{
    colorlinks=true,
    linkcolor=blue,
    citecolor=blue,
    filecolor=magenta,      
    urlcolor=blue,
}
\usepackage{mathtools}

\renewcommand\makeLineNumber{} % turn off linenumbers for arXiv

\usepackage{listings}
\newcommand{\rhomax}{\ensuremath{\rho_{max}}\xspace}
\newcommand{\Rp}{\ensuremath{\mathbb{R}_{\geq0}}}

\newcommand{\del}{\partial}
\let\epsilon\varepsilon

\begin{document}

%%%%%%%%%%%%%%%%%%%%%%%%%%%%%%%%%%%%%%%%%%%%%%%%%%%%%%%%%%%%%%%%%%%%%%%%%%%%%%%%%%%%%%
%%%%%%%%%%%%%%%%%%%%%%%%%%%%%%%%%%%%%%%%%%%%%%%%%%%%%%%%%%%%%%%%%%%%%%%%%%%%%%%%%%%%%%

\title{Towards an agnostic algorithm for sampling empirical structure models}

\subtitle{The case of Uranus and Neptune}

\author{
Stefano Wirth\inst{1,2} \and Luca Morf \inst{2}\thanks{Corresponding author (luca.morf@uzh.ch), Stefano Wirth and Luca Morf contributed equally to this work.} \and Ravit Helled \inst{2}
}

\institute{
Department of Mathematics, Swiss Federal Institute of Technology Zurich (ETHZ), Rämistrasse 101, 8092 Zürich, Switzerland \and
Department of Astrophysics, University of Zurich, Winterthurerstrasse 190, 8057 Zürich, Switzerland
}

\date{Received 08 October 2025, Accepted 18 December 2025}

%%%%%%%%%%%%%%%%%%%%%%%%%%%%%%%%%%%%%%%%%%%%%%%%%%%%%%%%%%%%%%%%%%%%%%%%%%%%%%%%%%%%%%
%%%%%%%%%%%%%%%%%%%%%%%%%%%%%%%%%%%%%%%%%%%%%%%%%%%%%%%%%%%%%%%%%%%%%%%%%%%%%%%%%%%%%%

\abstract{
We present an algorithm to efficiently sample the full space of planetary interior density profiles. 
Our approach uses as few assumptions as possible to pursue an agnostic algorithm. 
The algorithm avoids the common Markov chain Monte Carlo method and instead uses an optimisation-based gradient-descent approach designed for computational efficiency. 
In this work, we use Uranus and Neptune as test cases and obtain empirical models that provide density and pressure profiles consistent with the observed physical properties (total mass, radius, and gravitational moments). 
We compared our findings to other work and find that while other studies are generally in line with our findings, they do not cover the entire space of solutions faithfully. 
Furthermore, we present guidance for modellers that construct Uranus or Neptune interior models with a fixed number of layers. 
We provide a statistical relation between the steepness classifying a density discontinuity and the resulting number of discontinuities to be expected.
For example, if one classifies a discontinuity as a density gradient larger than 0.02 kg\,m$^{-4}$, then most solutions should have at most one such discontinuity. 
Finally, we find that discontinuities, if present, are concentrated around a planetary normalised radius of 0.65 for Uranus and 0.7 for Neptune. 
Our algorithm to efficiently and faithfully investigate the full space of possible interior density profiles can be used to study all planetary objects with gravitational field data.
}

%%%%%%%%%%%%%%%%%%%%%%%%%%%%%%%%%%%%%%%%%%%%%%%%%%%%%%%%%%%%%%%%%%%%%%%%%%%%%%%%%%%%%%


\maketitle

%%%%%%%%%%%%%%%%%%%%%%%%%%%%%%%%%%%%%%%%%%%%%%%%%%%%%%%%%%%%%%%%%%%%%%%%%%%%%%%%%%%%%%%%%%%%%%%%%%%%%%%%%%%%%%%%%%%%%%%%%%%%%%%%%%%%%%%%%%%%%%%%%%%%%%%%%%%%%%%%%%%%%%%%%%%%

\section{Introduction}

%%%%%%%%%%%%%%%%%%%%%%%%%%%%%%%%%%%%%%%%%%%%%%%%%%%%%%%%%%%%%%%%%%%%%%%%%%%%%%%%%%%%%%

Uranus and Neptune have always been key to understanding the formation and evolution history of the Solar System. 
As the investigation of our own planetary system continues to expand, the study of exoplanets is gaining more attention.
Although the characterisation of exoplanets remains limited, it is now clear that in terms of mass and size, many exoplanets in the galaxy resemble Uranus and Neptune \citep{Zhu2021}. 
By revealing the internal structures of Uranus and Neptune, we can improve our knowledge of these two planetary objects in our own neighbourhood and the many observed intermediate-mass exoplanets orbiting distant stars.

%%%%%%%%%%%%%%%%%%%%%%%%%%%%%%%%%%%%%%%%%%%%%%%%%%%%%%%%%%%%%%%%%%%%%%%%%%%%%%%%%%%%%%

To study the internal structure of a planet, we need to know its density profile.
In the absence of direct measurements, the profile can be inferred from other observations.
Density profiles can be inferred by varying all values constituting a profile until the result matches the available observational data (for example total mass, radius, and the gravitational field).
The number of such observational parameters is typically small; for Uranus and Neptune, there are fewer than ten.
In contrast, density profiles typically consist of orders of magnitude more data points.
Inferring density profiles is therefore severely degenerate, as many density profiles can fit the same observational data. 

%%%%%%%%%%%%%%%%%%%%%%%%%%%%%%%%%%%%%%%%%%%%%%%%%%%%%%%%%%%%%%%%%%%%%%%%%%%%%%%%%%%%%%

To address this problem, one can simplify the approach used to infer density profiles by relying on a multitude of assumptions. 
However, caution must be taken.
Making unsound simplifying assumptions could exclude a large portion (or all) of the valid density profiles.
Therefore, obtaining a complete overview of all possible solutions is paramount before simplifications are even considered.

%%%%%%%%%%%%%%%%%%%%%%%%%%%%%%%%%%%%%%%%%%%%%%%%%%%%%%%%%%%%%%%%%%%%%%%%%%%%%%%%%%%%%%

To gain such an overview, we must sample profiles with an approach that is as unbiased as possible.
This kind of unbiased investigation is called empirical structure modelling.
Empirical structure modelling has been used before to investigate the internal structure of Uranus and Neptune.
The approach harkens back to earlier foundational works such as \cite{Marley1995} and \cite{Podolak2000}.
More recently, \cite{Neuenschwander2022} presented empirical structure models parametrised by polytropes that directly relate density to interior pressure.
In comparison, \cite{Movshovitz2022} used models that employed a combination of polynomials and sigmoid functions.

%%%%%%%%%%%%%%%%%%%%%%%%%%%%%%%%%%%%%%%%%%%%%%%%%%%%%%%%%%%%%%%%%%%%%%%%%%%%%%%%%%%%%%

The current numerical method of choice for empirical structure modelling is often the Markov chain Monte Carlo (MCMC) method \citep[for example][]{Marley1995, Movshovitz2022}.
The MCMC approach starts by generating initial density profiles and then randomly alters their parameters until they fit the observational data.
This poses challenges due to the aforementioned large number of data points that make up a density profile.
Therefore, one must either tolerate the slow convergence of the MCMC method when handling countless parameters or reduce the dimensionality of the parameter space to accelerate the search.
Previous studies have taken the latter route by parametrising density profiles with polytropes, polynomials, or sigmoid functions.
This reduces the number of parameters to an order of magnitude comparable to the number of observational parameters, but it can introduce biases, which is precisely what we aim to avoid here. 

%%%%%%%%%%%%%%%%%%%%%%%%%%%%%%%%%%%%%%%%%%%%%%%%%%%%%%%%%%%%%%%%%%%%%%%%%%%%%%%%%%%%%%

We replace the common MCMC method with a new optimisation-based approach that uses an algorithm to efficiently investigate the full space of possible interior density profiles that fit the known observational data of a planet.
The assumptions made for the profiles are reduced to an absolute minimum:
We only require that there are no density plateaus or inversions and that the maximal density does not exceed the density of any feasible material.
Otherwise, (with a resolution of $N=1024$) the $N$ density values are all allowed to change freely.
Because we treat all $N$ density values as free parameters, the method avoids biases introduced by the parametrisations of previous MCMC studies.
However, this alone does not ensure that the results are free from bias. 
We must further ensure that the generation of starting profiles is unbiased, and --- because we no longer use an MCMC algorithm --- we must additionally ensure that the way we find the density profiles is unbiased.
For the first point, if the generation of starting profiles favours certain types of density profiles, for instance, by initially under-representing profiles with sharp density discontinuities, an adjustment that increases the sampling of such profiles is required.
For the second point, we need to verify that our optimisation approach does not overlook harder-to-find local solutions in favour of dominant global ones that are easily spotted.

%%%%%%%%%%%%%%%%%%%%%%%%%%%%%%%%%%%%%%%%%%%%%%%%%%%%%%%%%%%%%%%%%%%%%%%%%%%%%%%%%%%%%%

Our paper is structured as follows. 
Section \ref{sec:methods} provides an overview of the algorithm. 
In Section \ref{sec:results} we present and interpret the results. In Section \ref{sec:discussion} we discuss the results, highlight caveats, and provide an outlook for future work. Section \ref{sec:summaryandconclusions} presents a summary of the key findings of this study. 
We note that our methodology is based on the theory of figures \citep[ToF,][]{ToF}. 
For the purposes of notation, we follow the elaboration presented in Appendix A of \cite{Morf2024}.
We employed the open-source code \texttt{PyToF}\footnote{See \url{https://doi.org/10.5281/zenodo.17592269}} \citep{PyToF} to implement the ToF numerically.

%%%%%%%%%%%%%%%%%%%%%%%%%%%%%%%%%%%%%%%%%%%%%%%%%%%%%%%%%%%%%%%%%%%%%%%%%%%%%%%%%%%%%%
%%%%%%%%%%%%%%%%%%%%%%%%%%%%%%%%%%%%%%%%%%%%%%%%%%%%%%%%%%%%%%%%%%%%%%%%%%%%%%%%%%%%%%

\section{Methods}
\label{sec:methods}

%%%%%%%%%%%%%%%%%%%%%%%%%%%%%%%%%%%%%%%%%%%%%%%%%%%%%%%%%%%%%%%%%%%%%%%%%%%%%%%%%%%%%%

\subsection{Algorithm}

%%%%%%%%%%%%%%%%%%%%%%%%%%%%%%%%%%%%%%%%%%%%%%%%%%%%%%%%%%%%%%%%%%%%%%%%%%%%%%%%%%%%%%

Planets are flattened due to their rotation. 
Surfaces of both constant density and potential are not spherical shells defined by a single radius, $r$, but are rather spheroidal in shape.
Spheroidal surfaces are described by a function, $r_l(\vartheta)$, that depends on the polar angle, $\vartheta$.
The variable $l$ uniquely identifies a spheroid and represents its volumetric mean radius:
\begin{equation}
    \frac{4\pi}{3}l^3 = 2\pi \int_{-1}^{1}\mathrm{d}\cos\vartheta \int_{0}^{r_l(\vartheta)} {r^\prime}^2 \mathrm{d}r^\prime. 
\end{equation}
We also refer to $l$ as the level surface of a spheroid.
A planet with such a spheroidal structure has a multipolar (non-spherical) external gravity field
\begin{equation}
    V_\text{ext}(r,\vartheta)=\frac{Gm}{r}\left(1-\sum_{n=1}^\infty\left(\frac{R_\text{ref}}{r}\right)^{2n}J_{2n}P_{2n}\left(\cos\vartheta\right)\right),
\end{equation}
where $R_\text{ref}$ is a normalisation constant, $m$ denotes the planetary mass, and $P_{2n}$ are the Legendre polynomials.
The $J_{2n}$ are called gravitational moments and can be calculated with the ToF.

%%%%%%%%%%%%%%%%%%%%%%%%%%%%%%%%%%%%%%%%%%%%%%%%%%%%%%%%%%%%%%%%%%%%%%%%%%%%%%%%%%%%%%

For the numerical implementation, density profiles consist of $N$ density values, $\rho_k$, at evenly spaced level surfaces, $l_k$.
We restricted the maximum density to some threshold, \rhomax, and enforced a minimum density increase per step, $\Delta_\text{min}$.
The density values are derived from $N-1$ parameters, $p_i$, as
\begin{equation}\rho_k = \sum\limits_{i=0}^{k-1} \frac{e^{p_i}}{w_i}\ \alpha. \label{eq:parameters}\end{equation}
The weights, $w_i$, were added to normalise the parameter values and are defined as
\begin{equation}w_i=\sum\limits_{j=i+1}^{N-1}l_j^2.\label{eq:weights}\end{equation}
The scaling factor $\alpha$ was added to ensure the density profile mass, $m_{calc}$, agrees with the planetary mass, $m$. 
It is defined as
\begin{equation}\alpha = \frac{m}{m_{calc}},\end{equation}
where the calculated mass, $m_{calc}$, was obtained by numerically integrating the density values over the level surfaces via
\begin{equation}m_{calc} = 4\pi \int_{l_\text{min}}^{l_\text{max}} \sum\limits_{i=0}^{k-1} \frac{e^{p_i}}{w_i} l^2 \ \mathrm{d}l.\end{equation}
Listing \ref{lst:algorithm} gives a high-level overview of the optimisation algorithm.

%%%%%%%%%%%%%%%%%%%%%%%%%%%%%%%%%%%%%%%%%%%%%%%%%%%%%%%%%%%%%%%%%%%%%%%%%%%%%%%%%%%%%%

\begin{lstlisting}[caption={The optimisation algorithm.}, label={lst:algorithm}]
input: int epochs
output: tuple results
begin
	params $\gets$ create_start_params()
	for epoch in epochs do
		check convergence
		for steps in epochsize do
			params $\gets$ project(opt_step(params))
	return results
end
\end{lstlisting}

%%%%%%%%%%%%%%%%%%%%%%%%%%%%%%%%%%%%%%%%%%%%%%%%%%%%%%%%%%%%%%%%%%%%%%%%%%%%%%%%%%%%%%

Given the number of epochs to run the algorithm for, we first created some suitable starting parameters.
Then, for each epoch, we checked whether the optimisation converged to a stable solution.
If not, we performed \texttt{epochsize} for many optimisation steps.
For each optimisation step, we calculated the gradient and moved along it according to the Adam gradient descent scheme \citep[][]{adam}.
Then, we projected the new parameters to enforce our minimum density increase.
Once convergence was detected, the final result and other data, was returned.
Further details on the algorithm can be found in the Appendix \ref{app:algorithm}. 
Explanations regarding the requirements of an unbiased generation method and optimisation method are given in Appendix \ref{app:bias}. 
The method's limitations are discussed in Section \ref{sec:discussion}. 

%%%%%%%%%%%%%%%%%%%%%%%%%%%%%%%%%%%%%%%%%%%%%%%%%%%%%%%%%%%%%%%%%%%%%%%%%%%%%%%%%%%%%%

\begin{table*}[htpb!]
    \centering
    \caption{Physical values for Uranus and Neptune.}
    \begin{tabular}{lllll}
    \hline
    \hline
         & Uranus & & Neptune &  \\
        \hline
        Mass & 8.68099 $\cdot$ 10$^{25}$ kg  & \cite{Jacobson2025} & 1.02409 $\cdot$ 10$^{26}$ kg & \cite{Jacobson2009}\\
        Equatorial radius & 25559 km & \cite{Lindal1987}& 24766 km & \cite{Lindal1992}\\
        Reference radius & 25559 km & \cite{French2024}& 25225 km &\cite{Wang2023}\\
        $J_2$ & $(3509.291\pm 0.412) 10^{-6}$ & \cite{French2024}& $(3401.655\pm 3.994) 10^{-6}$ & \cite{Wang2023}\\
        $J_4$ & $(-35.522\pm 0.466) 10^{-6}$ & \cite{French2024} & $(-33.294\pm 10) 10^{-6}$ & \cite{Wang2023}\\
        $\text{cov}(J_2,J_4)$ & 0.9861$\sigma_{J_2}\sigma_{J_4}$ & \cite{French2024} & -- & -- \\
        Rotational period & 62064 s& \cite{Desch1986}& 57479 s & \cite{Karkoschka2011}\\
    \hline
    \hline
    \end{tabular}
    \label{tab:values}
\end{table*}

%%%%%%%%%%%%%%%%%%%%%%%%%%%%%%%%%%%%%%%%%%%%%%%%%%%%%%%%%%%%%%%%%%%%%%%%%%%%%%%%%%%%%%

\subsection{Simulation setup}

%%%%%%%%%%%%%%%%%%%%%%%%%%%%%%%%%%%%%%%%%%%%%%%%%%%%%%%%%%%%%%%%%%%%%%%%%%%%%%%%%%%%%%

The simulation was performed on an AMD EPYC 7702 Hypervisor CPU with 96 cores over 120 hours.
We dedicated 64 cores to Uranus and 32 towards Neptune.
We used twice as many cores for Uranus in order to run correlated ($\text{cov}(J_2,J_4)=0.9861\sigma_{J_2}\sigma_{J_4}$; see Table \ref{tab:values}) and uncorrelated ($\text{cov}(J_2,J_4)=0$) simulations in parallel.
All physical values for Uranus and Neptune are given in Table \ref{tab:values}. 
The number of layers, $N$, was set to 1024, and the epoch size was 25 steps.
We chose a maximum of 120 epochs (3000 steps).
The scalar factor, $c$, introduced in Equation \ref{eq:cost_factor} was set to $100$.
This factor ensures that the optimisation cost function does not consider extreme values for $\alpha$ (far away from $\alpha=1$) while optimising for gravitational field data.
As a convergence criterion within the ToF algorithm \texttt{PyToF}, we employed $\left|J_{2n}^{i+1}-J_{2n}^i\right| \leq 10^{-10}J_{2n}^i$ to check if  convergence was achieved after the $i$-th iteration.
The maximum acceptable density $\rhomax$ was set to 20000 kg\,m$^{-3}$.
The minimum acceptable density increase per unit length, $\Delta_\text{min}$, was set to 1 $\Delta \ell^{-1}$ kg\,m$^{-3}$, where $\Delta \ell = R/N$.
Explicitly, this resulted in a value of $\Delta \ell=$ 24960 m for Uranus and $\Delta \ell=$ 24186 m for Neptune (rounded to the nearest meter).
This yielded $\Delta_\text{min}=$ 4.01 $\cdot$ 10$^{-5}$ kg\,m$^{-4}$ for Uranus and $\Delta_\text{min}=$ 4.13 $\cdot$ 10$^{-5}$ kg\,m$^{-4}$ for Neptune (rounded to two significant figures).

%%%%%%%%%%%%%%%%%%%%%%%%%%%%%%%%%%%%%%%%%%%%%%%%%%%%%%%%%%%%%%%%%%%%%%%%%%%%%%%%%%%%%%

Overall, 694465 (791901) runs were completed for the correlated (uncorrelated) Uranus data, and 731407 runs were completed for Neptune. 
For Uranus, 90.39\% (90.48\%) of the runs respected \rhomax, and 18.77\% (18.56\%) achieved gravitational moments, $J_{2n}$, within the multivariate three-standard-deviation threshold, which is 20.77\% (20.56\%) of those that respected \rhomax.
For Neptune, the numbers (in the same order) are 89.22\% and 88.83\%, which is 99.57\% of those that respected \rhomax.
Our optimisation-based approach and the developed algorithm are highly efficient: 
The average time to complete a run was approximately 20 seconds. 

%%%%%%%%%%%%%%%%%%%%%%%%%%%%%%%%%%%%%%%%%%%%%%%%%%%%%%%%%%%%%%%%%%%%%%%%%%%%%%%%%%%%%%

We note that throughout all subsequent results, the parameter space of the possible density profiles for Neptune is larger in comparison to Uranus.
This is because the measurements we have for the gravitational moments of Uranus are about one order of magnitude more precise than those for Neptune (Table \ref{tab:values}).
Consequently, Uranian runs had a $\sim$19\% success rate, compared to $\sim$89\% of the Neptunian runs.

%%%%%%%%%%%%%%%%%%%%%%%%%%%%%%%%%%%%%%%%%%%%%%%%%%%%%%%%%%%%%%%%%%%%%%%%%%%%%%%%%%%%%%

\begin{figure*}
    \centering

    \begin{subfigure}{0.5\textwidth}
        \centering
        \includegraphics[width=\linewidth]{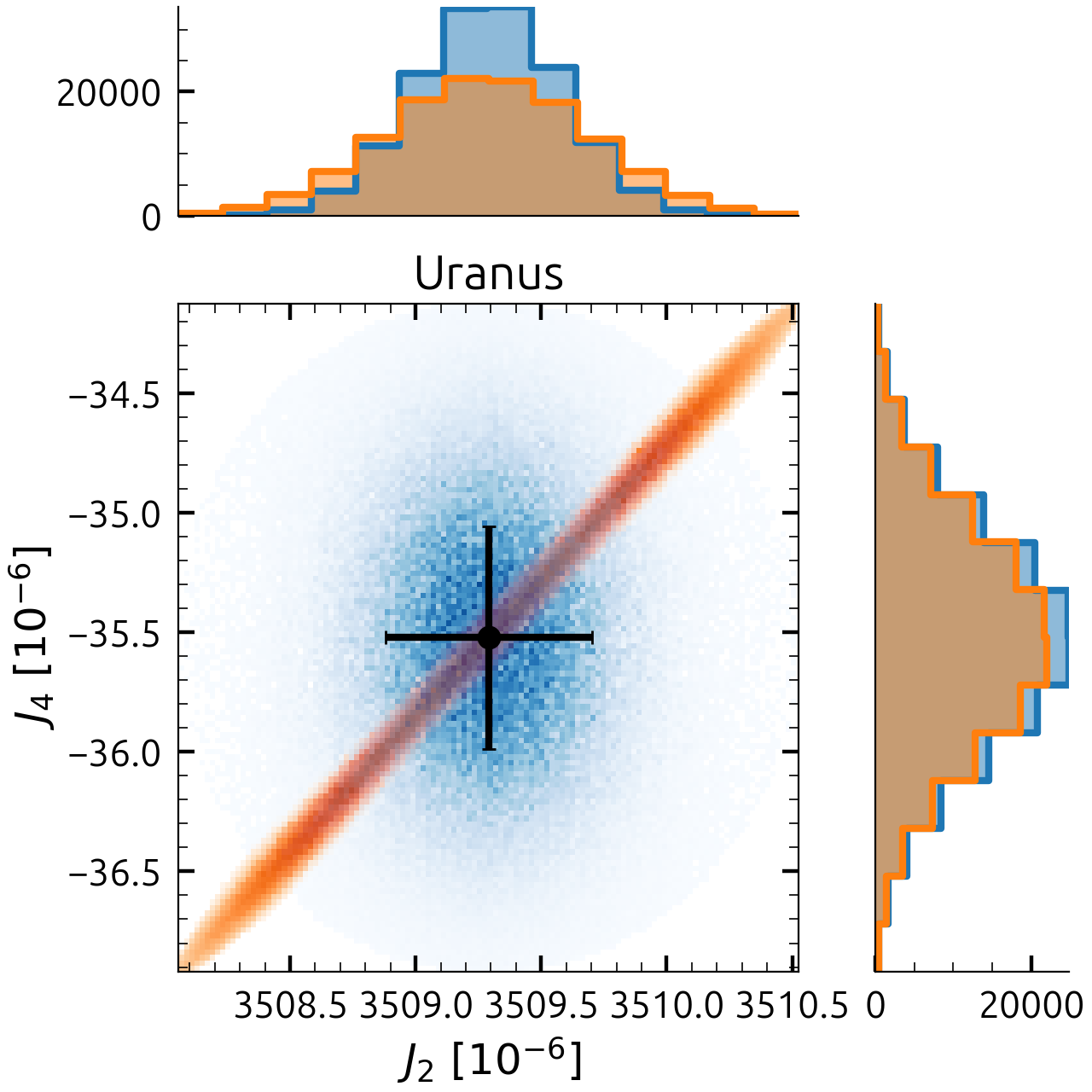}
        \caption{}
        \label{fig:Jsu}
    \end{subfigure}%
    \begin{subfigure}{0.5\textwidth}
        \centering
        \includegraphics[width=\linewidth]{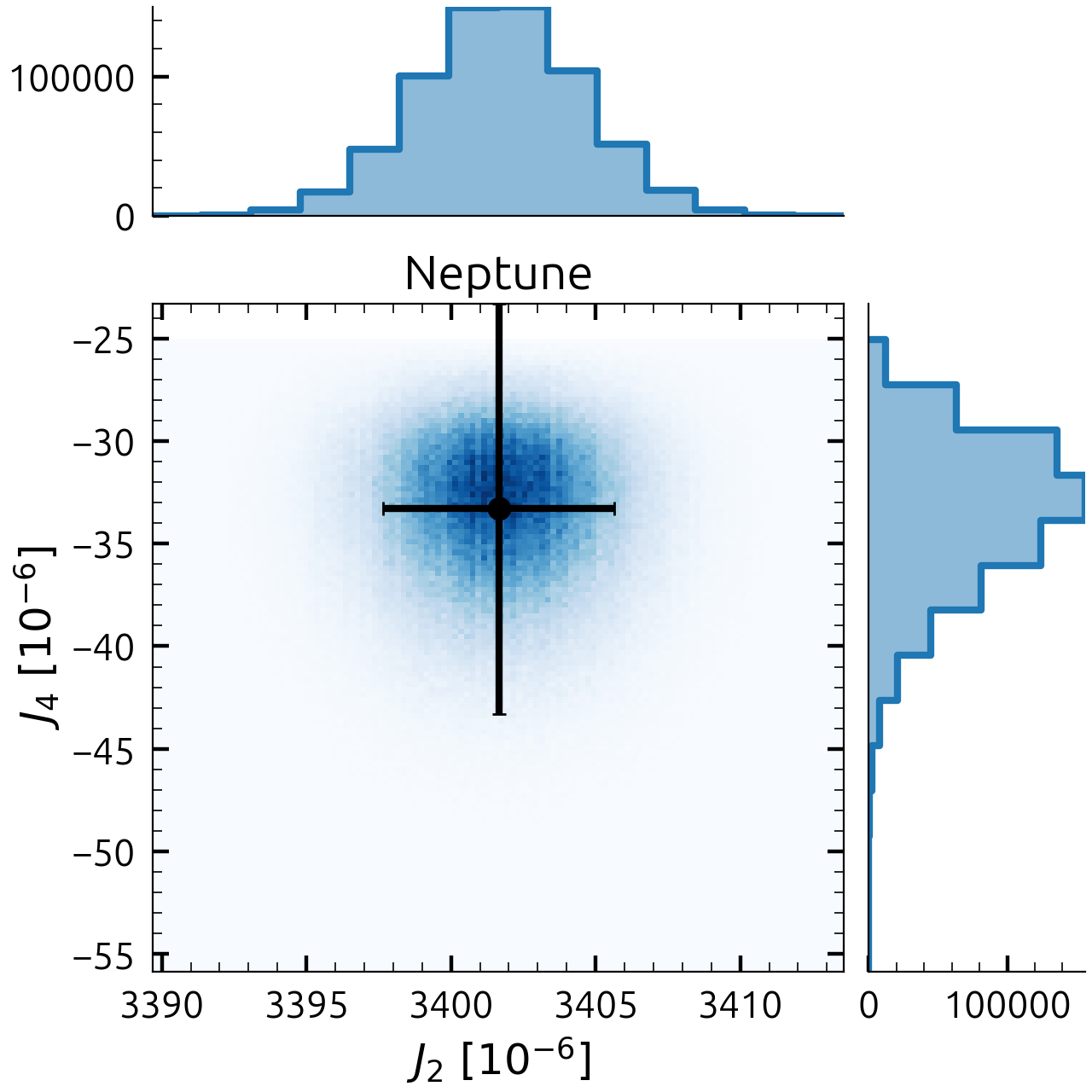}
        \caption{}
        \label{fig:Jsn}
    \end{subfigure}

    \caption{
        Gravitational moment distribution for successful Uranus
        (\subref{fig:Jsu}) and Neptune (\subref{fig:Jsn}) profiles.
        Both panels include a 2D histogram where bins more
        frequently occupied by solutions are shown with a relatively darker colour.
        Additionally, 1D histograms display the weighted number
        of solutions along each axis.
        The weights were assigned to solutions based on a multivariate Gaussian
        likelihood, with means, standard deviations, and covariance taken from
        Table \ref{tab:values}.
        In the left panel \subref{fig:Jsu}, an orange distribution is shown in addition
        to the blue distribution.
        Both blue distributions in \subref{fig:Jsu} and \subref{fig:Jsn}
        assume $\mathrm{cov}(J_2,J_4)=0$, whereas the orange distribution for
        Uranus takes $\mathrm{cov}(J_2,J_4)=0.9861\,\sigma_{J_2}\sigma_{J_4}$.
    }
    \label{fig:Js}
\end{figure*}

%%%%%%%%%%%%%%%%%%%%%%%%%%%%%%%%%%%%%%%%%%%%%%%%%%%%%%%%%%%%%%%%%%%%%%%%%%%%%%%%%%%%%%

\begin{figure*}
    \centering

    \begin{subfigure}{0.5\textwidth}
        \centering
        \includegraphics[width=\linewidth]{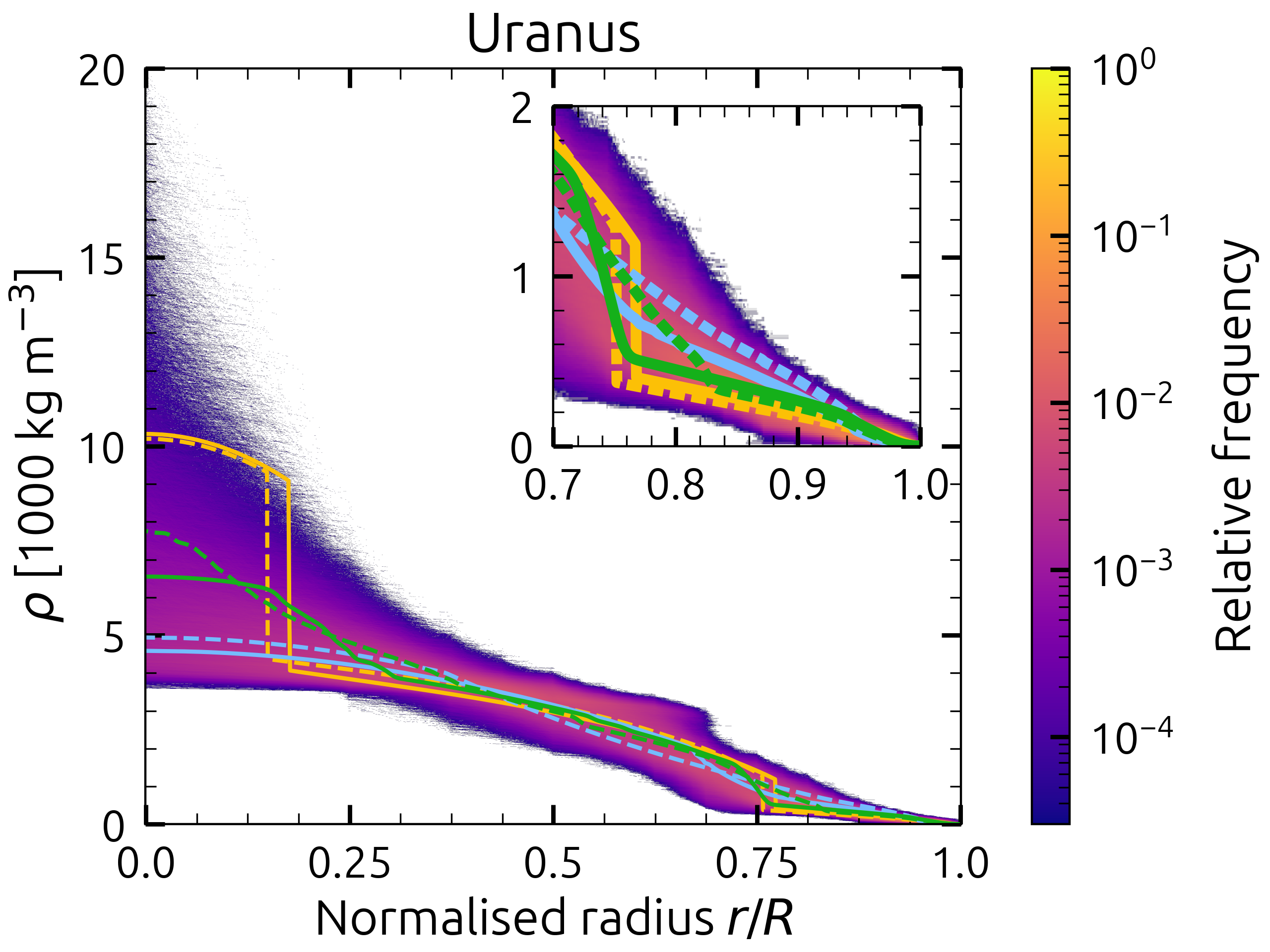}
        \caption{}
        \label{fig:densprofu}
    \end{subfigure}%
    \begin{subfigure}{0.5\textwidth}
        \centering
        \includegraphics[width=\linewidth]{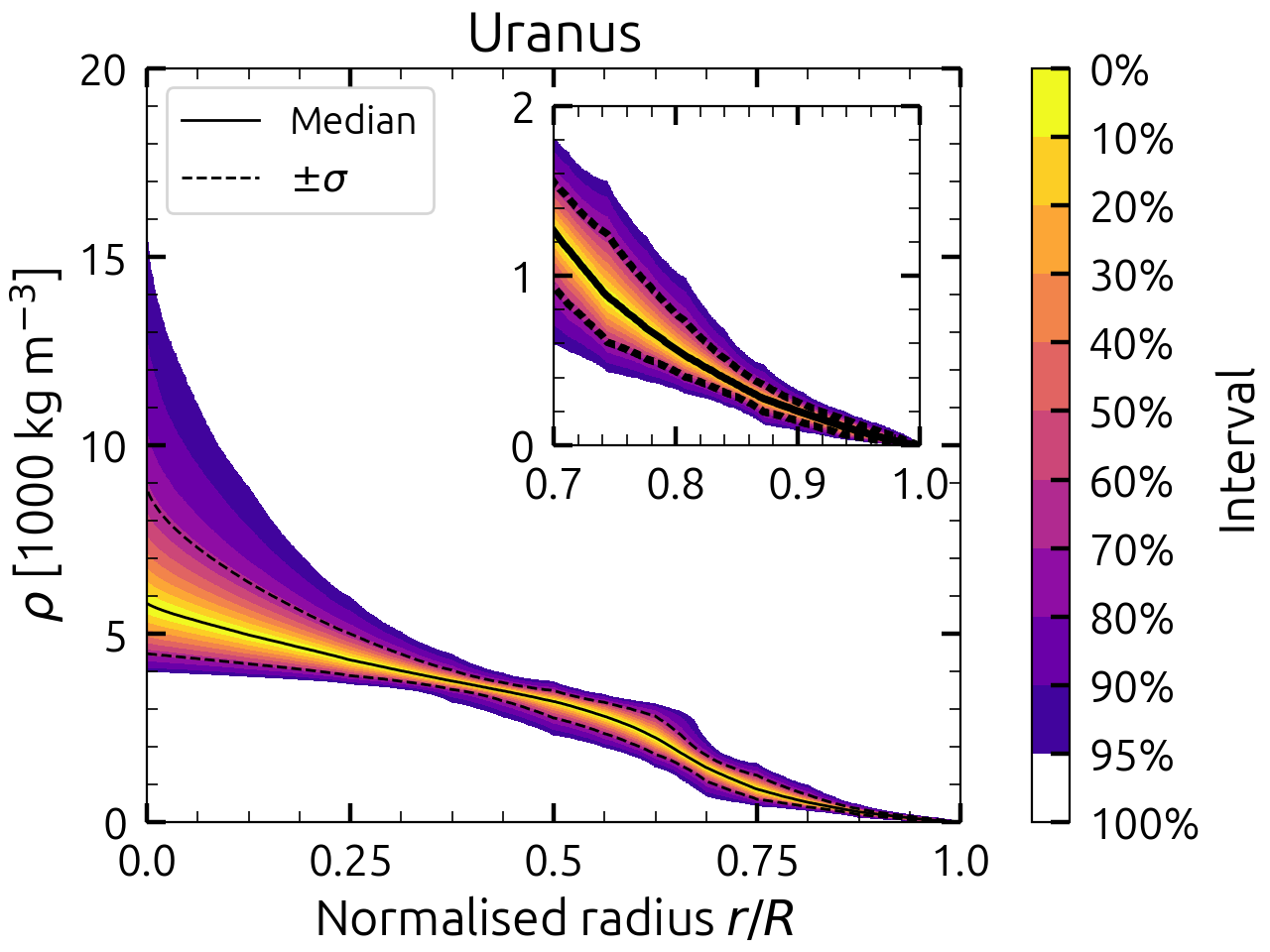}
        \caption{}
        \label{fig:contouru}
    \end{subfigure}

    \caption{
        Solution space distribution (\subref{fig:densprofu}) and contour of the distribution (\subref{fig:contouru}) for Uranus.
        The left panel \subref{fig:densprofu} shows the weighted distribution of all successful uncorrelated Uranus density profiles.
        The inset shows a zoomed-in view of the outermost region, from $r/R=1$ to $r/R=0.7$.
        Note that the zoom is not aspect-preserving.
        The colour scale shows the relative frequency over all distributions of a certain density value at a given radius.
        Coloured lines show density profiles from previous studies for comparison.
        Solid and dashed lines correspond to models U1 and U2 from \cite{Nettelmann2013} (orange); U1 and U3 from \cite{Morf2025} (green); and V2 and V3 from \cite{Vazan2020} (blue), respectively.
        The colour scale in the right panel \subref{fig:contouru} shows the percentile intervals in 10\% increments, except the last interval, where a 95\% interval was added.
        The black line shows the median.
        The dashed lines show the 16th (lower) and 84th (upper) percentiles.
    }
    \label{fig:distru}
\end{figure*}

%%%%%%%%%%%%%%%%%%%%%%%%%%%%%%%%%%%%%%%%%%%%%%%%%%%%%%%%%%%%%%%%%%%%%%%%%%%%%%%%%%%%%%

\begin{figure*}
    \centering

    \begin{subfigure}{0.5\textwidth}
        \centering
        \includegraphics[width=\linewidth]{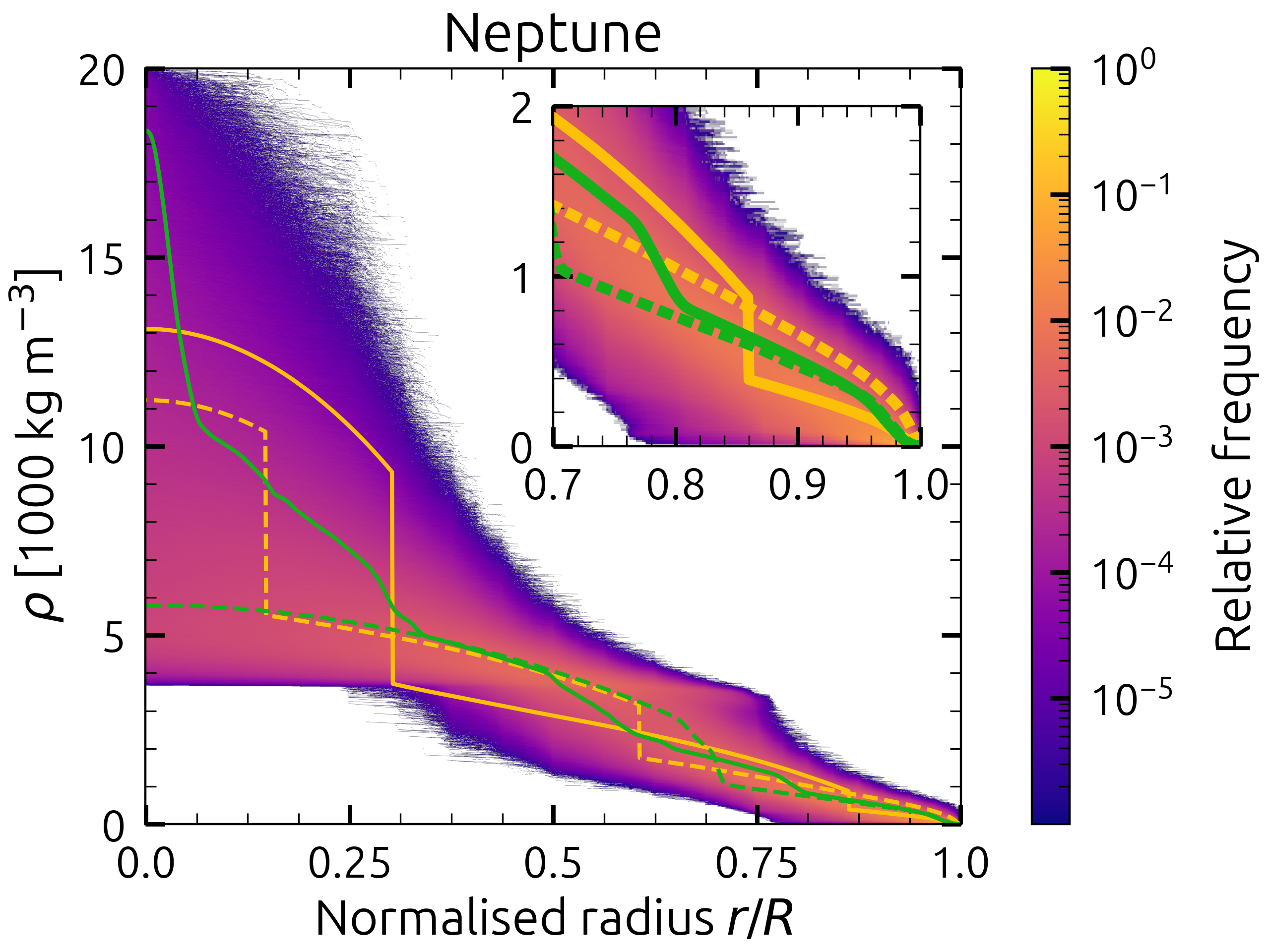}
        \caption{}
        \label{fig:densprofn}
    \end{subfigure}%
    \begin{subfigure}{0.5\textwidth}
        \centering
        \includegraphics[width=\linewidth]{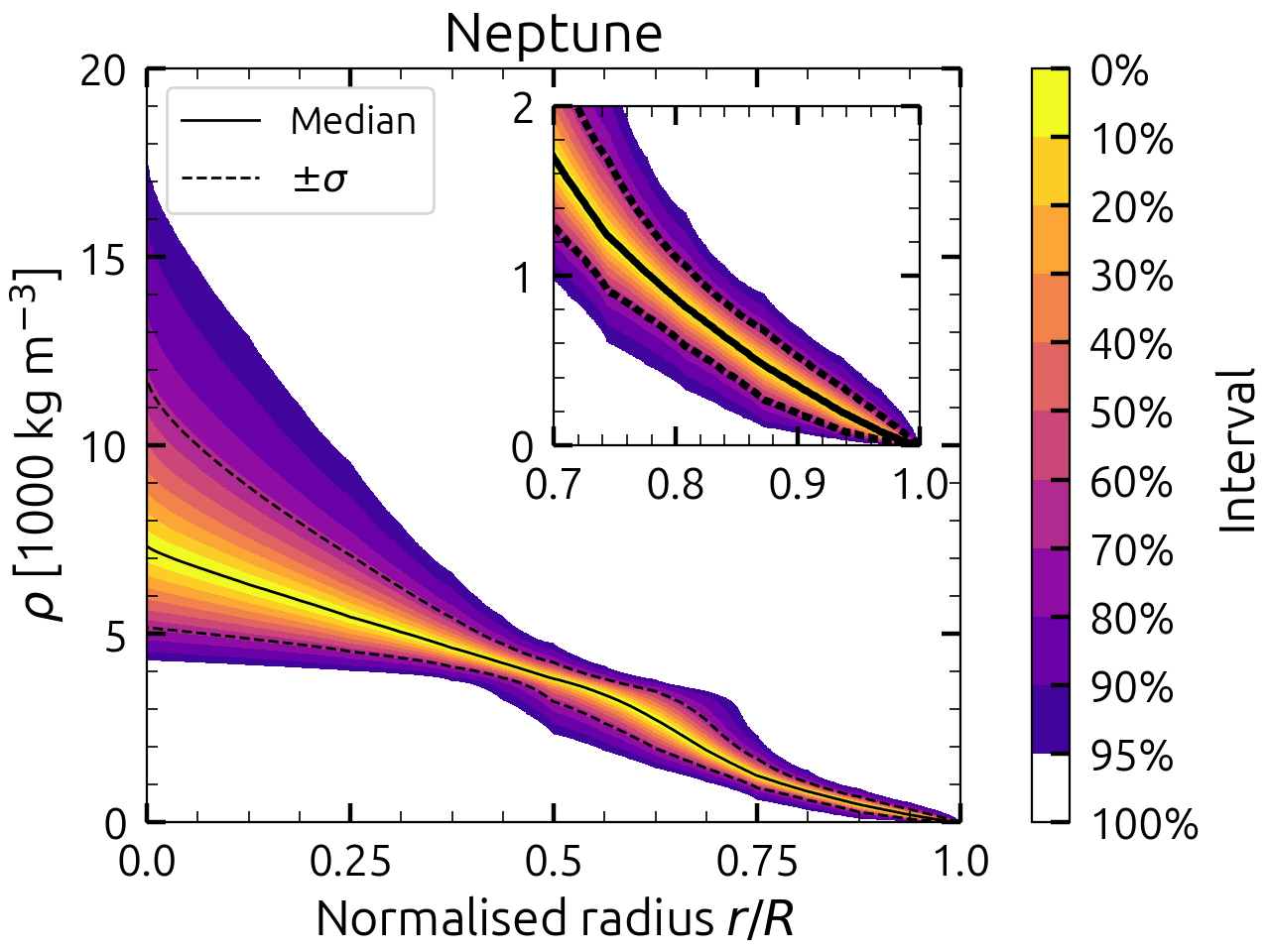}
        \caption{}
        \label{fig:contourn}
    \end{subfigure}

    \caption{
        Same as Figure \ref{fig:distru} but for Neptune.
        The solid and dashed line in the left panel \subref{fig:densprofn} correspond to the models N1 and N2b from \cite{Nettelmann2013} (orange) and to models N1 and N3 from \cite{Morf2025} (green), respectively.
    }
    \label{fig:distrn}
\end{figure*}

%%%%%%%%%%%%%%%%%%%%%%%%%%%%%%%%%%%%%%%%%%%%%%%%%%%%%%%%%%%%%%%%%%%%%%%%%%%%%%%%%%%%%%
%%%%%%%%%%%%%%%%%%%%%%%%%%%%%%%%%%%%%%%%%%%%%%%%%%%%%%%%%%%%%%%%%%%%%%%%%%%%%%%%%%%%%%

\section{Results}
\label{sec:results}

%%%%%%%%%%%%%%%%%%%%%%%%%%%%%%%%%%%%%%%%%%%%%%%%%%%%%%%%%%%%%%%%%%%%%%%%%%%%%%%%%%%%%%

Below we present the inferred density profiles and their characteristics. 
For simplicity, we refer to the normalised volumetric mean radius of a spheroid as `the normalised radius $r/R$'.

%%%%%%%%%%%%%%%%%%%%%%%%%%%%%%%%%%%%%%%%%%%%%%%%%%%%%%%%%%%%%%%%%%%%%%%%%%%%%%%%%%%%%%

\subsection{Density profiles}
\label{subsec:densityprofiles}

%%%%%%%%%%%%%%%%%%%%%%%%%%%%%%%%%%%%%%%%%%%%%%%%%%%%%%%%%%%%%%%%%%%%%%%%%%%%%%%%%%%%%%

\begin{figure}
	\centering
	\includegraphics[width=0.833\columnwidth]{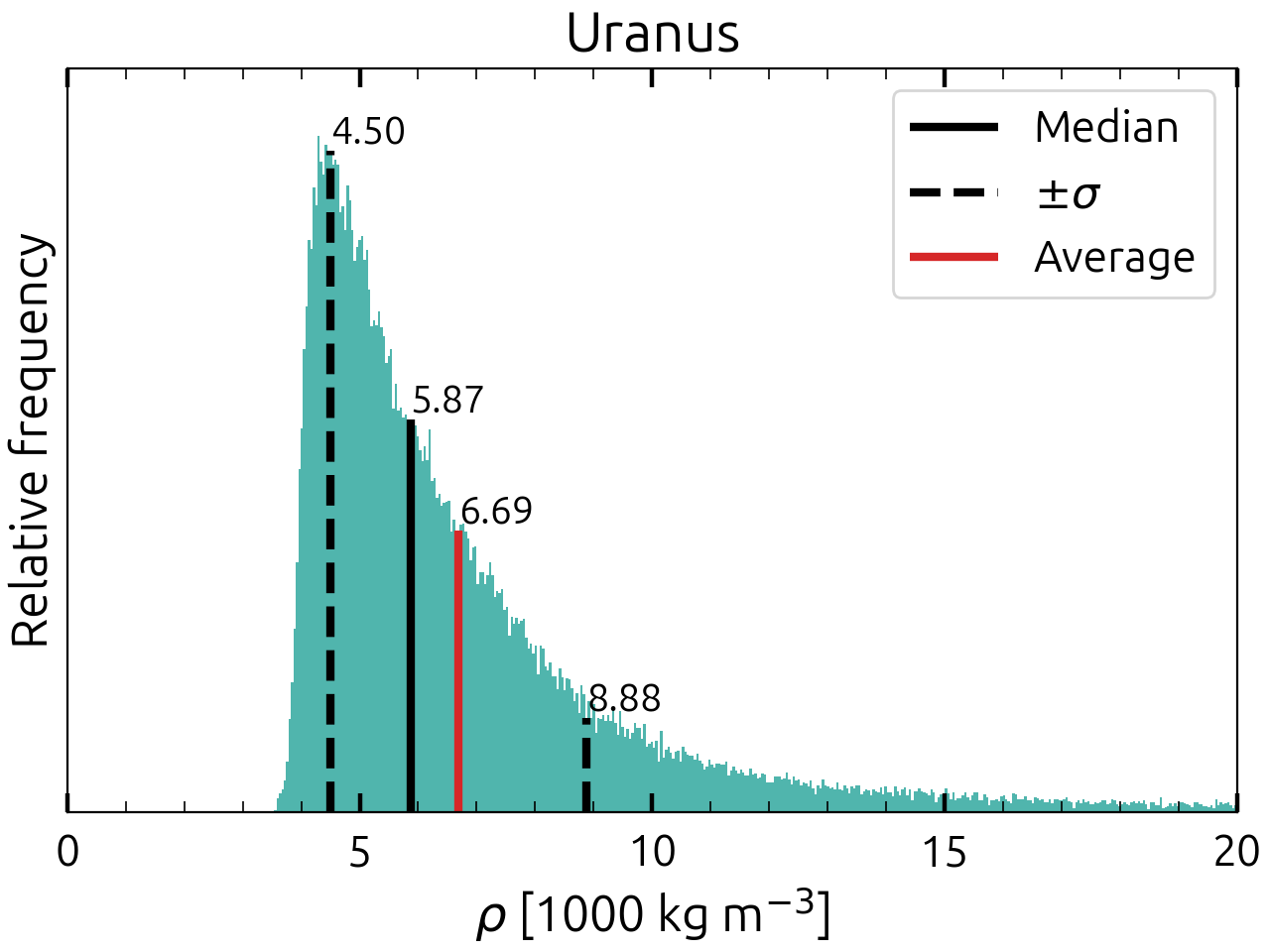}
	\includegraphics[width=0.833\columnwidth]{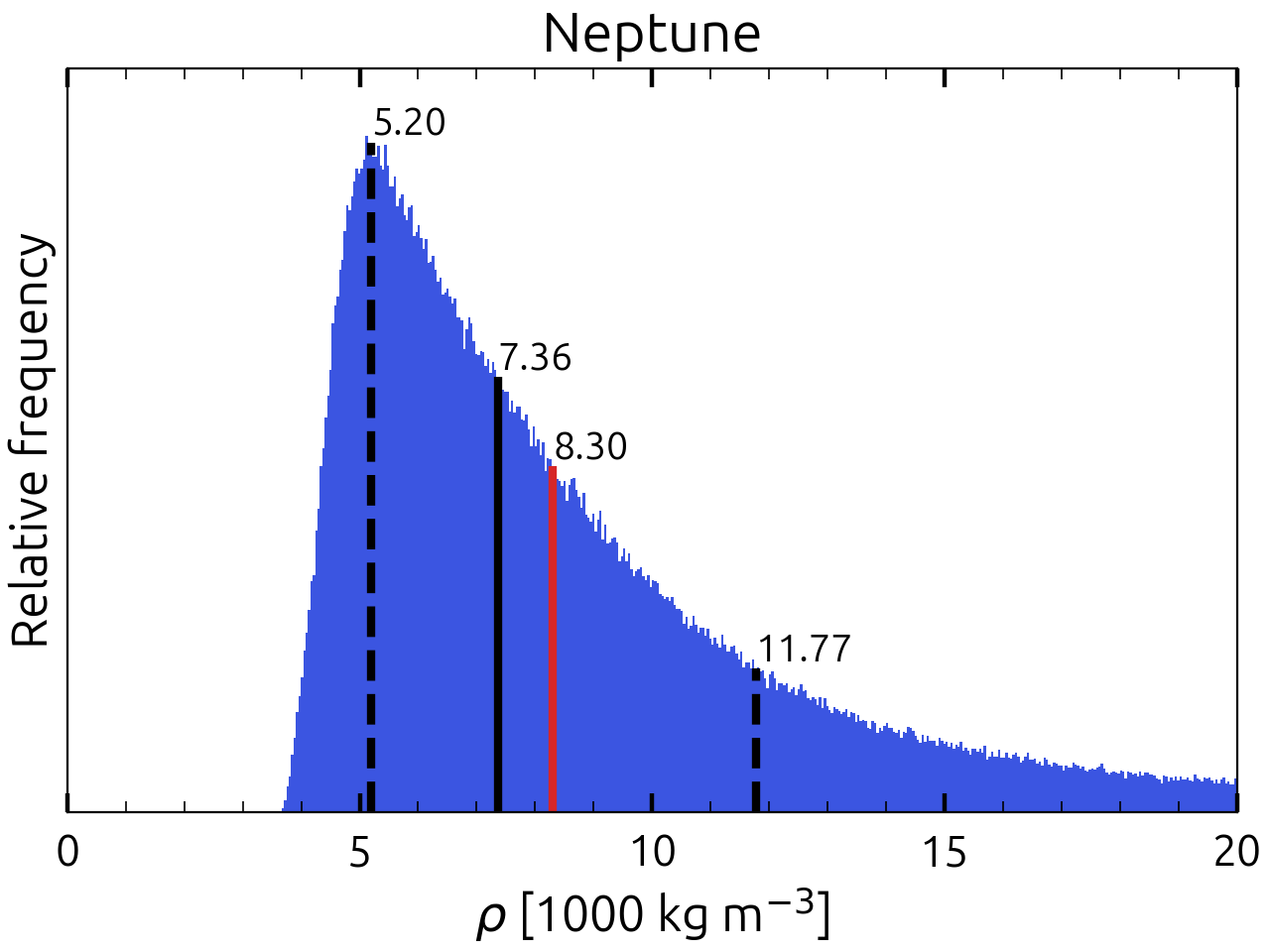}
	\caption{
    Central densities.
    The black lines are as before in Figures \ref{fig:contouru} and \ref{fig:contourn}.
    The red line denotes the average.
    This figure is equivalent to a side-on view of the density profile distributions (Figures \ref{fig:densprofu} and \ref{fig:densprofn}) at radius 0, except in linear scale.
    Note that the distribution is cut off at $\rho_{max}=$ 20000 kg\,m$^{-3}$.
    }
	\label{fig:coredens}
\end{figure}

\begin{figure}
	\centering
	\includegraphics[width=0.833\columnwidth]{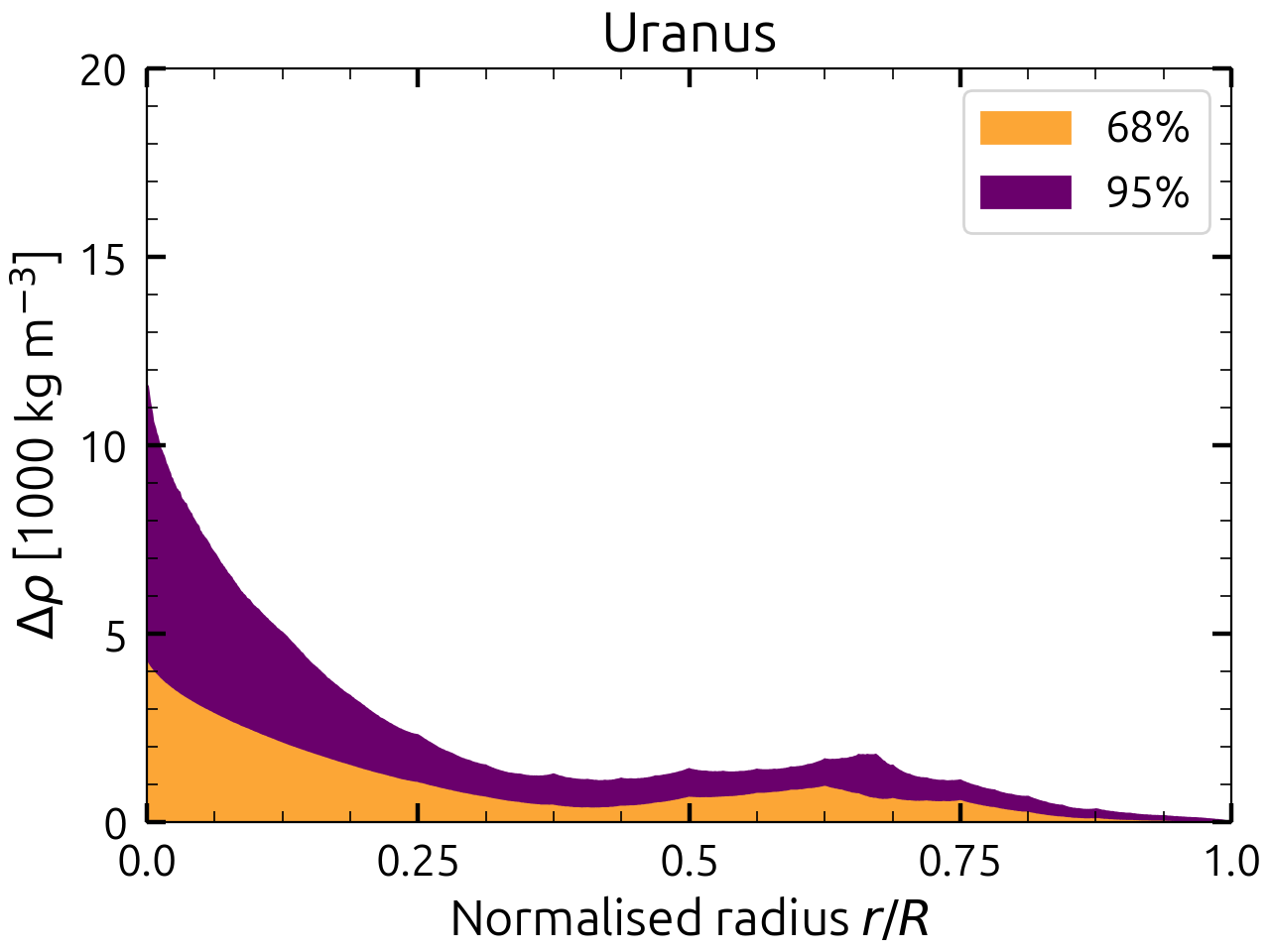}
	\includegraphics[width=0.833\columnwidth]{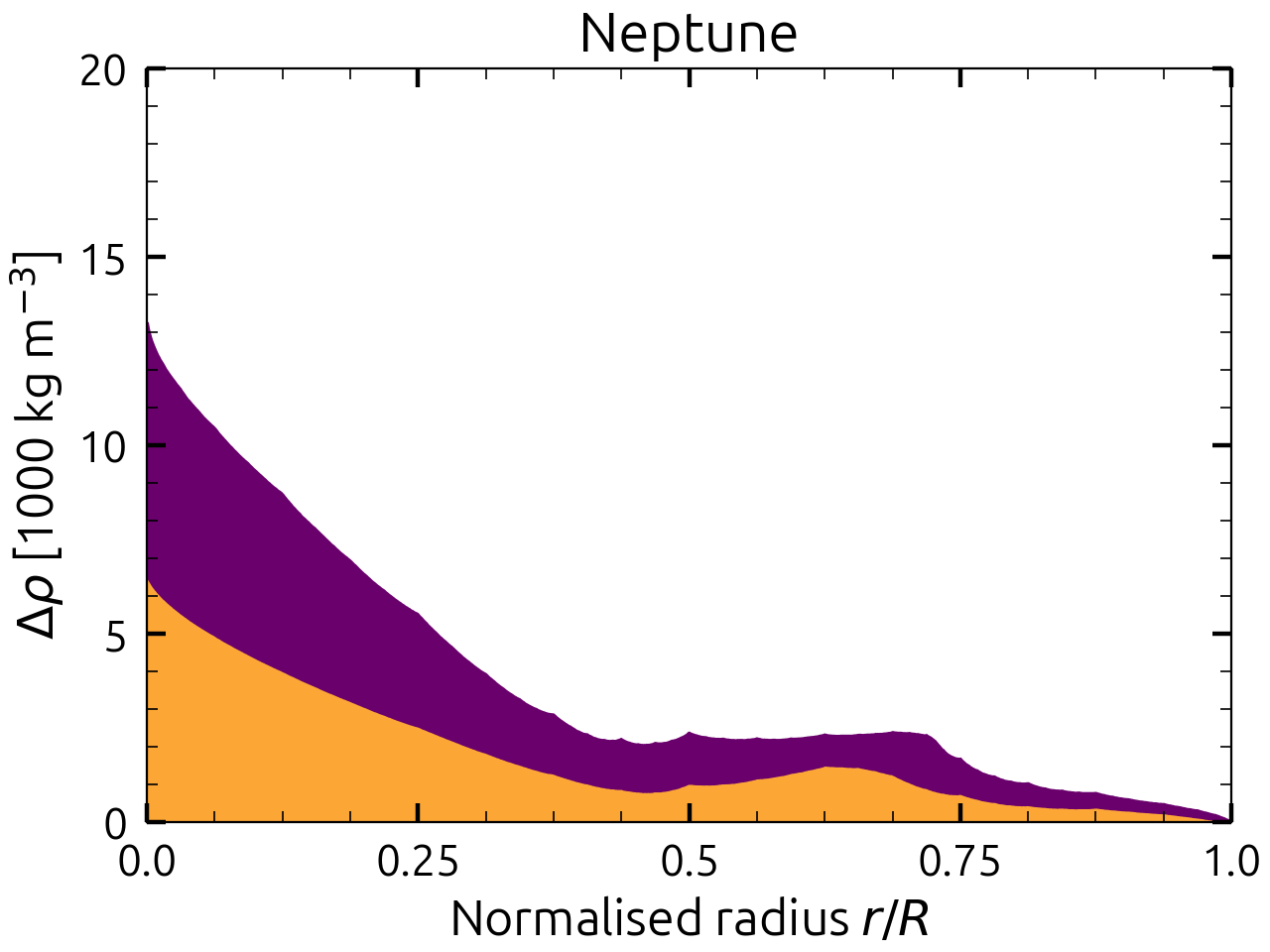}
	\caption{
    Confidence intervals.
    Size of the 68\% and 95\% confidence intervals in Figures \ref{fig:contouru} and \ref{fig:contourn} for the successful density profile distributions.
    The plot is cumulative, where the size of the 95\% confidence interval is the purple and the orange area combined.}
	\label{fig:confint}
\end{figure}

First, we present an aggregate view of the inferred profiles from the successful runs.
A run is considered to be successful if its resulting density profile did not exceed \rhomax, and it achieved the observed gravitational moments within the multivariate three standard deviation threshold.
All runs within this threshold were then weighted according to their multivariate Gaussian likelihood to recover a distribution compatible with the data of Table \ref{tab:values}.
Figure \ref{fig:Js} shows the gravitational moment distributions, and Figures \ref{fig:distru} and \ref{fig:distrn} respectively show the density profiles as a distribution in logarithmic scale and contour.
Figure \ref{fig:coredens} shows their central densities, while Figure \ref{fig:confint} shows the size of the confidence intervals (68\% and 95\%).

%%%%%%%%%%%%%%%%%%%%%%%%%%%%%%%%%%%%%%%%%%%%%%%%%%%%%%%%%%%%%%%%%%%%%%%%%%%%%%%%%%%%%%

The gravitational moment distributions shown in Figure \ref{fig:Js} demonstrate that our results cover the whole $J_2-J_4$ space faithfully, especially for Uranus.
For Neptune, the uncertainties are so large in comparison that our algorithm struggled to find a significant amount of solutions outside the multivariate threshold of one standard deviation, especially for $J_4$.
Since \cite{French2024} also provide a covariance estimate in their results, we additionally show a correlated distribution (orange) in panel \ref{fig:Jsu} for Uranus.
While the change between the uncorrelated and correlated distribution looks significant in this panel, the impact on the subsequent results is negligible.
All subsequent plots hence exclusively display results for the uncorrelated distributions to enable a fair comparison between Uranus and Neptune.
The correlated Uranus results produce visually indistinguishable plots.

%%%%%%%%%%%%%%%%%%%%%%%%%%%%%%%%%%%%%%%%%%%%%%%%%%%%%%%%%%%%%%%%%%%%%%%%%%%%%%%%%%%%%%

Figures \ref{fig:distru} and \ref{fig:distrn} show the density solution space for Uranus and Neptune, respectively. We find that successful density profiles are least constrained in the planetary deep interior. 
The outermost regions are highly constrained, with many distributions being affected by the minimum density increase.
Around $\sim$85\% of the planetary radius ($r/R=0.85$), a more diverse regime takes over for both planets.
Still, the general shape continues to show a relatively tight constriction of the possible space up to a normalised radius of $\sim0.4R$. 
This is most apparent in Figure \ref{fig:confint}.
Notably, there is a bump at around $r/R=0.65$ for Uranus (Figure \ref{fig:distru}) and $r/R=0.7$ for Neptune (Figure \ref{fig:distrn}).
In the innermost region, a large increase in the space of possible distributions develops.
The logarithmic scale is somewhat misleading, as can be seen by consulting the contour plots \ref{fig:contouru} and \ref{fig:contourn}.
Though many distributions fall outside the 95\% region (especially above it), their relative frequency is small, on the order of $10^{-4}$ or less.
The median density (as a function of radius) for Uranus and Neptune shows gradual increases in density, consistent with composition gradients.
We note that the median density shown in panels \ref{fig:contouru} and \ref{fig:contourn} is neither an actual density profile from the solution ensemble nor a maximum-likelihood indicator.

%%%%%%%%%%%%%%%%%%%%%%%%%%%%%%%%%%%%%%%%%%%%%%%%%%%%%%%%%%%%%%%%%%%%%%%%%%%%%%%%%%%%%%

Next we compare our results to previous work, focusing first on models in the literature that are based on equations of state and hence infer composition and temperature alongside density (and pressure).
We find that the central densities of profiles from \cite{Helled2011} and \cite{Vazan2020} are concentrated around the peak of the distribution, not its median.
Profiles from \cite{Nettelmann2013} have core densities higher than the 84th percentile.
The more recent work by \cite{Militzer2024} is similar, predicting core densities that are close to those inferred by \cite{Nettelmann2013}. 
The Uranus models U1 and U3 with extensive composition gradients from \cite{Morf2025} are close to the median result.
Their convective models U2 and U4 are more similar to \cite{Vazan2020} and are situated at the peak of the distribution.
We note that although their U2 and U4 profiles possess the same comparatively low central density of $\sim 4.5$ g/cm$^3$, their inferred total rock-to-water mass fractions vary significantly (0.64 vs. 3.92).
We can therefore conclude that the density-pressure relation alone is insufficient for determining the planetary bulk composition.
Interestingly, the N1 profile for Neptune by \cite{Morf2025} falls outside the 95th percentile with a significantly higher central density.
This occurs because their approach can also probe less standard solutions.
On the other hand, their N3 solution is close to the median result.

%%%%%%%%%%%%%%%%%%%%%%%%%%%%%%%%%%%%%%%%%%%%%%%%%%%%%%%%%%%%%%%%%%%%%%%%%%%%%%%%%%%%%%

Studies that, as in this work, analyse large ensembles of interior profiles enable a more direct and systematic comparison.
For example, Figure 4 in \cite{Helled2020}, Figure 10 in \cite{Movshovitz2022}, and Figure 2 in \cite{Neuenschwander2022} are directly comparable to Figures \ref{fig:contouru} and \ref{fig:contourn}.
\cite{Helled2020} employed eighth-order polynomials, and \cite{Movshovitz2022} additionally incorporated sigmoid functions.
They obtained profiles with generally higher densities in the deep interior for both Uranus and Neptune, especially in the case of \cite{Helled2020}, below $0.3R$.
Their median profiles also flatten off below $0.1R$, whereas our results show no such trend.
Our medians maintain a consistent slope below $0.3R$ and even curve up slightly.
Moreover, the distribution of \cite{Helled2020} is symmetric around the median, whereas our results show a wider spread for higher-density solutions.
The inclusion of sigmoid terms in \cite{Movshovitz2022} produced extended flat regions that do not appear in our models due to the imposed minimum density increment per layer, $\Delta_{\min}$.
In the work of \cite{Neuenschwander2022}, interior structures were constructed using three consecutive polytropes.
This imprinted distinct features on the confidence intervals, including sharp discontinuities and narrow choke points, which are absent in our results.
Similar to our results, their approach yielded a distribution with a wider spread around the median for higher central densities.

%%%%%%%%%%%%%%%%%%%%%%%%%%%%%%%%%%%%%%%%%%%%%%%%%%%%%%%%%%%%%%%%%%%%%%%%%%%%%%%%%%%%%%

%%%%%%%%%%%%%%%%%%%%%%%%%%%%%%%%%%%%%%%%%%%%%%%%%%%%%%%%%%%%%%%%%%%%%%%%%%%%%%%%%%%%%%

The wider spread for higher central densities is further illustrated in Figure \ref{fig:coredens}. 
We find that the central densities are very unequally distributed, rising to a peak rapidly and gradually descending as density increases for both planets.
For Uranus, the peak is very tight at $\rho \sim$ 4500 kg\,m$^{-3}$, but the median is still somewhat higher.
The solutions for Neptune are distributed more broadly.
Hence, a significant portion of solutions exceeded the imposed maximum density \rhomax and were excluded.
The peak of the distribution is at $\rho \sim$ 5200 kg\,m$^{-3}$.

%%%%%%%%%%%%%%%%%%%%%%%%%%%%%%%%%%%%%%%%%%%%%%%%%%%%%%%%%%%%%%%%%%%%%%%%%%%%%%%%%%%%%%

%%%%%%%%%%%%%%%%%%%%%%%%%%%%%%%%%%%%%%%%%%%%%%%%%%%%%%%%%%%%%%%%%%%%%%%%%%%%%%%%%%%%%%

Figure \ref{fig:confint} reflects the span of our results. 
The uncertainty of the gravitational moment measurements of Neptune is a factor of ten larger than that of Uranus (Table \ref{tab:values}).
One could therefore expect that our results are considerably less constrained for Neptune as well.
However, Figure \ref{fig:confint} shows that this is not the case.
Our density profile results are only slightly more widely spread for Neptune than they are for Uranus.
Tighter bounds in the $J_2$-$J_4$ space do not linearly translate to tighter bounds in the density solution space.
Even if the $J_2$ and $J_4$ values were perfectly known, the density profile solution space would remain degenerate.
Only a measurement of all the gravitational moments $\left\{J_{2n} \mid n\in\mathbb{N} \right\}$ would provide a unique solution for a north-south and azimuthally symmetric planet.

%%%%%%%%%%%%%%%%%%%%%%%%%%%%%%%%%%%%%%%%%%%%%%%%%%%%%%%%%%%%%%%%%%%%%%%%%%%%%%%%%%%%%%

Finally, we note that in the contour (Figures \ref{fig:contouru} and \ref{fig:contourn}) and confidence interval (Figure \ref{fig:confint}) plots, small `spikes' are visible concentrated around inverses of multiples of two, most notably at $r/R=0.5$.
These are artefacts of the initial density profile generation method and are addressed in Section \ref{sec:discussion} as a caveat.

%%%%%%%%%%%%%%%%%%%%%%%%%%%%%%%%%%%%%%%%%%%%%%%%%%%%%%%%%%%%%%%%%%%%%%%%%%%%%%%%%%%%%%

\subsection{Derived quantities}

%%%%%%%%%%%%%%%%%%%%%%%%%%%%%%%%%%%%%%%%%%%%%%%%%%%%%%%%%%%%%%%%%%%%%%%%%%%%%%%%%%%%%%

\begin{figure}
	\centering
	\includegraphics[width=0.98\columnwidth]{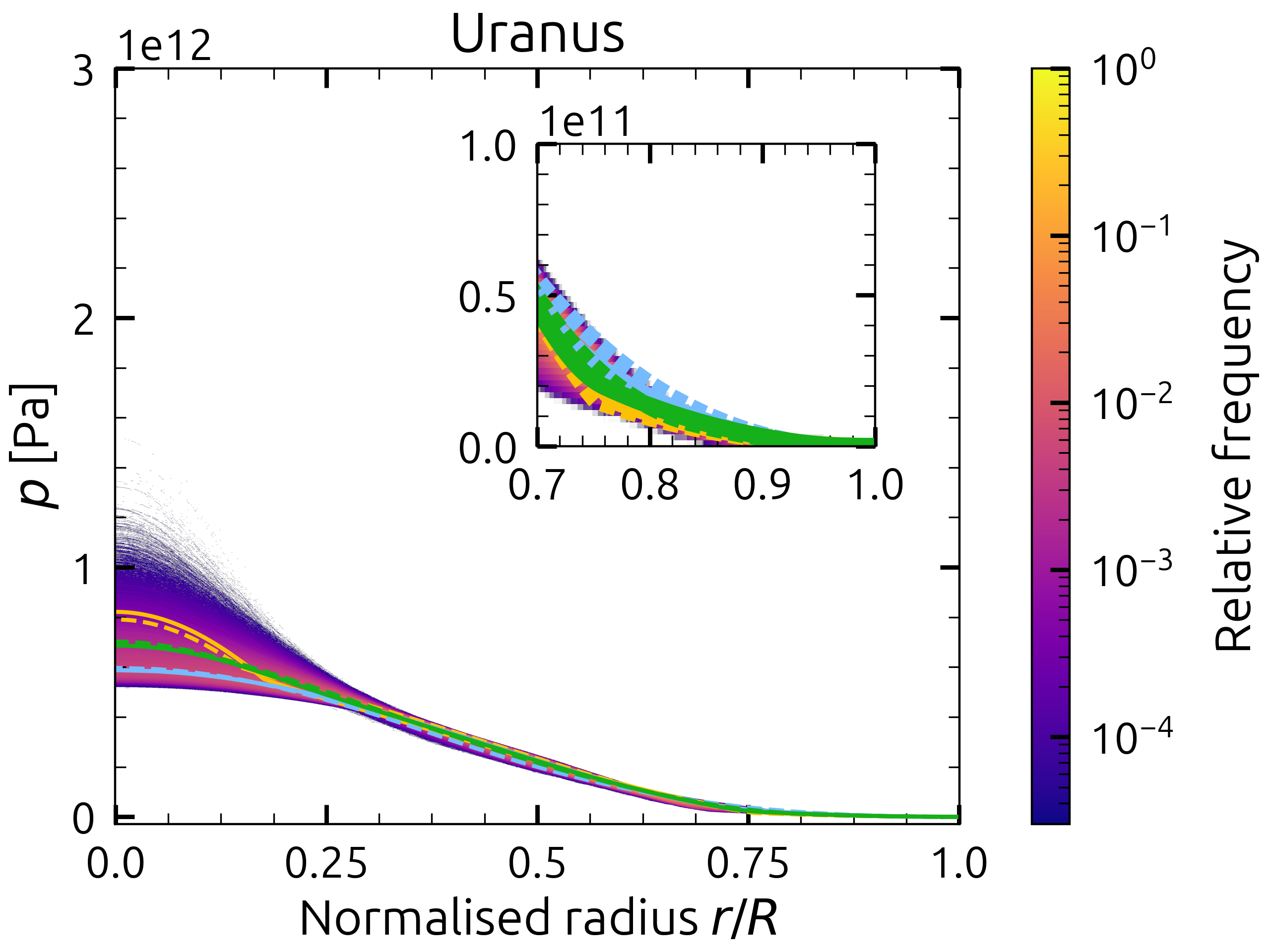}
	\includegraphics[width=0.98\columnwidth]{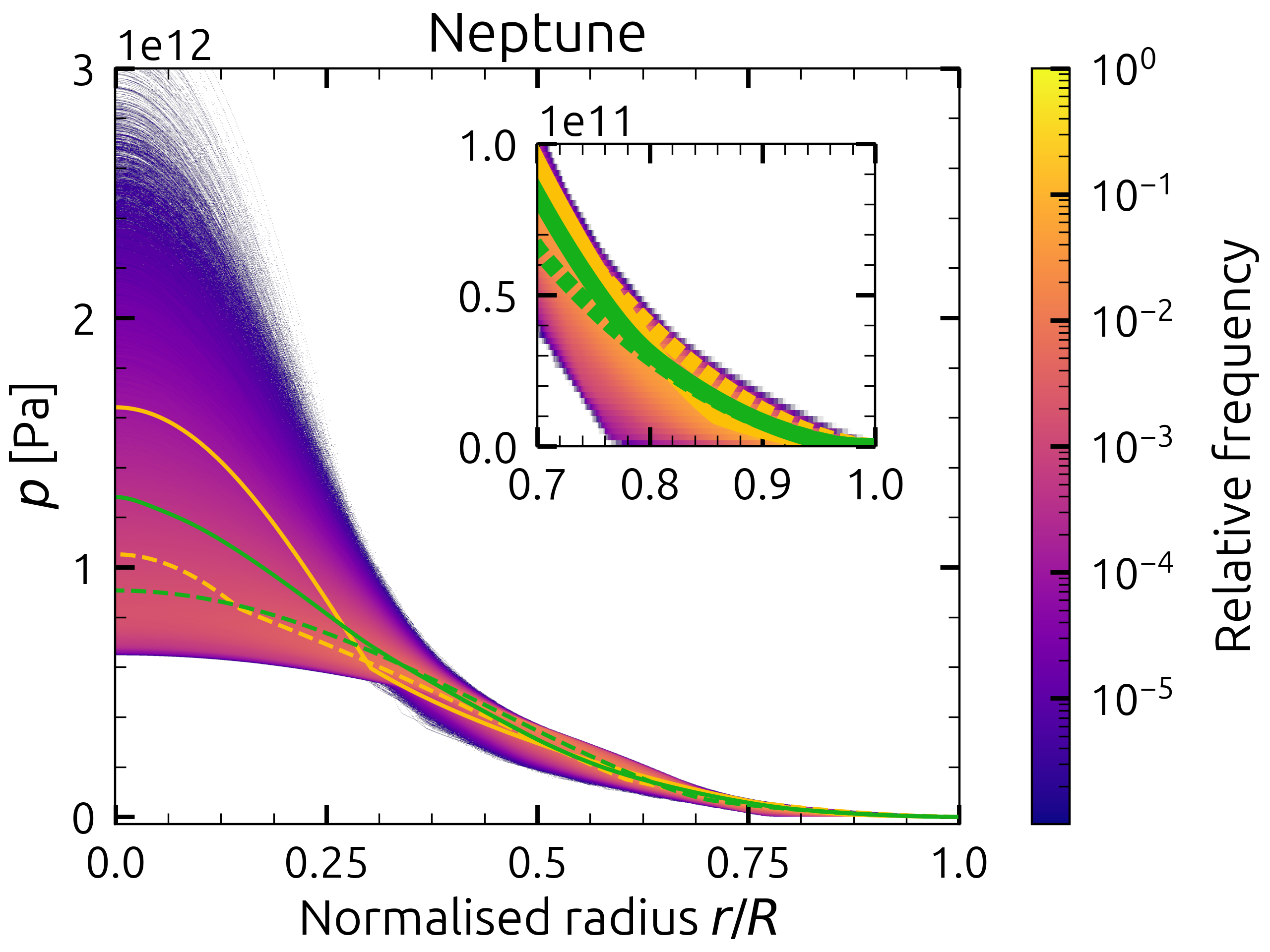}
	\caption{
    Pressure profiles.
    Weighted distribution of pressure profiles derived from all successful density profiles.
    The inset shows a zoomed-in view of the outermost region, from $r/R=1$ to $r/R=0.7$.
    Note that the zoom is not aspect-preserving.
    The colour scale shows the relative frequency over all distributions of a certain pressure value at a given radius.
    The coloured lines are as before in Figures \ref{fig:densprofu} and \ref{fig:densprofn}.
    }
	\label{fig:pdistr}
\end{figure}

%%%%%%%%%%%%%%%%%%%%%%%%%%%%%%%%%%%%%%%%%%%%%%%%%%%%%%%%%%%%%%%%%%%%%%%%%%%%%%%%%%%%%%

\begin{figure}
	\centering
	\includegraphics[width=0.833\columnwidth]{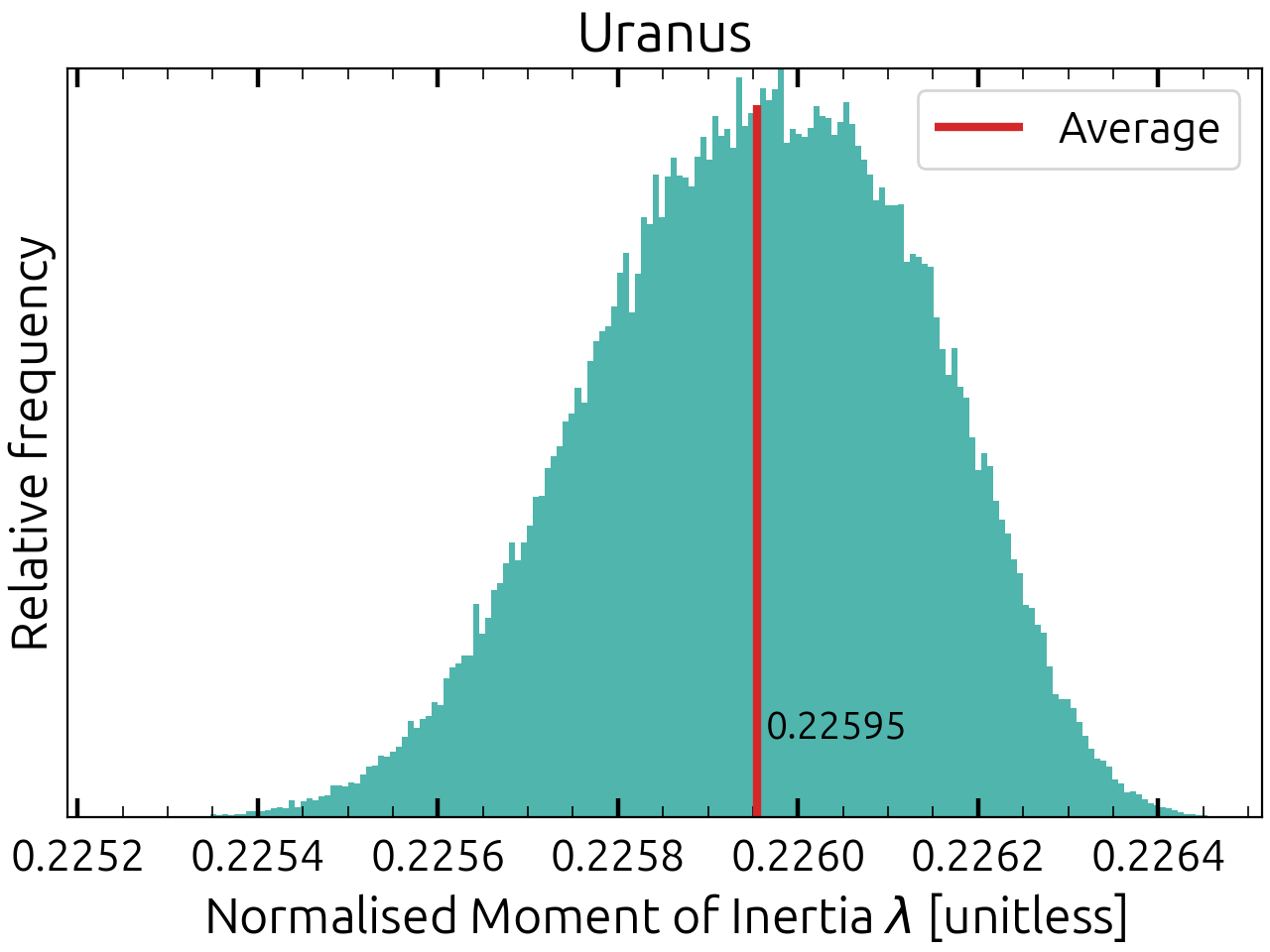}
	\includegraphics[width=0.833\columnwidth]{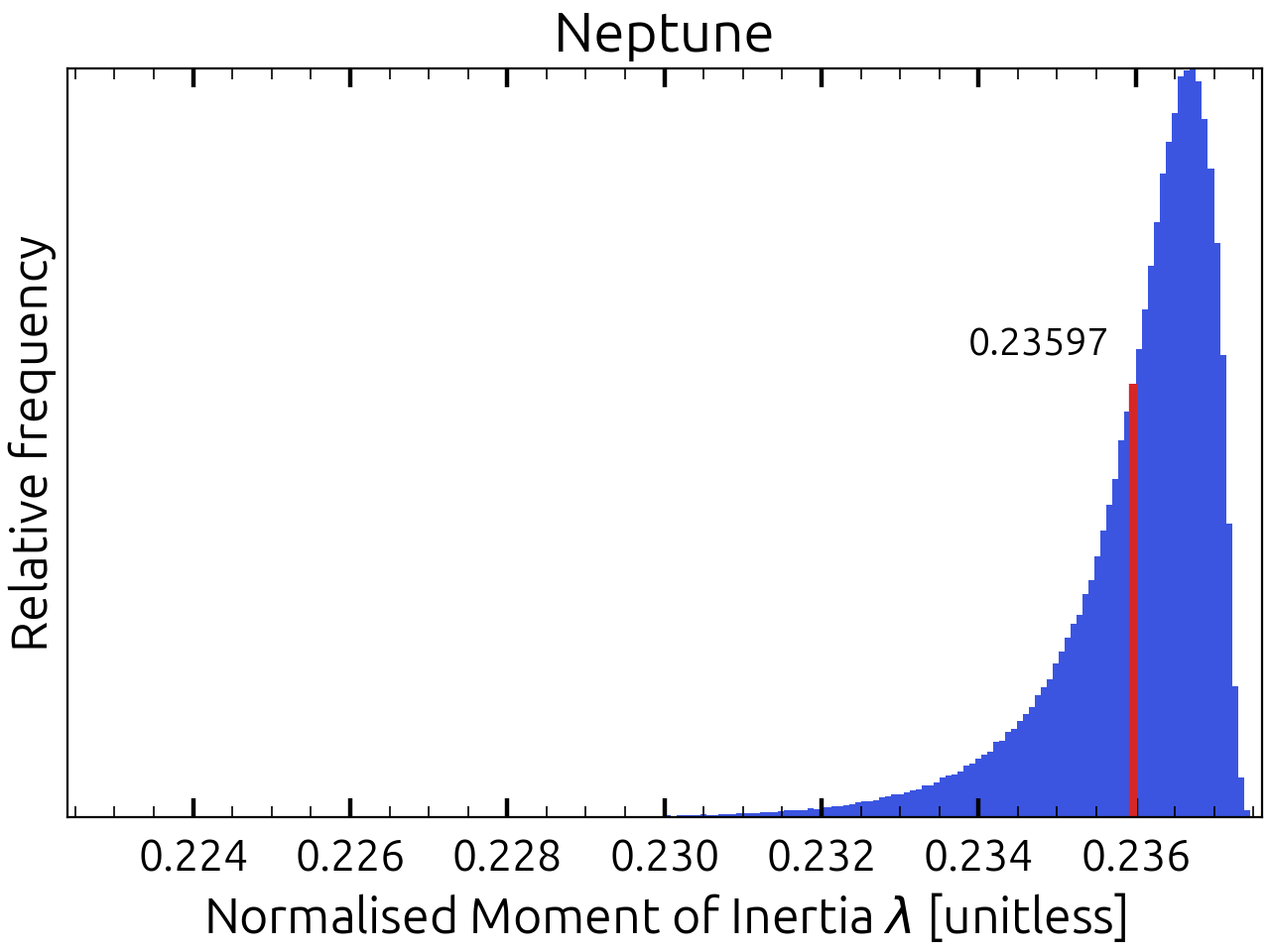}
	\caption{
    Normalised moments of inertia.
    Weighted distribution of the normalised moments of inertia of successful density profiles.
    The red line denotes the average.
    }
	\label{fig:nmoi}
\end{figure}

%%%%%%%%%%%%%%%%%%%%%%%%%%%%%%%%%%%%%%%%%%%%%%%%%%%%%%%%%%%%%%%%%%%%%%%%%%%%%%%%%%%%%%

\begin{figure}
	\centering
	\includegraphics[width=0.833\columnwidth]{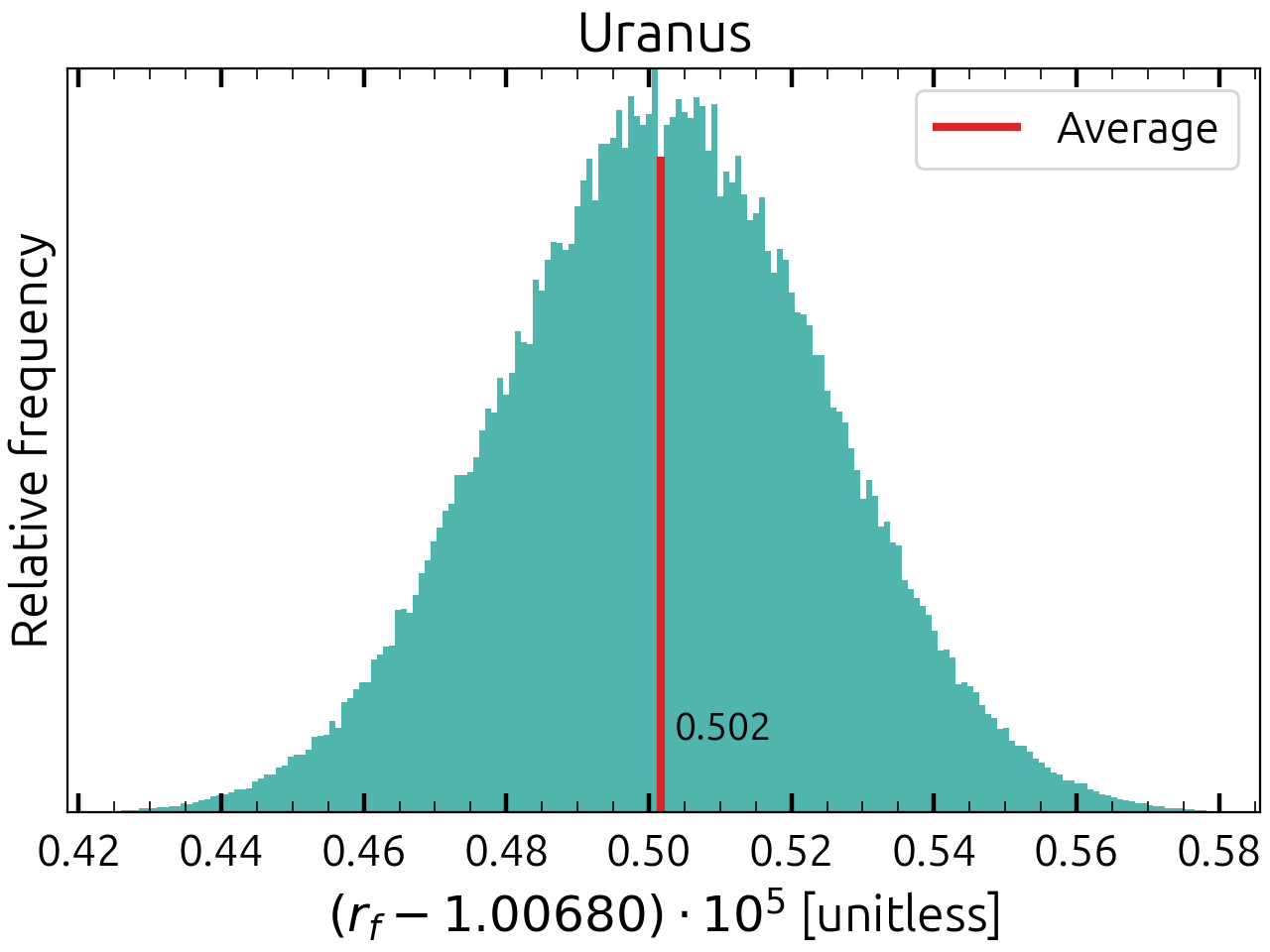}
	\includegraphics[width=0.833\columnwidth]{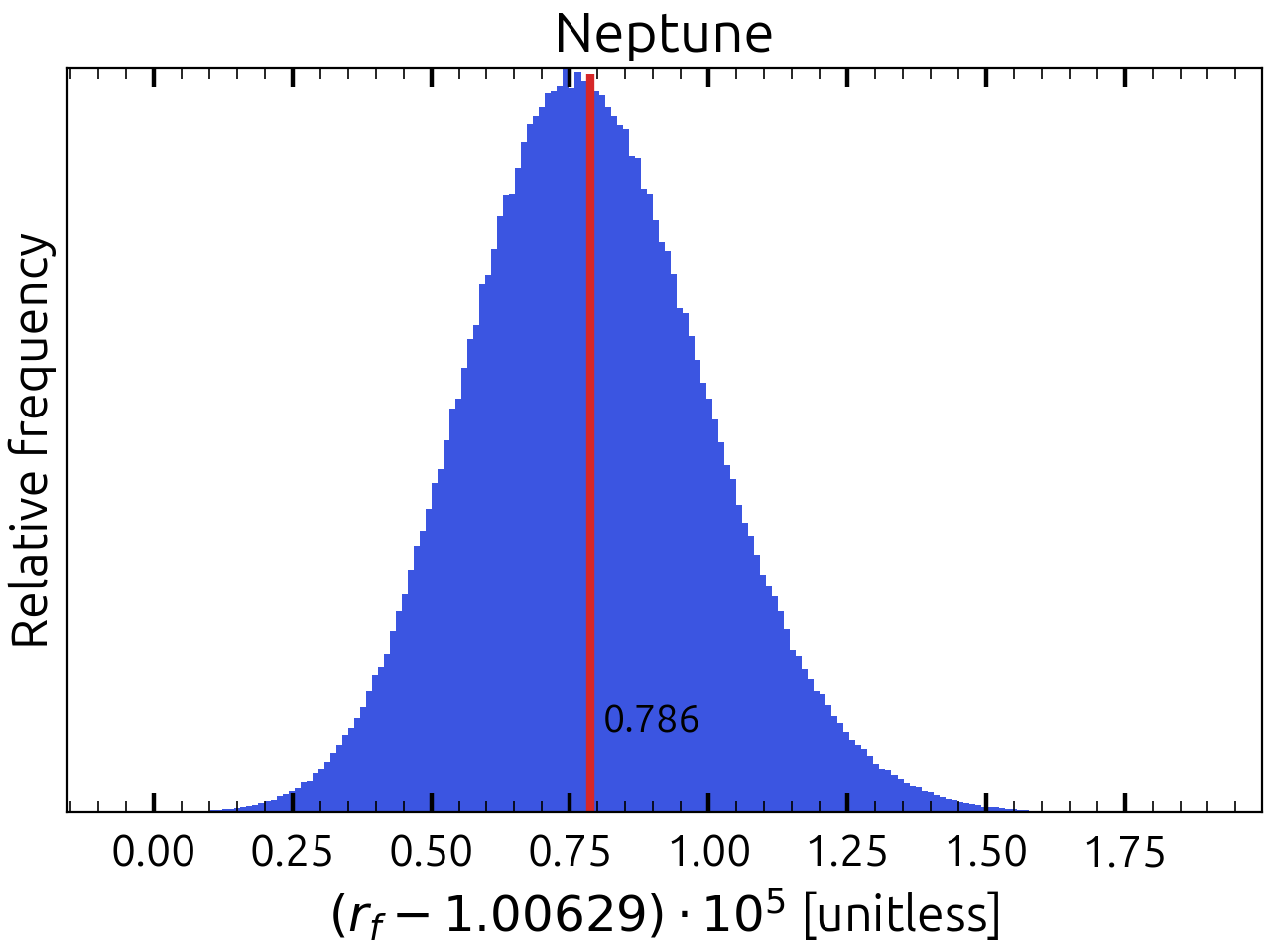}
	\caption{
    Flattening ratios.
    Weighted distribution of the flattening ratios of successful density profiles.
    Note the scale is given as offsets from the constant 1.00680 for Uranus and 1.00629 for Neptune. 
    }
	\label{fig:fr}
\end{figure}

%%%%%%%%%%%%%%%%%%%%%%%%%%%%%%%%%%%%%%%%%%%%%%%%%%%%%%%%%%%%%%%%%%%%%%%%%%%%%%%%%%%%%%

In this Section, we present quantities derived from the successful density profiles via the ToF.
Figure \ref{fig:pdistr} shows the distribution of pressure profiles with a logarithmic frequency scale.
Figure \ref{fig:nmoi} shows the distribution of the normalised moments of inertia, and Figure \ref{fig:fr} shows the distribution of the flattening ratios.

%%%%%%%%%%%%%%%%%%%%%%%%%%%%%%%%%%%%%%%%%%%%%%%%%%%%%%%%%%%%%%%%%%%%%%%%%%%%%%%%%%%%%%

We find that the pressure profiles in Figure \ref{fig:pdistr} agree well with previous studies and are more constrained for Uranus compared to Neptune.
The pressure profiles were calculated from the density profiles under the assumption of hydrostatic equilibrium, $\nabla p = \rho \nabla U$, where $U$ is the total (gravitational and rotational) potential of the planet. 

%%%%%%%%%%%%%%%%%%%%%%%%%%%%%%%%%%%%%%%%%%%%%%%%%%%%%%%%%%%%%%%%%%%%%%%%%%%%%%%%%%%%%%

The normalised moments of inertia depicted in Figure \ref{fig:nmoi} are defined as
$\lambda = I/(m R_\text{mean}^2)$, where $I$ is the axial moment of inertia,
\begin{equation} I = \int_V \rho r^2_\perp \mathrm{d}V, \end{equation}
and $r_\perp$ denotes the perpendicular radial component with respect to the given axis, usually the axis of rotation. Further,
$R_\text{mean}$ is equal to $l_\text{max}$, the outermost spheroid volumetric mean radius (see Section \ref{sec:methods}).
We find that our average value for Uranus of 0.22595 agrees well with previous findings from \cite{Neuenschwander2022}, who found a range of 0.22594--0.22670.
Our average value for Neptune of 0.23597 is lower than their finding of 0.23727--0.23900 for the standard model, though our more frequent values agree with the lower end of their range.

%%%%%%%%%%%%%%%%%%%%%%%%%%%%%%%%%%%%%%%%%%%%%%%%%%%%%%%%%%%%%%%%%%%%%%%%%%%%%%%%%%%%%%

The flattening ratios depicted in Figure \ref{fig:fr} are defined as the ratio between the equatorial and mean radii: 
\begin{equation}r_f = \frac{R_\text{eq}}{R_\text{mean}}.\end{equation}
To compare our findings, we calculated the equatorial-to-mean volumetric radii ratios based on the values from \cite{Seidelmann2007}. 
They are $(25559 \pm 4)/(25362 \pm 7)$ for Uranus and $(24764 \pm 15)/(24622 \pm 19)$ for Neptune. 
The resulting flattening ratios are $r_f=1.0078 \pm 0.0003$ and $r_f=1.0058 \pm 0.0010$ for Uranus and Neptune, respectively.  
Our inferred average value of 1.0068 in Figure \ref{fig:fr} for Uranus is lower compared to \cite{Seidelmann2007}, while our average of 1.0063 for Neptune is within their uncertainties. 
Our ToF implementation is benchmarked against \cite{Wisdom2016}, which shows that our flattening ratios should be accurate up to a relative error of $10^{-3}-10^{-4}$.
This is significant given the uncertainties from \cite{Seidelmann2007} for Uranus and can therefore explain this deviation.
Furthermore, the uncertainties given by \cite{Seidelmann2007} are quite conservative, especially given the uncertain rotation rates of these planets \citep[for example][]{Helled2010}.

%%%%%%%%%%%%%%%%%%%%%%%%%%%%%%%%%%%%%%%%%%%%%%%%%%%%%%%%%%%%%%%%%%%%%%%%%%%%%%%%%%%%%%

\subsection{Discontinuities}
\label{subsec:jumps}

%%%%%%%%%%%%%%%%%%%%%%%%%%%%%%%%%%%%%%%%%%%%%%%%%%%%%%%%%%%%%%%%%%%%%%%%%%%%%%%%%%%%%%

\begin{figure*}
    \centering

    \begin{subfigure}{0.5\textwidth}
        \centering
        \includegraphics[width=\linewidth]{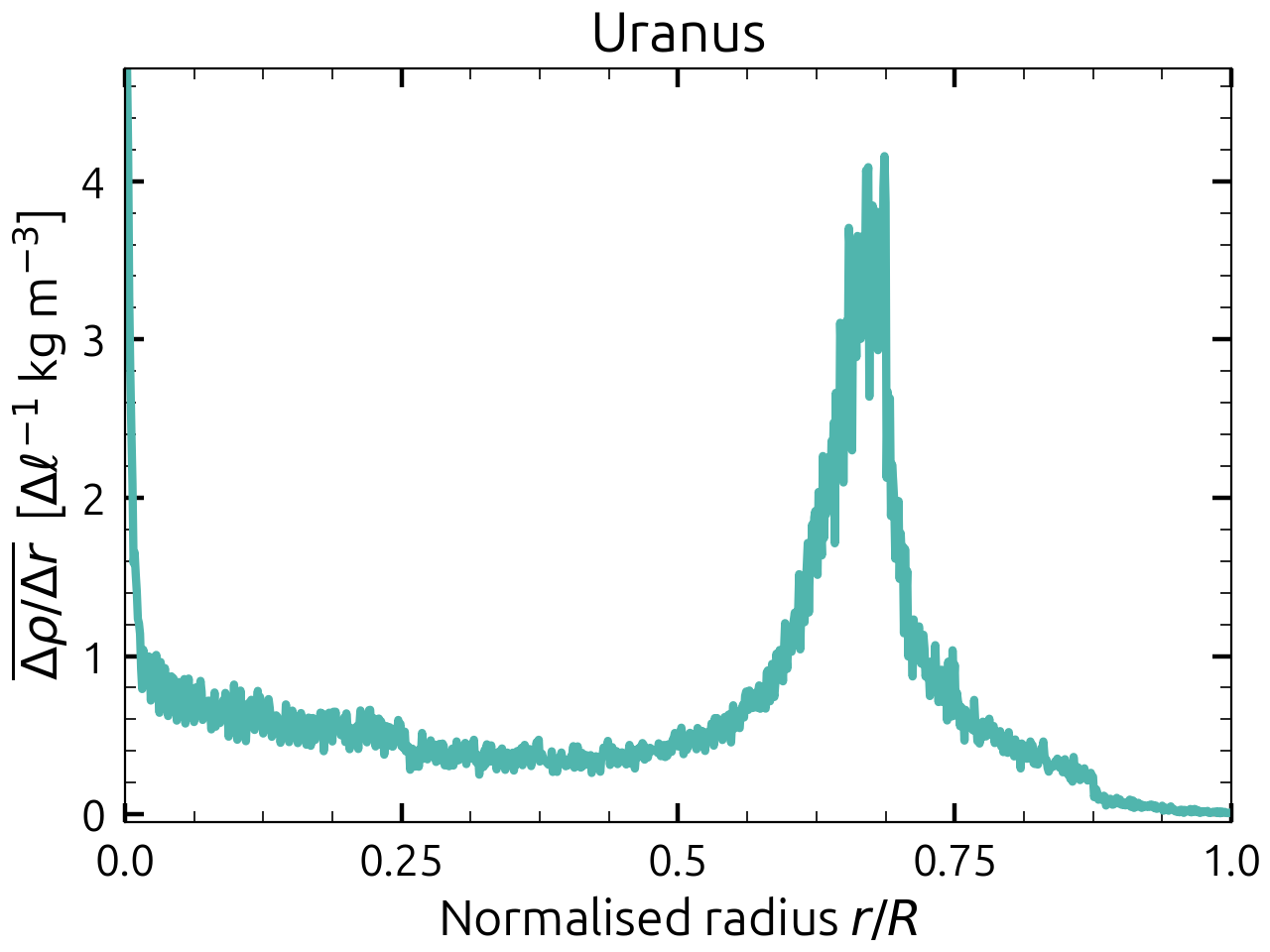}
        \caption{}
        \label{fig:jumpsu}
    \end{subfigure}%
    \begin{subfigure}{0.5\textwidth}
        \centering
        \includegraphics[width=\linewidth]{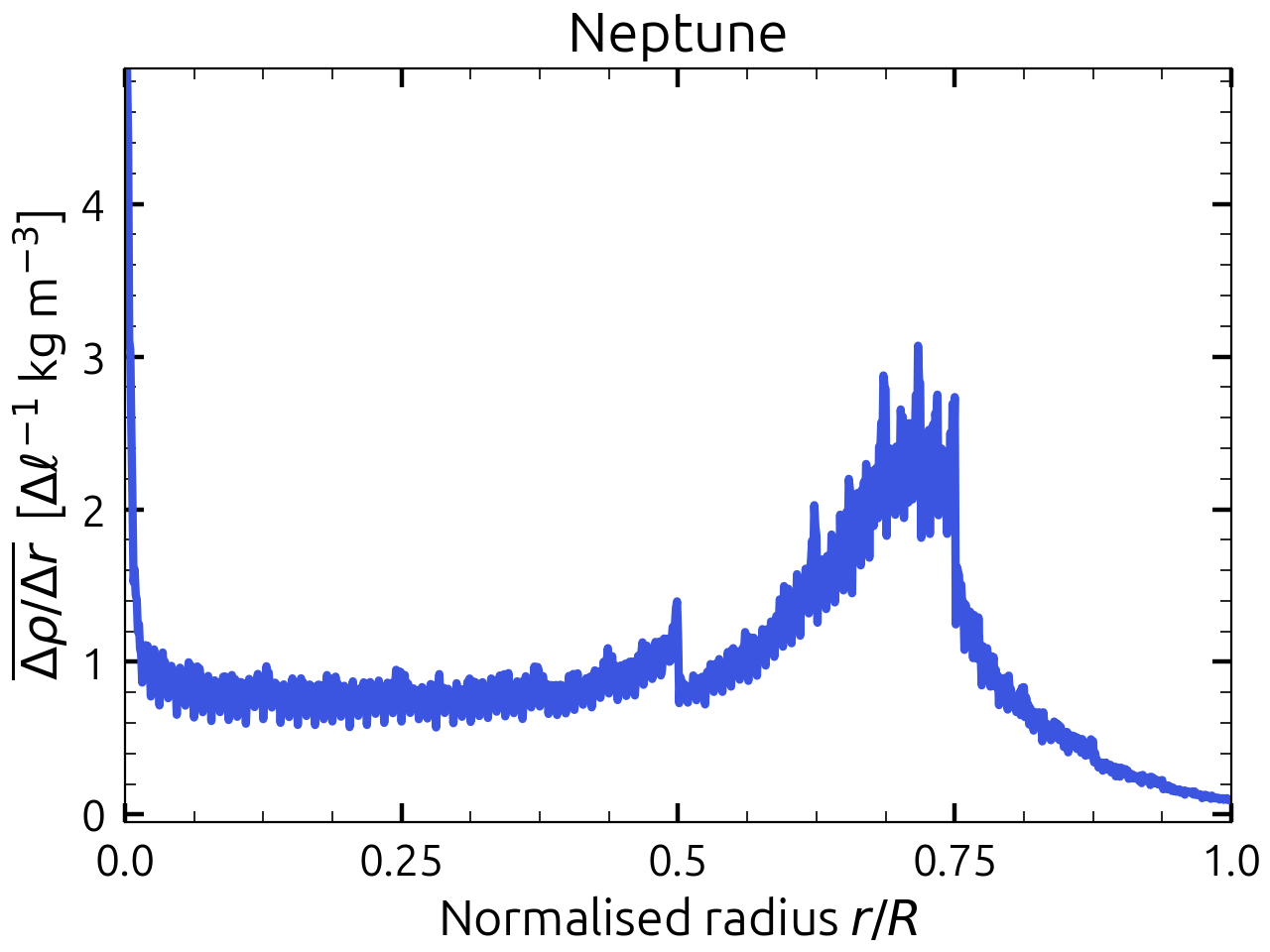}
        \caption{}
        \label{fig:jumpsn}
    \end{subfigure}

    \caption{
        Weighted discontinuity `intensity' for Uranus (\subref{fig:jumpsu}) and Neptune (\subref{fig:jumpsn}).
        The figures depict the average height, $\overline{\Delta\rho}$, divided by the average width, $\overline{\Delta r}$, of the discontinuities.
        For example, Uranus discontinuities with an average location of $r/R=0.5$ possess $\overline{\Delta\rho/\Delta r}\sim 0.5 \Delta\ell^{-1}$ kg\,m$^{-3}$.
        Profiles with no discontinuities also contribute to the averages with a value zero.
    }
    \label{fig:jumps}
\end{figure*}

%%%%%%%%%%%%%%%%%%%%%%%%%%%%%%%%%%%%%%%%%%%%%%%%%%%%%%%%%%%%%%%%%%%%%%%%%%%%%%%%%%%%%%

In this section, we present an analysis on discontinuities of successful density profiles.
First, a `discontinuity criterion' ($c_D$) must be chosen in order to distinguish discontinuities from regular slopes.
Unless otherwise specified, all results were obtained with $c_D=$ 100 $\Delta \ell^{-1}$ kg\,m$^{-3}$.
As a reminder, $\Delta \ell$ is equal to 24960 m for Uranus and 24186 m for Neptune (rounded to the nearest meter).
Therefore, $c_D$ has a value of 4.01 $\cdot$ 10$^{-3}$ kg\,m$^{-4}$ for Uranus and 4.13 $\cdot$ 10$^{-5}$ kg\,m$^{-4}$ for Neptune (rounded to two significant figures).
As a second distinguishing feature, we only considered discontinuities that are `minimally maximal'.
A precise definition of the term minimally maximal is given in Appendix \ref{app:jumps}.
In a nutshell, minimally maximal specifies that the (radial) width of a discontinuity can neither be extended (because it would add sections of the density profile below $c_D$) nor shortened (because it would remove sections of the density profile exceeding $c_D$).

%%%%%%%%%%%%%%%%%%%%%%%%%%%%%%%%%%%%%%%%%%%%%%%%%%%%%%%%%%%%%%%%%%%%%%%%%%%%%%%%%%%%%%

Figure \ref{fig:jumps} shows the average discontinuity `intensity' as a function of radius for all successful density profiles for both Uranus and Neptune.
More precisely, it shows the average discontinuity height, $\overline{\Delta\rho}$, divided by the average discontinuity width, $\overline{\Delta r}$, for discontinuities that are minimally maximal and in agreement with $c_D$ at their respective average normalised radius, $r/R$.
For example, Figure \ref{fig:jumpsu} shows that minimally maximal and $c_D$-consistent discontinuities in Uranus with an average location of $r/R=0.5$ possess an average slope of $\overline{\Delta\rho/\Delta r}\sim 0.5 \Delta\ell^{-1}$ kg\,m$^{-3}$.
We note that profiles that have no minimally maximal and $c_D$-consistent discontinuity at any given location also contribute to the averages with a discontinuity intensity value of zero.
Separately, we show $\overline{\Delta\rho}$ and $\overline{\Delta r}$ at their respective average normalised radii, $r/R$, in Figures \ref{fig:jumpsdrhodru} for Uranus and \ref{fig:jumpsdrhodrn} for Neptune. 

%%%%%%%%%%%%%%%%%%%%%%%%%%%%%%%%%%%%%%%%%%%%%%%%%%%%%%%%%%%%%%%%%%%%%%%%%%%%%%%%%%%%%%

\begin{figure}
	\centering
	\includegraphics[width=0.833\columnwidth]{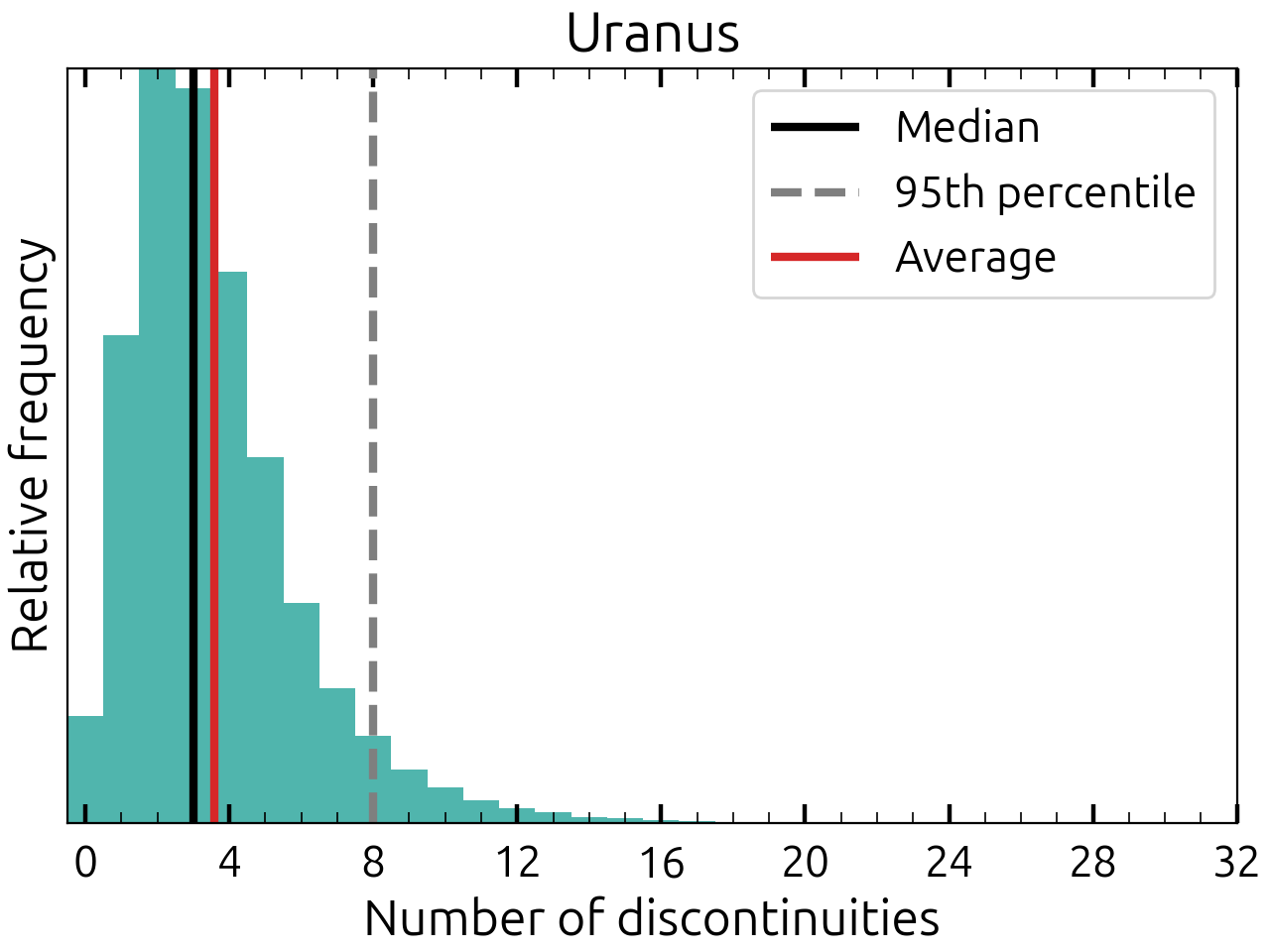}
	\includegraphics[width=0.833\columnwidth]{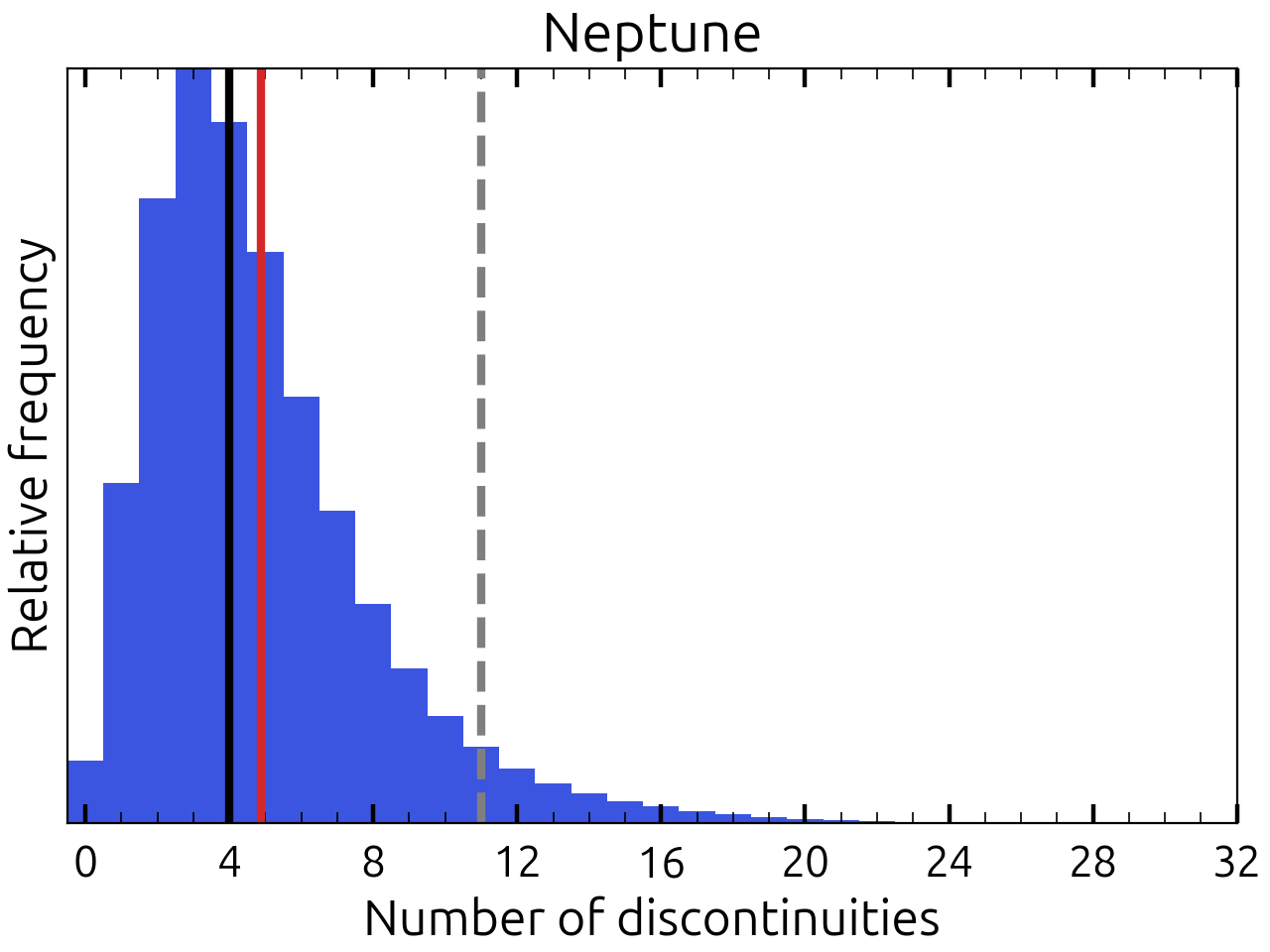}
	\caption{
    Number of discontinuities.
    Weighted distribution of the total number of discontinuities detected in a successful density profile for $c_D=$ 100 $\Delta \ell^{-1}$ kg\,m$^{-3}$.
    }
	\label{fig:njumps}
\end{figure}

%%%%%%%%%%%%%%%%%%%%%%%%%%%%%%%%%%%%%%%%%%%%%%%%%%%%%%%%%%%%%%%%%%%%%%%%%%%%%%%%%%%%%%

For Uranus (Figure \ref{fig:jumpsu}), we observed a discontinuity intensity peak at $\sim0.65R$, which explains the bump in the density profile distribution seen in panel \ref{fig:densprofu}.
A discontinuity in this location could mark the transition point between the outer atmosphere dominated by hydrogen and helium and deeper layers dominated by, for example, water or rocks. 
There are no other significant discontinuity regions, although the innermost region tends to have bigger slopes in general, and thus the discontinuity intensity also increases inward.
In the same plot for Neptune (\ref{fig:jumpsn}) the effects are reduced in visibility, but the same general trends exist, with the peak being located slightly further out at $\sim0.7R$.
We note that here as well artefacts of the initial density profile generation method are visible, but this time as small cliffs around the inverses of multiples of two (we discuss this further in Section \ref{sec:discussion}).

%%%%%%%%%%%%%%%%%%%%%%%%%%%%%%%%%%%%%%%%%%%%%%%%%%%%%%%%%%%%%%%%%%%%%%%%%%%%%%%%%%%%%%

\begin{figure}
	\centering
	\includegraphics[width=0.833\columnwidth]{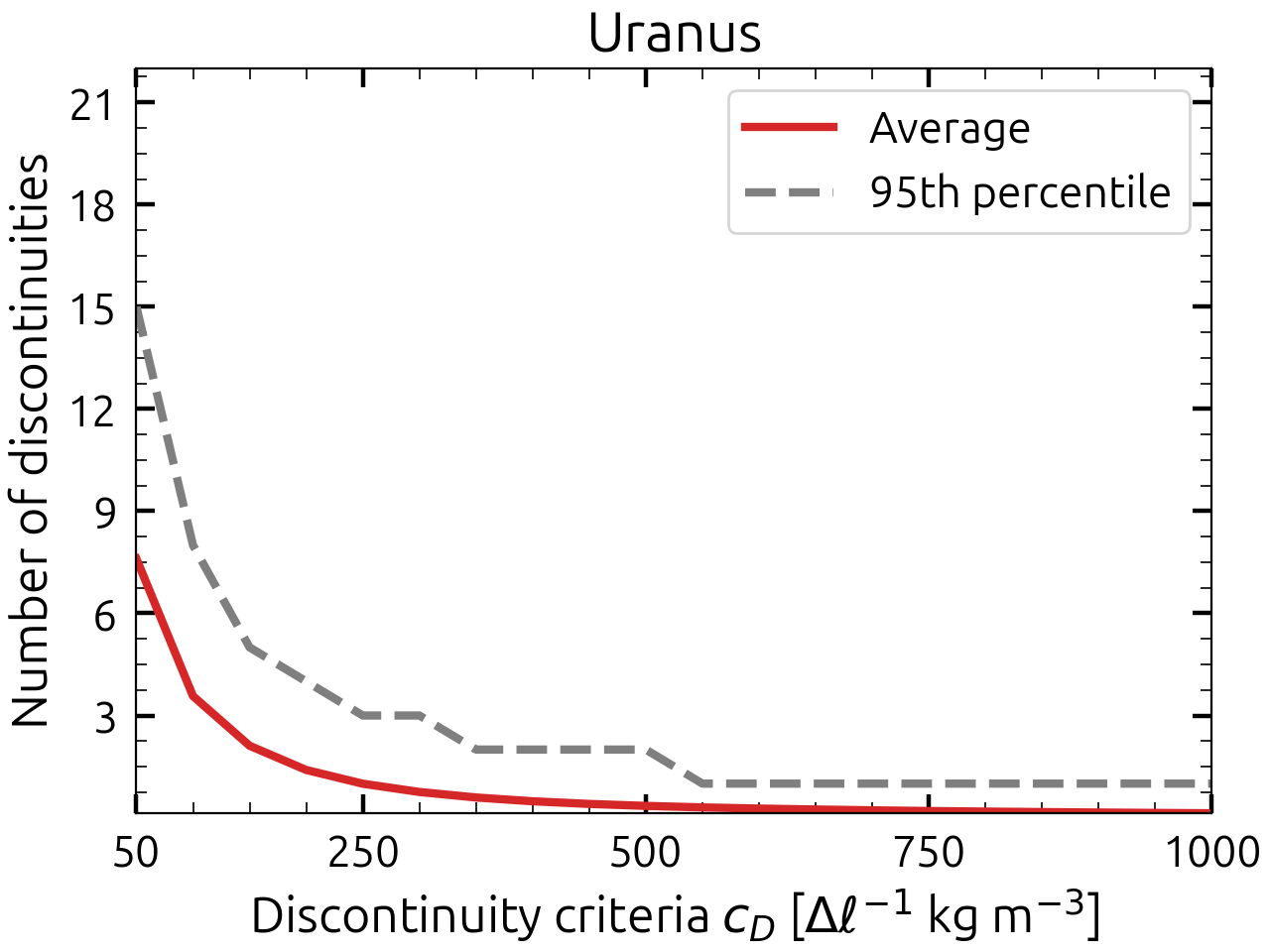}
	\includegraphics[width=0.833\columnwidth]{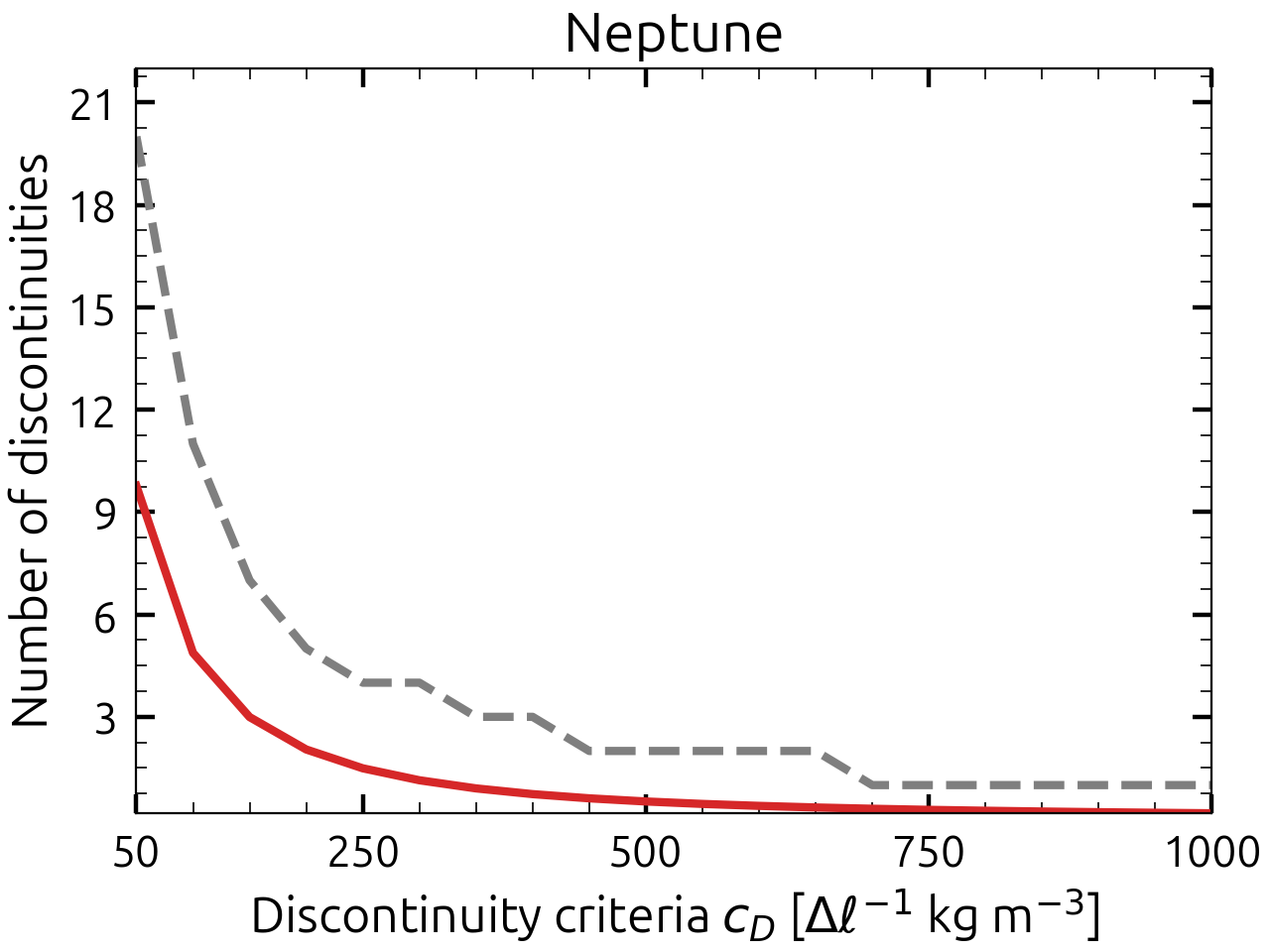}
	\caption{
    Number of discontinuities versus $c_D$.
    The real valued average (red) and the integer valued upper 95th percentile (black) are depicted.
    All criteria from 50 $\Delta \ell^{-1}$ kg\,m$^{-3}$ to 1000 $\Delta \ell^{-1}$ kg\,m$^{-3}$ in steps of 50 $\Delta \ell^{-1}$ kg\,m$^{-3}$ were considered.
    }
	\label{fig:jumpspercrit}
\end{figure}

%%%%%%%%%%%%%%%%%%%%%%%%%%%%%%%%%%%%%%%%%%%%%%%%%%%%%%%%%%%%%%%%%%%%%%%%%%%%%%%%%%%%%%

Modellers often assume a fixed number of density discontinuities \citep[for example][]{Nettelmann2013, Neuenschwander2022, Morf2024, Militzer2024}.
However, this leads to the question of how many of these discontinuities should be chosen to account for all possible density profiles (given a certain criterion $c_D$).
To answer this, we analysed the number of discontinuities compatible with a given $c_D$.
Figure \ref{fig:njumps} shows the distribution of the total number of (minimally maximal and $c_D$-consistent) discontinuities in a given profile over all successful density profiles. 
We find that the number of discontinuities in one solution can be quite large, particularly in comparison to the above-mentioned published models, which set the number of discontinuities to three.
However, the results in Figure \ref{fig:njumps} clearly depend on the discontinuity criterion, $c_D$.
We hence investigated the influence of $c_D$ on the number of discontinuities in Figure \ref{fig:jumpspercrit}.
We show the average and upper 95th percentile number of discontinuities detected per successful density profile as a function of the criterion $c_D$.
In addition, Table \ref{tab:appjumpspercrit} lists all values from this figure numerically.
Figure \ref{fig:jumpspercrit} and Table \ref{tab:appjumpspercrit} should be read as follows: 
If we consider, for example, $c_D=$ 100 $\Delta \ell^{-1}$ kg\,m$^{-3}$ to be a discontinuity, then the average number of discontinuities on Uranus is four, and 95\% of all solutions have eight discontinuities or fewer.
Alternatively, the statement `95\% of Uranian density profiles have at most one discontinuity' is only correct if we set $c_D \gtrsim$ 550 $\Delta \ell^{-1}$ kg\,m$^{-3}$. 
Finally, we note that we consistently tend to find slightly more jumps for Neptune than Uranus, probably due to the larger uncertainty in Neptune's gravity data, which allows for a larger parameter space of solutions. 

%%%%%%%%%%%%%%%%%%%%%%%%%%%%%%%%%%%%%%%%%%%%%%%%%%%%%%%%%%%%%%%%%%%%%%%%%%%%%%%%%%%%%%
%%%%%%%%%%%%%%%%%%%%%%%%%%%%%%%%%%%%%%%%%%%%%%%%%%%%%%%%%%%%%%%%%%%%%%%%%%%%%%%%%%%%%%

\section{Discussion}
\label{sec:discussion}

%%%%%%%%%%%%%%%%%%%%%%%%%%%%%%%%%%%%%%%%%%%%%%%%%%%%%%%%%%%%%%%%%%%%%%%%%%%%%%%%%%%%%%

We begin by noting that our results depend on the measured values listed in Table \ref{tab:values}.
Some of these values remain uncertain. The rotation rates and the contributions of zonal winds, for example, remain unclear.
Such fundamental uncertainties persist regardless of the method used.
Below we focus on the limitations exclusive to our approach.

%%%%%%%%%%%%%%%%%%%%%%%%%%%%%%%%%%%%%%%%%%%%%%%%%%%%%%%%%%%%%%%%%%%%%%%%%%%%%%%%%%%%%%

While our algorithm is able to efficiently and faithfully investigate the full space of planetary interior density profiles with an absolute minimum amount of assumptions, some caveats remain.
Most importantly, our method for generating starting profiles has a higher variance (or degree of exploration) at some normalised radii compared to others. 
Listing \ref{lst:startgen} shows our generation method: the binary subdivision algorithm.
The binary subdivision algorithm starts by setting the density value at $r/R=1/2$.
It does so by uniformly choosing at random from the entire range of possible values and subdividing the range into two halves in the process.
It then recursively calls itself on those halves. 
\begin{lstlisting}[caption={The binary subdivision algorithm.}, label={lst:startgen}]
input: int pivot, int step, float $\underline{y}$, float $\overline{y}$
output: array rand_decr
define subdivide_binary(pivot, step, $\underline{y}$, $\overline{y}$):
	y $\gets$ uniform($\underline{y}$, $\overline{y}$)
	rand_decr[pivot] $\gets$ y
	if step = 0 do
		break # base case; no further subdivision
	subdivide_binary(pivot-step, step//2, y, $\overline{y}$)
	subdivide_binary(pivot+step, step//2, $\underline{y}$, y)
\end{lstlisting}
As a result, some binary points in the range from $r/R=0$ to $r/R=1$ (for example 1/2, 1/4, 3/4) are decided earlier, and their bounds, $\underline{y}$ and $\overline{y}$, for the (uniform) random generation are larger.
Therefore, the variance at these points is higher than at others, and the generated starting density profiles tend to have kinks at these locations.
These kinks can influence the final results, as can be seen in the small spikes that appear in Figure \ref{fig:confint} and the cliffs in Figure \ref{fig:jumps}.
We nonetheless decided to use this method because it generates profiles sampled from the space of all possible density profiles such that the likelihood of generating a profile similar to another reference profile is equal everywhere.
Appendix \ref{app:bias_generation} provides more details and a mathematically sound justification of the former statement.
More practically speaking, we find that this limitation is acceptable because many of the starting profiles do not have these kinks and because the algorithm is able to remove the kinks after the initial generation.
This is most apparent in Figure \ref{fig:jumpsu}, where the algorithm has almost completely removed the cliffs for Uranus because of the tighter gravitational moment constraints that are incompatible with density discontinuities at the inverse multiples of two in units of $r/R$.
This can also be seen by directly comparing Figure \ref{fig:jumpsdrhodru} (Uranus), where there are fewer kinks or cliffs visible at inverse multiples of two, with Figure \ref{fig:jumpsdrhodrn} (Neptune), where these kinks or cliffs are very prominent.
We provide a discussion on the bias possibly introduced through the gradient method in Appendix \ref{app:bias_gradient}, which shows that our gradient descent method can be considered unbiased for the purpose of this work.

%%%%%%%%%%%%%%%%%%%%%%%%%%%%%%%%%%%%%%%%%%%%%%%%%%%%%%%%%%%%%%%%%%%%%%%%%%%%%%%%%%%%%%

While the linear approximation (see Appendix \ref{app:gradcalc}) for the gradient was sufficient here, we find that there are limitations in performance on harder problems.
This is especially true if there are more measured gravitational moments or more precise measurements.
There are several ways in which the algorithm could be modified accordingly.
First, the optimisation scheme is always readily exchangeable.
We used the Adam \citep{adam} scheme, but any number of other descent schemes could be implemented.
Due to the structure of the ToF, a full linear gradient was the most amenable choice, but depending on the method used to calculate gravitational moments, the problem could be approached with a stochastic gradient.
This would alter the requirements, though Adam is still likely to perform very well.
Second, the density profile parametrisation is also a viable candidate for modifications.
Our parametrisation has the advantage of being simple to implement and easy to analyse for the gradient.
However, it cannot structurally change the density profile in a meaningful way, as the gradient cannot 'see' directions of changes very well.
Density profiles that are very similar to each other can have parameters ($p_i$ in Equation \ref{eq:parameters}) that are very different.
Different parametrisations could be preferred for this task, such as using orthogonal bases of polynomials, for example.

%%%%%%%%%%%%%%%%%%%%%%%%%%%%%%%%%%%%%%%%%%%%%%%%%%%%%%%%%%%%%%%%%%%%%%%%%%%%%%%%%%%%%%

Finally, we note that a more comprehensive approach could eliminate the difficulties with the ToF and the need for a linear approximation entirely.
Our algorithm was performance-bound exclusively by the implementation of the ToF we used; therefore, a reduction in time usage there would have direct equal benefits.
If a differentiable model could be trained to learn the ToF to within a certain acceptable loss, one would gain significant advantages in the accuracy of the gradients (and, presumably, the calculation of a single gradient would be quicker too).
Alternatively, a non-differentiable machine model of the ToF would still have benefits in performance time but would require zero-order methods.
Additionally, an iterative approach where the average starting point moves towards the best performant point every so often could mean more time is spent resolving the space of acceptable solutions.
However, this would not have served this work since our goal was the exploration of the entire space, not just the successful portion. 

%%%%%%%%%%%%%%%%%%%%%%%%%%%%%%%%%%%%%%%%%%%%%%%%%%%%%%%%%%%%%%%%%%%%%%%%%%%%%%%%%%%%%%

Regarding future work, the analysis done on the solution space could be expanded further.
One could include higher-order or odd-gravitational moments, especially in connection to dedicated Uranus and/or Neptune missions such as in \cite{Parisi2024, Mankovich2025}.
Chiefly, an analysis of the compositional and thermal structure of the planets based on the found density profiles could be performed.
Special attention should be given to investigating the compositional structure around the hypothesised transition point between the hydrogen-helium rich outer layer and the below layer consisting of denser elements such as water or rocks.
\cite{Morf2025} perform such a compositional and thermal analysis for profiles of Uranus and Neptune based on random starting profiles generated in a fashion similar to our models.
However, their methods are limited by computational cost and hence yield only a handful of models (four for each planet).
Hence, the key challenge would lie within ensuring a high enough computational efficiency that does not come at the cost of physical self-consistency.

%%%%%%%%%%%%%%%%%%%%%%%%%%%%%%%%%%%%%%%%%%%%%%%%%%%%%%%%%%%%%%%%%%%%%%%%%%%%%%%%%%%%%%
%%%%%%%%%%%%%%%%%%%%%%%%%%%%%%%%%%%%%%%%%%%%%%%%%%%%%%%%%%%%%%%%%%%%%%%%%%%%%%%%%%%%%%

\section{Summary and conclusions}
\label{sec:summaryandconclusions}

%%%%%%%%%%%%%%%%%%%%%%%%%%%%%%%%%%%%%%%%%%%%%%%%%%%%%%%%%%%%%%%%%%%%%%%%%%%%%%%%%%%%%%

We have presented a novel algorithm for efficiently and faithfully sampling planetary interior density profiles.
Unlike conventional MCMC approaches, our method employs a new optimisation-based strategy.
We used this algorithm to construct interior models of Uranus and Neptune, and we analysed the results from an extensive suite of simulations and compared them to prior studies.

%%%%%%%%%%%%%%%%%%%%%%%%%%%%%%%%%%%%%%%%%%%%%%%%%%%%%%%%%%%%%%%%%%%%%%%%%%%%%%%%%%%%%%

Our results are broadly consistent with previous studies, whose solutions typically fall within the 95\% confidence interval of our model ensemble. 
However, the earlier models do not capture the full diversity of possible interior configurations, particularly in the deep interior. 
We also discussed the limitations related to uneven sampling variance in our generation method and outlined potential future directions, including refining the algorithm in various ways and expanding the model scope to include compositional analyses.
Our key conclusions are summarised as follows:

%%%%%%%%%%%%%%%%%%%%%%%%%%%%%%%%%%%%%%%%%%%%%%%%%%%%%%%%%%%%%%%%%%%%%%%%%%%%%%%%%%%%%%

\begin{itemize}

    \item For Uranus and Neptune, literature models with sharp density discontinuities are biased towards a high central density.
    Literature models without these features are biased towards the peak of the solution space. 
    Models around the median central density have only recently been proposed by \cite{Morf2025}.

    \item The median density (as a function of radius) for Uranus and Neptune shows gradual increases in density consistent with composition gradients. 
    In particular, the medians maintain a consistent slope towards the centre and do not flatten off, with moderate central densities of $\sim$ 6 g\,cm$^{-3}$ for Uranus and $\sim$ 7.5 g\,cm$^{-3}$ for Neptune.

    \item The solution space for Uranus and Neptune is least well constrained for the innermost region of the planets, with the 95\% confidence interval reaching sizes of $\Delta\rho\sim$ 11 g\,cm$^{-3}$ for Uranus and $\Delta\rho\sim$ 13 g\,cm$^{-3}$ for Neptune.

    \item Accounting for the covariance between $J_2$ and $J_4$, such as given in \cite{French2024} for Uranus, has a negligible effect on our results.
    
    \item Discontinuities with a slope greater than 0.02 kg\,m$^{-4}$ are generally rare. 
    
    \item Discontinuities with more moderate slopes (0.002 - 0.02 kg\,m$^{-4}$) are frequent. 
    The literature generally tends to only consider three or fewer discontinuities, which excludes a significant part of the solution space.
    If discontinuities are considered, we recommend adopting the values provided in Figure \ref{fig:jumpspercrit} or equivalently those in Table \ref{tab:appjumpspercrit}.

    \item If at least one discontinuity is present, it is most likely to be found around $\sim0.65R$ for Uranus and $\sim0.7R$ for Neptune.
    
\end{itemize}

%%%%%%%%%%%%%%%%%%%%%%%%%%%%%%%%%%%%%%%%%%%%%%%%%%%%%%%%%%%%%%%%%%%%%%%%%%%%%%%%%%%%%%

By embracing uncertainty rather than constraining it, our approach faithfully captures the complexity of planetary interiors. 
Our ensemble of solutions does not converge on a single solution but defines the space within which physical reality must lie. 
This empirical and agnostic perspective reaffirms that progress in planetary science depends not only on more precise data but also on broader, more flexible ways of interpreting it.

%%%%%%%%%%%%%%%%%%%%%%%%%%%%%%%%%%%%%%%%%%%%%%%%%%%%%%%%%%%%%%%%%%%%%%%%%%%%%%%%%%%%%%
%%%%%%%%%%%%%%%%%%%%%%%%%%%%%%%%%%%%%%%%%%%%%%%%%%%%%%%%%%%%%%%%%%%%%%%%%%%%%%%%%%%%%%

\begin{acknowledgements}

We thank the anonymous referee for valuable comments that greatly improved the manuscript. 
We thank user Kaolay for making their Python implementation of Adam available for use on GitHub. 
This work was supported by the Swiss National Science Foundation (SNSF) through a grant provided as a part of project number 215634: \url{https://data.snf.ch/grants/grant/215634}.

\end{acknowledgements}

%%%%%%%%%%%%%%%%%%%%%%%%%%%%%%%%%%%%%%%%%%%%%%%%%%%%%%%%%%%%%%%%%%%%%%%%%%%%%%%%%%%%%%
%%%%%%%%%%%%%%%%%%%%%%%%%%%%%%%%%%%%%%%%%%%%%%%%%%%%%%%%%%%%%%%%%%%%%%%%%%%%%%%%%%%%%%

\bibliographystyle{aa}
\bibliography{literature.bib}

%%%%%%%%%%%%%%%%%%%%%%%%%%%%%%%%%%%%%%%%%%%%%%%%%%%%%%%%%%%%%%%%%%%%%%%%%%%%%%%%%%%%%%
%%%%%%%%%%%%%%%%%%%%%%%%%%%%%%%%%%%%%%%%%%%%%%%%%%%%%%%%%%%%%%%%%%%%%%%%%%%%%%%%%%%%%%

\begin{appendix}

%%%%%%%%%%%%%%%%%%%%%%%%%%%%%%%%%%%%%%%%%%%%%%%%%%%%%%%%%%%%%%%%%%%%%%%%%%%%%%%%%%%%%%
%%%%%%%%%%%%%%%%%%%%%%%%%%%%%%%%%%%%%%%%%%%%%%%%%%%%%%%%%%%%%%%%%%%%%%%%%%%%%%%%%%%%%%

\section{Algorithm}
\label{app:algorithm}

%%%%%%%%%%%%%%%%%%%%%%%%%%%%%%%%%%%%%%%%%%%%%%%%%%%%%%%%%%%%%%%%%%%%%%%%%%%%%%%%%%%%%%

We give further details regarding the implementation of the optimisation algorithm.
For the reader's convenience, listing \ref{lst:algorithm} presenting a high-level overview of the optimisation algorithm is reproduced here.

%%%%%%%%%%%%%%%%%%%%%%%%%%%%%%%%%%%%%%%%%%%%%%%%%%%%%%%%%%%%%%%%%%%%%%%%%%%%%%%%%%%%%%

\begin{lstlisting}[caption={The optimisation algorithm (reprinted).}]
input: int epochs
output: tuple results
begin
	params $\gets$ create_start_params()
	for epoch in epochs do
		check convergence
		for steps in epochsize do
			params $\gets$ project(opt_step(params))
	return results
end
\end{lstlisting}

%%%%%%%%%%%%%%%%%%%%%%%%%%%%%%%%%%%%%%%%%%%%%%%%%%%%%%%%%%%%%%%%%%%%%%%%%%%%%%%%%%%%%%

\subsubsection*{\texttt{create\_start\_params()}}

%%%%%%%%%%%%%%%%%%%%%%%%%%%%%%%%%%%%%%%%%%%%%%%%%%%%%%%%%%%%%%%%%%%%%%%%%%%%%%%%%%%%%%

A random decreasing function is generated to serve as the  starting distribution.
To do this, we use a recursive subdivision method described in listing \ref{lst:startgen}  which is reprinted below for the reader's convenience. 

%%%%%%%%%%%%%%%%%%%%%%%%%%%%%%%%%%%%%%%%%%%%%%%%%%%%%%%%%%%%%%%%%%%%%%%%%%%%%%%%%%%%%%

\begin{lstlisting}[caption={The binary subdivision algorithm.} (reprinted)]
input: int pivot, int step, float $\underline{y}$, float $\overline{y}$
output: array rand_decr
define subdivide_binary(pivot, step, $\underline{y}$, $\overline{y}$):
	y $\gets$ uniform($\underline{y}$, $\overline{y}$)
	rand_decr[pivot] $\gets$ y
	if step = 0 do
		break # base case; no further subdivision
	subdivide_binary(pivot-step, step//2, y, $\overline{y}$)
	subdivide_binary(pivot+step, step//2, $\underline{y}$, y)
\end{lstlisting}

%%%%%%%%%%%%%%%%%%%%%%%%%%%%%%%%%%%%%%%%%%%%%%%%%%%%%%%%%%%%%%%%%%%%%%%%%%%%%%%%%%%%%%

Note that \texttt{uniform(a,b)} denotes a function that generates a number uniformly at random in the given interval and that \texttt{//} denotes floor integer division.
Each call is supplied with a pivot (where the algorithm is currently operating), a step size, and bounding values.
We generate a value $y$ uniformly at random between $\underline{y}$ and $\overline{y}$ and set the value of the function at the pivot to $y$.
These bounds ensure that the resulting array \texttt{rand\_decr} is decreasing.
If we have reached step size 0, the recursion breaks, as this is the base case and there are no other pivots to set.
Otherwise, we must still process the $y$-values of the pivots that lay \texttt{step} to the left and right of the pivot (and all values in-between later on, recursively).
In that case, we call the function again at those pivots, with halved step size, and with $y$ as the new lower and upper bound respectively.

%%%%%%%%%%%%%%%%%%%%%%%%%%%%%%%%%%%%%%%%%%%%%%%%%%%%%%%%%%%%%%%%%%%%%%%%%%%%%%%%%%%%%%

To generate a random decreasing function of $N$ values between $0$ and $1$, the method is called as \texttt{subdivide\_binary($N$/2,$N$/4, 0, 1)}.
The resulting array \texttt{rand\_decr} is then rescaled by $\alpha$ to ensure the resulting density profile has the correct mass.
Then, the profile is used to calculate the starting parameters.

%%%%%%%%%%%%%%%%%%%%%%%%%%%%%%%%%%%%%%%%%%%%%%%%%%%%%%%%%%%%%%%%%%%%%%%%%%%%%%%%%%%%%%

\subsubsection*{\texttt{opt\_step}}

%%%%%%%%%%%%%%%%%%%%%%%%%%%%%%%%%%%%%%%%%%%%%%%%%%%%%%%%%%%%%%%%%%%%%%%%%%%%%%%%%%%%%%

To calculate \texttt{opt\_step}, we determine goal functions $F_1$ and $F_2$ and a gradient.
The first goal function we wished to minimise was the sum of the costs of being far away from the $d$ observed gravitational moments $J_{2n}$:
\begin{align} 
    F_1(\rho) &= \frac{1}{d}\left[\sum\limits_{n=1}^d \frac{1}{2} \frac{(J_{2n,\text{ToF}}(\rho)-J_{2n}^*)^2}{\sigma\mkern-3mu J_{2n}^2} \right. \\
    &+ \left. \sum_{\substack{n,m=1\\n\neq m}}^{d}\frac{1}{2} \text{cov}\left(J_{2n}^*,J_{2m}^*\right)\left(J_{2n,\text{ToF}}(\rho)-J_{2n}^*\right)\left(J_{2m,\text{ToF}}(\rho)-J_{2m}^*\right)\right]. \nonumber
\end{align}
The gravitational moments calculated by the ToF algorithm \texttt{PyToF} \citep{PyToF} are denoted as $J_{2n,\text{ToF}}(\rho)$.
The observed gravitational moments are denoted as $J_{2n}^*$ and their standard deviations as $\sigma\mkern-3mu J_{2n}$.
For numerical stability, we include a second cost function for deviating from the planet's mass:
\begin{equation}
F_2(\rho)=c\frac{1}{2}\left({\frac{m_{calc}}{m}-1}\right)^2. \label{eq:cost_factor} \end{equation}
The numerical constant $c$ is chosen so that the mass term is not dominated by the cost term and was set to $100$ based on experimental experience.

%%%%%%%%%%%%%%%%%%%%%%%%%%%%%%%%%%%%%%%%%%%%%%%%%%%%%%%%%%%%%%%%%%%%%%%%%%%%%%%%%%%%%%

We set the gradient for the optimisation $\nabla_p F$ as
\begin{equation}
    \nabla_p F = \nabla_p F_1 + \nabla_p F_2 \Vert\nabla_p F_1\Vert,
\end{equation}
where $p$ denote the parameters introduced in Equation \ref{eq:parameters}.
This choice ensures that $\nabla_p F_2$ never gets dominated by $\nabla_p F_1$, but makes it hard to find an explicit form for $F$.
The gradients $\nabla_p F_1$ and $\nabla_p F_2$ are calculated in Appendix \ref{app:gradcalc}.
For the case of $\text{cov}(J_{2n}^*,J_{2m}^*)=0$ for all $m,n$, the results read
\begin{align}
\label{eq:gradient}
    \nabla_p F_1 &= \frac{1}{d}\sum\limits_{i=1}^d \nabla_p J_{2n,\text{ToF}}(\rho)  \frac{(J_{2n,\text{ToF}}(\rho)-J_{2n}^*)}{\sigma\mkern-3mu J_{2n}^2}  \\
	&= \frac{1}{d}\sum\limits_{i=1}^d \frac{(J_{2n,\text{ToF}}(\rho)-J_{2n}^*)}{\sigma\mkern-3mu J_{2n}^2} \left[ -\left(\frac{R_m}{R_{eq}}\right)^{2n} \left(S_{2n,\text{ToF}}^{(-1)}-S_{2n,\text{ToF}}^{(-2)}\right)\right. \nonumber \\
	& \qquad \left. \mkern-16mu\left(\mathbf{\frac{e^p}{w}} \odot \overset{\leftarrow}{\Sigma} \mathbf{\frac{w}{e^p}}
- \gamma\cdot \mathbf{e^p} \cdot \left(\left(\Sigma \mathbf{\frac{e^p}{w}}\right)^{\mathsf{T}}\mathbf{\frac{w}{e^p}}\right)\right)\right], \nonumber \\
\nabla_p F_2 &= c\ \nabla_p m_{calc} \left(\frac{m_{calc}}{m}-1\right) = c\ 4\pi\Delta l \mathbf{e^{p}} \left(\frac{m_{calc}}{m}-1\right), \nonumber
\end{align}
where $R_m$ and $R_{eq}$ are the mean and equatorial radius, respectively.
$S_{2n,\text{ToF}}^{(-1)}$ denotes the last (innermost) and $S_{2n,\text{ToF}}^{(-2)}$ the second to last result of $S_{2n}(z)$ as defined by the ToF \citep[see Appendix A of][]{Morf2024} and calculated by \texttt{PyToF}.
$\mathbf{e^p}$ denotes the element-wise exponential of the parameter vector.
$\mathbf{{e^p}/{w}}$ and $\mathbf{{w}/{e^p}}$ are defined analogously.
$\Sigma \mathbf{{e^p}/{w}}$ denotes the cumulative sum vector of $\mathbf{{e^p}/{w}}$ and $\overset{\leftarrow}{\Sigma} \mathbf{{w}/{e^p}}$ denotes the reverse cumulative sum vector of $\mathbf{{w}/{e^p}}$ (a cumulative sum starting with a sum of all elements and ending with the last element only).
Lastly, $^{\mathsf{T}}$ denotes the usual vector product and $\odot$ the element-wise product.

%%%%%%%%%%%%%%%%%%%%%%%%%%%%%%%%%%%%%%%%%%%%%%%%%%%%%%%%%%%%%%%%%%%%%%%%%%%%%%%%%%%%%%

With the gradient, we update the parameters in an optimisation step according to the Adam gradient descent scheme \citep{adam}.
We chose Adam because of its excellent performance in practice.
Hyperparameters were tuned experimentally, the learning rate was set at 0.08 because a rapid convergence to a solution was desired.
All other parameters were found to perform well at default values.

%%%%%%%%%%%%%%%%%%%%%%%%%%%%%%%%%%%%%%%%%%%%%%%%%%%%%%%%%%%%%%%%%%%%%%%%%%%%%%%%%%%%%%

\subsubsection*{\texttt{project()}}

%%%%%%%%%%%%%%%%%%%%%%%%%%%%%%%%%%%%%%%%%%%%%%%%%%%%%%%%%%%%%%%%%%%%%%%%%%%%%%%%%%%%%%

After each optimisation step, the projection $\Pi$ is applied to enforce the minimum density increase $\Delta_\text{min}$ per step:
\begin{equation}\Pi(p_i) = \max(p_i,\ln( w_i\Delta_\text{min})).\end{equation}
Note that the algorithm treats the parameters $p_i, w_i$, and $\Delta_\text{min}$ as unitless, Equation \ref{eq:parameters} reads
\begin{equation}
    \rho_k = \left(\sum_{i=0}^{k-1} \frac{e^{p_i}}{w_i} \ \alpha \right) \frac{\text{kg}}{\text{m}^3}
\end{equation}
with units.
This implies that the minimal possible central density (where $p_i < \ln( w_i\Delta_\text{min})$ for all $i\in{0,\dots,N-2}$) is given by
\begin{equation}
    \rho_\text{min}^\text{centre} = \alpha N \Delta_\text{min} \frac{\text{kg}}{\text{m}^3} = 1024\alpha\frac{\text{kg}}{\text{m}^3}
\end{equation}
for $N=1024$ and $\Delta_\text{min}=1$.
$\alpha$ ensures that our density profiles always have the correct mass, regardless of the value of $\alpha$.
The introduction of the function $F_2$ ensures that $\alpha\neq1$ comes with a cost, but does not force $\alpha=1$.
Consequently, even profiles that lie below the minimal profile linearly increasing to 1024 kg/m$^3$ in the centre are considered by the algorithm.
This is best seen in the zoom panels of subfigures \ref{fig:densprofu} and \ref{fig:densprofn}, where some density profiles are quite flat in the outermost parts of the planets.

%%%%%%%%%%%%%%%%%%%%%%%%%%%%%%%%%%%%%%%%%%%%%%%%%%%%%%%%%%%%%%%%%%%%%%%%%%%%%%%%%%%%%%

\subsubsection*{\texttt{check convergence}}

%%%%%%%%%%%%%%%%%%%%%%%%%%%%%%%%%%%%%%%%%%%%%%%%%%%%%%%%%%%%%%%%%%%%%%%%%%%%%%%%%%%%%%

Before each epoch, we check for convergence.
We do this by comparing the change in cost.
If the exponential moving average of the relative change in cost is close enough to one (difference of less than 0.0005), we terminate and return the following results:
\begin{itemize}
    \item The starting density profile
    \item The final density profile
    \item The gravitational moments $J_2$ and $J_4$
    \item The pressure profile
    \item The normalised moment of inertia
    \item The flattening ratio
    \item Whether \rhomax has been respected
\end{itemize}

%%%%%%%%%%%%%%%%%%%%%%%%%%%%%%%%%%%%%%%%%%%%%%%%%%%%%%%%%%%%%%%%%%%%%%%%%%%%%%%%%%%%%%
%%%%%%%%%%%%%%%%%%%%%%%%%%%%%%%%%%%%%%%%%%%%%%%%%%%%%%%%%%%%%%%%%%%%%%%%%%%%%%%%%%%%%%

\section{Gradient calculation}
\label{app:gradcalc}

%%%%%%%%%%%%%%%%%%%%%%%%%%%%%%%%%%%%%%%%%%%%%%%%%%%%%%%%%%%%%%%%%%%%%%%%%%%%%%%%%%%%%%

Here, we calculate the gradients of $F_1$ and $F_2$:
\begin{align}
\nabla_p F_1 &= \frac{1}{d}\sum\limits_{i=1}^d \nabla_p J_{2n,\text{ToF}}(\rho)  \frac{(J_{2n,\text{ToF}}(\rho)-J_{2n}^*)}{\sigma\mkern-3mu J_{2n}^2} \\
\nabla_p F_2 &= c\ \nabla_p m_{calc} \left(\frac{m_{calc}}{m}-1\right). \nonumber
\end{align}
Since $J_{2n}=-\left(\frac{R_m}{R_{eq}}\right)^{2n}S_{2n}(1)$, we get
\begin{equation}\nabla_p J_{2n,\text{ToF}}(\rho)=-\left(\frac{R_m}{R_{eq}}\right)^{2n}\nabla_p S_{2n,\text{ToF}}(p)\Bigr|_{z=1}.\end{equation}
We avoid some notation for legibility and note that
\begin{equation}\nabla_p S = \frac{dS}{dp}=\frac{d\rho}{dp}\frac{dz}{d\rho}\frac{dS}{dz}\Bigr|_{z=1}.\end{equation}
Each of these terms can be readily calculated.
By plugging in $z=1$, the rightmost derivative is approximated as
\begin{equation}
\left(S_{2n,\text{ToF}}^{(-1)}-S_{2n,\text{ToF}}^{(-2)}\right)/\Delta z,
\end{equation}
where $S_{2n,\text{ToF}}^{(-i)}$ denotes the last/innermost ($i=-1$) and second to last ($i=-2$) value calculated by the ToF. 
The small change $\Delta z$ does not have to be further specified, it vanishes in the end result.
The middle derivative (denoted element-wise here, but actually a vector) is easily calculated by its inverse, which we can approximate as
\begin{equation}
\frac{d\rho_k}{dz} \approx \frac{e^{p_{k-1}}}{w_{k-1}}\alpha/\Delta z
\end{equation}
since the difference between $\rho_k$ and $\rho_{k-1}$ is exactly this term.
We calculated the leftmost derivative element-wise: 
Write $\del_i$ for the partial derivative by $p_i$, and therefore $\frac{\del}{\del p_i}$. 
The $k$-th entry of $\rho$ is $\rho_k = \sum_{j=0}^{k-1} \frac{e^{p_j}}{w_j}\alpha$. 
Thus,
\begin{equation}\partial_i\rho_k=\del_i \left(\sum_{j=0}^{k-1} \frac{e^{p_j}}{w_j}\right)\ \alpha + \sum_{j=0}^{k-1} \frac{e^{p_j}}{w_j}\ \del_i\alpha.\end{equation}
The first derivative is only non-zero if $k > i$, since $p_i$ only appears in the sum $\sum_{j=0}^{k-1} \frac{e^{p_j}}{w_j}\alpha$ for $k > i$. 
Therefore,
\begin{equation}\del_i \left(\sum_{j=0}^{k-1} \frac{e^{p_j}}{w_j}\right)=\begin{cases}\del_i \frac{e^{p_i}}{w_i}=\frac{e^{p_i}}{w_i} & k > i \\
0 & k\leq i
\end{cases}\end{equation}
since $w_i$ is constant with respect to $p_i$.
The second derivative, $\partial_i \alpha$, is
\begin{equation}\partial_i \alpha = \del_i \frac{m}{m_{calc}} =  -\frac{m}{m_{calc}^2} \del_i m_{calc}.\end{equation}
We remind that $m_{calc} = 4\pi \sum_{k=0}^{N-1}\tilde{\rho}_k l_k^2\ \Delta l $, where $\tilde{\rho}_k=\sum_{j=0}^{k-1} \frac{e^{p_j}}{w_j}$ and note that $l_k$ is constant with respect to the parameters. 
Thus, 
\begin{align}
\partial_i \alpha &= -\frac{m}{m_{calc}^2} \del_i \left( 4\pi \sum_{k=0}^{N-1} \sum_{j=0}^{k-1} \frac{e^{p_j}}{w_j} l_k^2\ \Delta l \right) \\
&=  -\frac{m}{m_{calc}^2} 4\pi \Delta l \sum_{k=0}^{N-1}l_k^2  \del_i\sum_{j=0}^{k-1}  \frac{e^{p_j}}{w_j}, \nonumber
\end{align}
where again only terms with $k > i$ survive (always yielding $e^{p_i}/w_i$), so
\begin{equation}\partial_i \alpha =  -\frac{m}{m_{calc}^2} 4\pi \Delta l \sum_{k= i + 1}^{N-1}l_k^2  \frac{e^{p_i}}{w_i}.\end{equation}
Rewriting $m/m_{calc}$ as $\alpha$ and arranging terms yields
\begin{equation}\partial_i \alpha  =  -\alpha \underbrace{\frac{4\pi \Delta l }{m_{calc}}}_{=:\gamma} e^{p_i}   \frac{\sum_{k= i + 1}^{N-1}l_k^2}{w_i}.\end{equation}
At this point, it becomes clear why the weights are defined as they are in Equation \ref{eq:weights}. 
If we remember $w_i=\sum_{k=i+1}^{N-1}l_k^2$, we can simplify the expression as
\begin{equation}\partial_i \alpha =  -\alpha \gamma e^{p_i}.\end{equation}
Putting everything together, we arrive at
\begin{equation}\del_i \rho_k = \begin{cases}
	\sum_{j=0}^{k-1} \frac{e^{p_j}}{w_j} \cdot (-\alpha \gamma e^{p_i}), & k \leq i\\
	\frac{e^{p_i}}{w_i}\alpha + \sum_{j=0}^{k-1} \frac{e^{p_j}}{w_j} \cdot (-\alpha \gamma e^{p_i}), & k > i.
\end{cases}\end{equation}
We can now combine everything to finally obtain $\nabla_p S$.
In summary, we have
\begin{align}
\frac{dS}{dz}\Bigr|_{z=1}\mkern-25mu &= S_{2n,\text{ToF}}^{(-1)}-S_{2n,\text{ToF}}^{(-2)}, \\
\frac{d\rho}{dz} &= \alpha \begin{pmatrix}
	0\\
	e^{p_0}/w_0\\
	\vdots\\
	e^{p_{N-2}}/w_{N-2}
\end{pmatrix} \implies \frac{dz}{d\rho} = \frac{1}{\alpha}
\begin{pmatrix}
0\\
w_0/e^{p_0}\\
\vdots\\
w_{N-2}/e^{p_{N-2}}
\end{pmatrix}, \nonumber \\
\frac{d\rho}{dp} &= \alpha
\begin{pmatrix}
	0 & \frac{e^{p_0}}{w_0} & \cdots & \frac{e^{p_{0}}}{w_{0}}\\
	\vdots & \ddots & \ddots & \vdots\\
	0 & \cdots & 0 & \frac{e^{p_{N-2}}}{w_{N-2}}
\end{pmatrix} - \alpha\gamma
\underbrace{
	\begin{pmatrix}
		& & \\
		& e^{p_i} \ \displaystyle\sum\limits_{j=0}^{k-1} \frac{e^{p_j}}{w_j} &\\
		& &
	\end{pmatrix}
}_{k\rightarrow} 
\left. 
    \vphantom{\begin{pmatrix}
		& & \\
		& e^{p_i} \ \displaystyle\sum\limits_{j=0}^{k-1} \frac{e^{p_j}}{w_j} &\\
		& &
	\end{pmatrix}} 
\right\}
\begin{array}{l}
    \scriptstyle i \\
    [-6pt]
    \scriptstyle \downarrow 
\end{array}. \nonumber
\end{align}
We combine the above as vector and matrix products. 
They may be conveniently written if we define $\mathbf{e^p}$ to be the element-wise exponentiation of the parameter vector, and analogously define $\mathbf{{e^p}/{w}}$ and $\mathbf{{w}/{e^p}}$. 
Then write $\Sigma \mathbf{{e^p}/{w}}$ for the cumulative sum vector of $\mathbf{{e^p}/{w}}$ and $\overset{\leftarrow}{\Sigma} \mathbf{{w}/{e^p}}$ for the reverse cumulative sum vector of $\mathbf{{w}/{e^p}}$. 
With $^{\mathsf{T}}$ the usual vector product and $\odot$ the element-wise product, we get
\begin{align}
	\nabla_p J_{2n,\text{ToF}}(\rho) &= -\left(\frac{R_m}{R_{eq}}\right)^{2n}\mkern-16mu\left(S_{2n,\text{ToF}}^{(-1)}-S_{2n,\text{ToF}}^{(-2)}\right) \\
    &\qquad \left(\mathbf{\frac{e^p}{w}} \odot \overset{\leftarrow}{\Sigma} \mathbf{\frac{w}{e^p}}
- \gamma\cdot \mathbf{e^p} \cdot \left(\left(\Sigma \mathbf{\frac{e^p}{w}}\right)^{\mathsf{T}}\mathbf{\frac{w}{e^p}}\right)\right). \nonumber
\end{align}
We are almost done and just need to consider the remaining $\nabla_p m_{calc}$.
Luckily, during our calculation of $\del_i\alpha$, we actually already calculated this expression.
Since we calculated
\begin{equation}\del_i\alpha = -\frac{\alpha}{m_{calc}}\del_i m_{calc} = -\alpha \frac{4\pi \Delta l}{m_{calc}}e^{p_i},\end{equation}
we can infer directly
\begin{equation}\del_i m_{calc} = 4\pi\Delta l e^{p_i}\end{equation}
or in vector form $\nabla_p m_{calc} = 4\pi \Delta l \mathbf{e^{p}}$. 
This finally yields the expressions presented in Appendix \ref{app:algorithm}, Equation \ref{eq:gradient}.

%%%%%%%%%%%%%%%%%%%%%%%%%%%%%%%%%%%%%%%%%%%%%%%%%%%%%%%%%%%%%%%%%%%%%%%%%%%%%%%%%%%%%%
%%%%%%%%%%%%%%%%%%%%%%%%%%%%%%%%%%%%%%%%%%%%%%%%%%%%%%%%%%%%%%%%%%%%%%%%%%%%%%%%%%%%%%

\section{Bias}
\label{app:bias}

In this section, we address potential biases of our algorithm in detail.

%%%%%%%%%%%%%%%%%%%%%%%%%%%%%%%%%%%%%%%%%%%%%%%%%%%%%%%%%%%%%%%%%%%%%%%%%%%%%%%%%%%%%%

The observational data together with their respective uncertainties allow us to decide whether a given density profile is plausible (fits the data) or not.
We call the set of all plausible density profiles the `space of possible interior density profiles that fit their known observational data' or the solution space for short.
Our algorithm samples this solution space.
This sampling is unbiased if all plausible density profiles in the solution space have equal likelihood of being generated by our algorithm.

%%%%%%%%%%%%%%%%%%%%%%%%%%%%%%%%%%%%%%%%%%%%%%%%%%%%%%%%%%%%%%%%%%%%%%%%%%%%%%%%%%%%%%

Formally, we may call the solution space $S$.
An algorithm $A$ is a random variable that produces density profiles $\rho\in S$.
It has a probability density function $f_A: S\to \Rp$, which fulfils
\begin{equation}
    \Pr[A \in R] = \int_Rf_A  \ \mathrm{d}\mu,
\end{equation}
for any observable event, $R$.
The probability measure on $S$ is $\mu$.
We call the algorithm $A$ unbiased if its probability density function $f_A$ is equal to $1$ almost everywhere.
This is the formal definition of all profiles having equal likelihood of being generated.
Note that this definition is stronger than just demanding $\Pr[A=\rho]$ be equal for all $\rho\in S$.
Every algorithm fulfils this weaker requirement, because the probability to generate exactly $\rho$ is $\mu(\{\rho\})=0$.
As a result, a more careful definition is needed:
We also require that all regions $R\subseteq S$ have proportional likelihood $\int_R f_A  \ \mathrm{d}\mu = \mu(R)$.

%%%%%%%%%%%%%%%%%%%%%%%%%%%%%%%%%%%%%%%%%%%%%%%%%%%%%%%%%%%%%%%%%%%%%%%%%%%%%%%%%%%%%%

To define a concept of distance between profiles, we employed the infinity norm
\begin{equation}
    \lVert\rho-\rho'\rVert_\infty = \max_{i\in\{1,\cdots,N\}} \lvert\rho_i-\rho'_i\rvert
\end{equation}
which allowed us to approximate any region in $S$ as the countable union of distinct neighbourhoods $N_{\rho}^{\epsilon}$ of radius $\epsilon/2$ around $\rho$:
\begin{equation}
    N^{\epsilon}_\rho=\left\{\rho'\mid \ \lVert\rho-\rho'\rVert_{\infty}<\frac{\epsilon}{2}\right\}.
    \label{eq:neighbourhood}
\end{equation}
We call two density profiles, $\rho$ and $\rho^\prime$, similar or $\epsilon$-close if $\rho^\prime \in N^{\epsilon}_\rho$.
From now on, we refer to the measure of $N^{\epsilon}_\rho$ as its volume, and write $V(N^{\epsilon}_\rho)$.

%%%%%%%%%%%%%%%%%%%%%%%%%%%%%%%%%%%%%%%%%%%%%%%%%%%%%%%%%%%%%%%%%%%%%%%%%%%%%%%%%%%%%%

The algorithm $A$ first generates a starting profile.
It subsequently evolves this starting profile through optimisation towards a result profile, implausible results are later discarded.
If our algorithm $A$ generates profiles such that all regions $N_\rho^\epsilon$ of equal radius have equal volume $V(N^{\epsilon}_\rho)$ (probability), and it then evolves these regions such that their outputs also all have the same volume, then the algorithm must be unbiased:
The probability that our algorithm ends up in $R$ depends only on the probability measure $\mu$ of $R$ in $S$, not on the algorithm $A$, because it treats all regions equally.
The relative frequency of successful runs in $R$ during our simulation is then a valid proxy for the size (probability) of $R$ in $S$, which allows us to calculate results such as confidence intervals.

%%%%%%%%%%%%%%%%%%%%%%%%%%%%%%%%%%%%%%%%%%%%%%%%%%%%%%%%%%%%%%%%%%%%%%%%%%%%%%%%%%%%%%

We will investigate the above steps in the next two sections of this appendix in detail.
The first step --- generating profiles with equal volume everywhere --- is treated analytically.
The second step --- evolving profiles without distorting volumes --- is treated using empirical evidence obtained by monitoring the behaviour of the descent.
Additionally, we analyse the distortions that do occur in the descent method to quantify their impact on our results.

%%%%%%%%%%%%%%%%%%%%%%%%%%%%%%%%%%%%%%%%%%%%%%%%%%%%%%%%%%%%%%%%%%%%%%%%%%%%%%%%%%%%%%

\subsection{Generation method}
\label{app:bias_generation}

%%%%%%%%%%%%%%%%%%%%%%%%%%%%%%%%%%%%%%%%%%%%%%%%%%%%%%%%%%%%%%%%%%%%%%%%%%%%%%%%%%%%%%

We can convert the space of all density profiles $\rho=(\rho_0,\cdots,\rho_{N-1})$ to its corresponding $N-1$-dimensional space of increases $\Delta\rho_k=\rho_{k+1}-\rho_k$ between values: $\Delta\rho=(\Delta\rho_0,\cdots,\Delta\rho_{N-2})$.
If we then rescale this space, we obtain the standard $(N-2)$-simplex.
The uniform distribution over the standard $(N-2)$-simplex is the flat Dirichlet distribution obtained from the general Dirichlet distribution
\begin{equation}
    f(\Delta \rho_0,\dots,\Delta \rho_{N-2}) = \frac{\Gamma(\alpha (N-1))}{\Gamma(\alpha)^{N-1}}\prod_{k=0}^{N-2}\Delta\rho_k^{\alpha-1}
\end{equation}
by setting $\alpha = 1$.
$f$ is a probability density function.
This forces $\Delta\rho_k^0 = 1$ and thus the distribution has constant value $f=(N-2)!$ and is therefore uniform.
This is natural, given that the volume of the simplex is $\frac{1}{(N-2)!}$.
This distribution is not, however, uniformly dense in the space of density profiles.
Profiles around the average (with each increase $\sim 1/(N-1)$) have many more neighbours than sparse profiles (with most increases close to 0).
This is because, given some small radius $\epsilon/2$, the volume of the neighbourhood $N_\rho^\epsilon$ can be bounded as 
\begin{align}
    &\prod_{k=1}^{N-2}\left[\min\left(\frac{\epsilon}{2},\frac{\Delta\rho_{k-1}}{2}\right)+\min\left(\frac{\epsilon}{2},\frac{\Delta\rho_{k}}{2}\right)\right] \leq V(N_\rho^\epsilon) \\
    &\mkern+100mu V(N_\rho^\epsilon) \leq
    \prod_{k=1}^{N-2}\left[\min\left(\frac{\epsilon}{2},\Delta\rho_{k-1}\right)+\min\left(\frac{\epsilon}{2},\Delta\rho_{k}\right)\right], \nonumber
\end{align}
which is a bound tighter (and hence better) compared to a bound relying on $\epsilon$ alone.
Using only half of each increase ($\Delta\rho_{k-1}/2$ and $\Delta\rho_{k}/2$) guarantees no overlap between increases but thus undercounts possible neighbours, whereas using the full increase ($\Delta\rho_{k-1}$ and $\Delta\rho_{k}$) both times has overlap and thus overcounts.
For the average case, it holds $\epsilon\leq\Delta\rho_k$ for sufficiently small $\epsilon$, and thus
\begin{equation}
    \min(\frac{\epsilon}{2},\frac{\Delta\rho_{k}}{2})=\frac{\epsilon}{2}
\end{equation}
always.
In that case the volume therefore equals $V(N_\rho^\epsilon)=\epsilon^{N-2}$.
For the sparse case however, this is smaller ($V(N_\rho^\epsilon)<\epsilon^{N-2}$), because in this case there exist $\Delta\rho_k$ smaller than $\epsilon/2$.

%%%%%%%%%%%%%%%%%%%%%%%%%%%%%%%%%%%%%%%%%%%%%%%%%%%%%%%%%%%%%%%%%%%%%%%%%%%%%%%%%%%%%%

We look for a probability density function $\hat{f}$ for generating initial profiles that is uniformly dense in the space of density profiles:
\begin{equation}
    V(\hat{f}(N_\rho^\epsilon)) = \int_{N_\rho^\epsilon}\hat{f}(\rho^\prime)  \ \mathrm{d}V=\text{const.},
    \label{eq:uniformly_dense}
\end{equation}
where the integration over $\mathrm{d}V$ includes all $\rho^\prime\in N_\rho^\epsilon$.
This is fulfilled by
\begin{equation}
    \hat{f}(\Delta\rho_0,\dots,\Delta\rho_{N-2})=\prod_{k=0}^{N-2}\frac{1}{\Delta\rho_k},
\end{equation}
equivalent to the Dirichlet distribution if one were to set $\alpha=0$ (which is impossible due to the gamma function scale factor).
$\hat{f}$ fulfils Equation \ref{eq:uniformly_dense} because it is the exact inverse of the volume distortion.

%%%%%%%%%%%%%%%%%%%%%%%%%%%%%%%%%%%%%%%%%%%%%%%%%%%%%%%%%%%%%%%%%%%%%%%%%%%%%%%%%%%%%%

We now show that our generation function $f_\text{BD}$ (the binary subdivision algorithm) takes on a very similar shape.
At the last phase of the recursion, conditioned on the fact that all higher-level calls successfully chose the correct density values, the probability to choose $\rho_k$ out of all possible values for index $k$ is
\begin{equation}
    \frac{1}{\Delta\rho_{k-1}+\Delta\rho_{k}}.
\end{equation}
This is because the uniform distribution has probability density equal to one over the volume of the interval, which in this case must be $\Delta\rho_{k-1}+\Delta\rho_{k}$.
At higher phases of the recursion, this is not the case.
The interval (and thus the variance) is larger and the probability is smaller.
This causes the artefacts discussed in Section \ref{sec:discussion}.
Barring these limitations, we find that $f_\text{BD}$ takes on the shape
\begin{equation}
    f_\text{BD}(\Delta\rho_0,\dots,\Delta\rho_{N-2})=\prod_{k=0}^{N-2}\frac{1}{\Delta\rho_{k-1}+\Delta\rho_{k}},
\end{equation}
which also fixes the volume distortion like $\hat{f}$ does and is actually computable.
With this, we find that our volume is now bounded as
\begin{align}
    &\prod_{k=1}^{N-2} \frac{\min\left(\frac{\epsilon}{2},\frac{\Delta\rho_{k-1}}{2}\right)+\min\left(\frac{\epsilon}{2},\frac{\Delta\rho_{k}}{2}\right)}{\Delta\rho_{k-1}+\Delta\rho_{k}}  \leq V(f_\text{BD}(N_\rho^\epsilon))\\
    & \mkern+100mu V(f_\text{BD}(N_\rho^\epsilon))\leq \prod_{k=1}^{N-2}\frac{\min\left(\frac{\epsilon}{2},\Delta\rho_{k-1}\right)+\min\left(\frac{\epsilon}{2},\Delta\rho_{k}\right)}{\Delta\rho_{k-1}+\Delta\rho_{k}}. \nonumber
\end{align}
For a given $\epsilon$ (that is, for a given level of stringency as to what two profiles are considered similar) and small enough $\Delta\rho_{k-1}, \Delta\rho_k < \epsilon$, the volume becomes constant (const. $\in[0.5,1]$), which is what was desired.

%%%%%%%%%%%%%%%%%%%%%%%%%%%%%%%%%%%%%%%%%%%%%%%%%%%%%%%%%%%%%%%%%%%%%%%%%%%%%%%%%%%%%%

In summary, by correcting for the uneven distribution of possible density profiles in the simplex through the use of a sampling method with the inverse probability distribution, we attain a method of generating density profiles that samples from the space of possible density profiles without bias. 
We emphasise again that our generation method is not the uniform distribution of the space of density profiles. 
We conjecture, however, that a uniform and unbiased generation method (formally, a method with uniform probability distribution and uniformly distributed probability measure) is, in fact, impossible.

%%%%%%%%%%%%%%%%%%%%%%%%%%%%%%%%%%%%%%%%%%%%%%%%%%%%%%%%%%%%%%%%%%%%%%%%%%%%%%%%%%%%%%

\subsection{Gradient method}
\label{app:bias_gradient}

%%%%%%%%%%%%%%%%%%%%%%%%%%%%%%%%%%%%%%%%%%%%%%%%%%%%%%%%%%%%%%%%%%%%%%%%%%%%%%%%%%%%%%

The usage of a gradient can result in local solutions being ignored in favour of a large and prominent global solution, which would introduce bias.
For our method to be unbiased, the search method must return close, local results.
To be more specific, call the local solution $\hat{\rho}$.
The gradient method $G$ transforms inputs into solutions.
The target region for $G$ is the neighbourhood $N_{\hat{\rho}}^{\epsilon}$.
Successful inputs are thus the set $G^{-1}(N_{\hat{\rho}}^{\epsilon})$.
The size ratio of $G$ at $\hat{\rho}$ with distance $\epsilon$ is thus
\begin{equation}
    r_G(\hat{\rho})=   \frac{V(G^{-1}(N_{\hat{\rho}}^{\epsilon}))}{V(N_{\hat{\rho}}^{\epsilon})},
\end{equation}
which ideally should be independent of $\epsilon$.
The universal size ratio $\Bar{r}_G$ of $G$ is equal to the ratio of the total input and output volumes and does not necessarily equal to one.
We divide all size ratios by this universal size ratio to obtain the relative local performance of $G$ at $\hat{\rho}$, which we call the discoverability $\mathfrak{d}$ of $\hat{\rho}$ by $G$:
\begin{equation}
    \mathfrak{d}(\hat{\rho})= \frac{r_G(\hat{\rho})}{\Bar{r}_G}.
\end{equation}
Our method is unbiased if and only if all solutions over all distances have discoverability 1.
Otherwise, a preferred dangerous global solution $\hat{\rho}'$ would exist with larger input domain such that
\begin{equation}
    V(G^{-1}(N_{\hat{\rho}'}^{\epsilon})) > V(G^{-1}(N_{\hat{\rho}}^{\epsilon})).
\end{equation}

%%%%%%%%%%%%%%%%%%%%%%%%%%%%%%%%%%%%%%%%%%%%%%%%%%%%%%%%%%%%%%%%%%%%%%%%%%%%%%%%%%%%%%

To test our results, we drew 512 successful solutions from the correlated Uranus run at random and used these as references $\rho^*$.
To calculate the discoverability of a given $\rho^*$, we considered all the $n$ results $\rho^{(i)}$ that are $\epsilon$-close to $\rho^*$ for a given $\epsilon$.
We then calculated the average distance between all of them:
\begin{equation}
    \binom{n}{2}^{-1}\sum_{\substack{i,j=1\\i<j}}^n \lVert\rho^{(i)}-\rho^{(j)}\rVert_\infty,
    \label{eq:avdistances}
\end{equation}
which scales with the volume of $V(N_{\rho^*}^\epsilon)$ and hence was used as a proxy for it.
For each reference solution $\rho^*$ that a given $\rho$ was close to, we added its input and output density profile to a list of neighbours of $\rho^*$.
Due to numerical constraints, only the most recent 128 neighbours could be saved.
Any additional $\rho$ overwrote earlier neighbours and the new average distance was estimated by calculating the average distance to the most recent 128 neighbours while scaling up the result proportionally.

%%%%%%%%%%%%%%%%%%%%%%%%%%%%%%%%%%%%%%%%%%%%%%%%%%%%%%%%%%%%%%%%%%%%%%%%%%%%%%%%%%%%%%

\begin{figure}
	\centering
	\includegraphics[width=0.98\columnwidth]{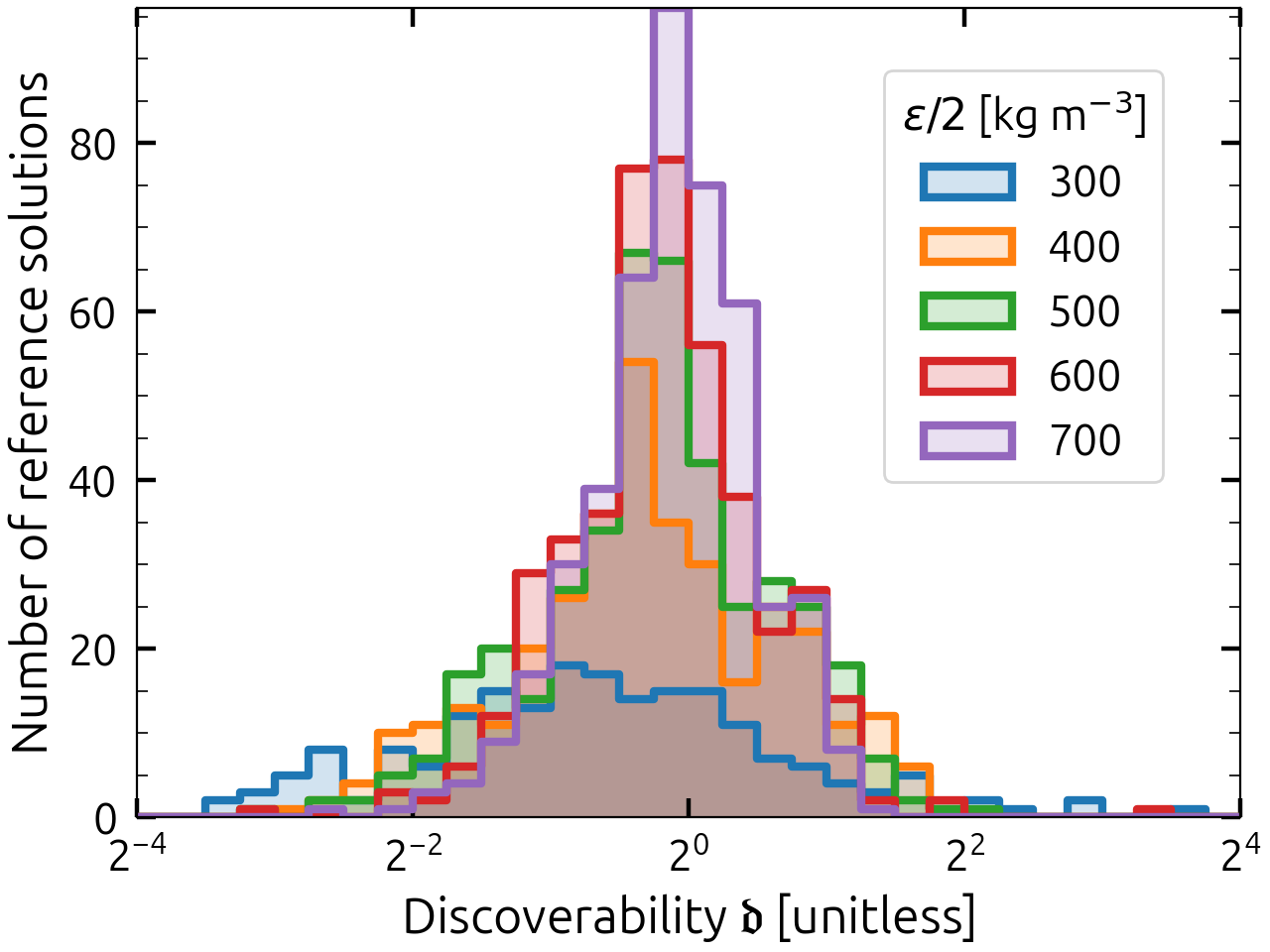}
	\caption{
    Discoverabilities.
    Distribution of the discoverabilities of randomly chosen reference solutions.
    Different colours indicate different $\epsilon$ distances according to equation \ref{eq:neighbourhood}.
    A discoverability of $\mathfrak{d}=1=2^0$ is equivalent to the employed gradient method being unbiased for the given reference solution.
    }
	\label{fig:disc}
\end{figure}

%%%%%%%%%%%%%%%%%%%%%%%%%%%%%%%%%%%%%%%%%%%%%%%%%%%%%%%%%%%%%%%%%%%%%%%%%%%%%%%%%%%%%%

Figure \ref{fig:disc} shows the discoverability distribution of the reference solutions $\rho^{*}$ for five different values of $\epsilon/2\in\{300,400,500,600,700\}$ kg\,m$^{-3}$.
Both ends of this $\epsilon$-range are motivated:
At $\epsilon/2 = 300$ kg\,m$^{-3}$, over half of all references already had to be discarded for lack of neighbours.
\footnote{To be precise: $\{316, 166, 101, 73, 52\}$ reference profiles $\rho^*$ had no neighbours for $\epsilon/2\in\{300,400,500,600,700\}$ kg\,m$^{-3}$. These were discarded because their average distances were undefined.}
At $\varepsilon/2 = 700$ kg\,m$^{-3}$, some reference solutions had upwards of
30000 neighbours, a significant portion of the total number of runs.

%%%%%%%%%%%%%%%%%%%%%%%%%%%%%%%%%%%%%%%%%%%%%%%%%%%%%%%%%%%%%%%%%%%%%%%%%%%%%%%%%%%%%%

Figure \ref{fig:disc} demonstrates that the discoverability distribution gets better (most reference solutions have $\mathfrak{d}\approx1$) for bigger values of $\epsilon$ and loses its bias towards low discoverability.
To ensure that the outliers with a discoverability $\mathfrak{d}$ not close to 1 are unproblematic, we compare the popularity of a solution (that is, its number of neighbours) against its discoverability.
If low discoverability solutions ($\mathfrak{d}\ll1$) were unpopular, but high discoverability solutions ($\mathfrak{d}\gg1$) popular, our gradient method would suffer from severe bias.

%%%%%%%%%%%%%%%%%%%%%%%%%%%%%%%%%%%%%%%%%%%%%%%%%%%%%%%%%%%%%%%%%%%%%%%%%%%%%%%%%%%%%%

\begin{figure}
	\centering
	\includegraphics[width=0.98\columnwidth]{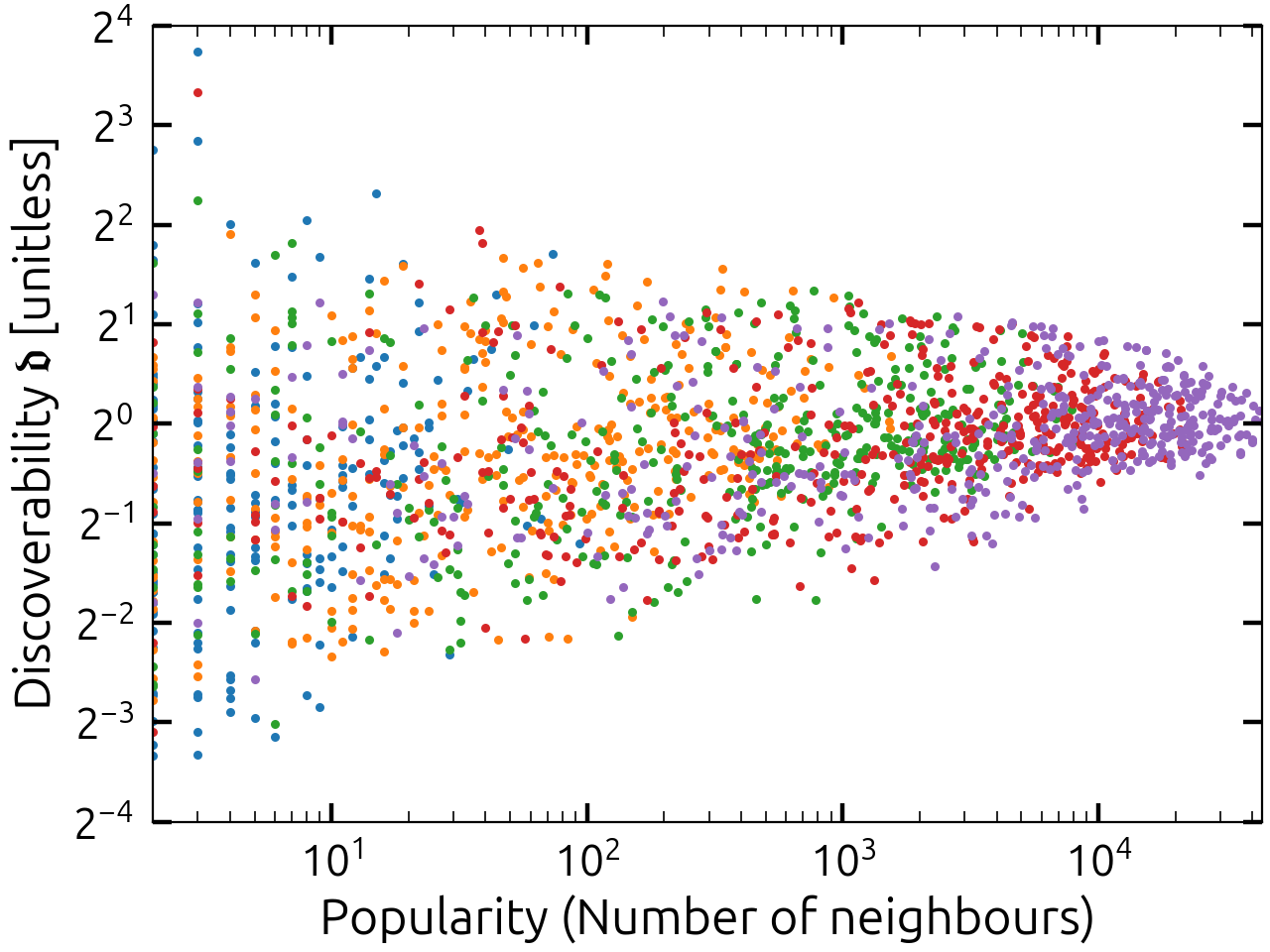}
	\caption{
    Discoverabilities vs popularities.
    Discoverabilities of the reference solutions (with neighbours) against their number of neighbours for different $\epsilon$ distances.
    The colours are analogous to Figure \ref{fig:disc}.
    }
	\label{fig:discvpop}
\end{figure}

%%%%%%%%%%%%%%%%%%%%%%%%%%%%%%%%%%%%%%%%%%%%%%%%%%%%%%%%%%%%%%%%%%%%%%%%%%%%%%%%%%%%%%

Figure \ref{fig:discvpop} shows the relationship of the discoverability of the reference solution vs their popularity.
We find Pearson correlation coefficients of $\{0.052, 0.137, 0.091, 0.058, 0.138\}$ for $\epsilon/2\in\{300,400,500,600,700\}$ kg\,m$^{-3}$.
The p-values are $\{0.467, 0.011, 0.065, 0.224, 0.003\}$.
The discoverabilities of the reference solution and their popularity are therefore only very weakly linearly correlated.
However, the two values are clearly non-linearly correlated:
The discoverability converges towards 1 as the popularity grows, our data is heteroscedastic.
This behaviour is expected, as the popularity is simultaneously the sample size of the discoverability measurement which hence should converge to its true value as we increase the sample size.

%%%%%%%%%%%%%%%%%%%%%%%%%%%%%%%%%%%%%%%%%%%%%%%%%%%%%%%%%%%%%%%%%%%%%%%%%%%%%%%%%%%%%%

We conclude that our gradient method is unbiased in practice based on the above presented empirical arguments.
We find that for large-scale statistics (like the confidence interval), where large distances $\epsilon$ apply, we do not underestimate relatively infrequent solutions.
However, caution must be taken in making statements about small subsections of our sample cohort.
The bias there is larger and its direction unknown.

%%%%%%%%%%%%%%%%%%%%%%%%%%%%%%%%%%%%%%%%%%%%%%%%%%%%%%%%%%%%%%%%%%%%%%%%%%%%%%%%%%%%%%
%%%%%%%%%%%%%%%%%%%%%%%%%%%%%%%%%%%%%%%%%%%%%%%%%%%%%%%%%%%%%%%%%%%%%%%%%%%%%%%%%%%%%%

\section{Discontinuities}
\label{app:jumps}

%%%%%%%%%%%%%%%%%%%%%%%%%%%%%%%%%%%%%%%%%%%%%%%%%%%%%%%%%%%%%%%%%%%%%%%%%%%%%%%%%%%%%%

\begin{table*}
    \caption{Same as Figure \ref{fig:jumpspercrit} but as a table for ease of reading.}
    \captionsetup[subfloat]{position = top}
    \centering
    \subfloat[\label{tab:appjumpspercritu} Uranus]{
        \resizebox{\textwidth}{!}{%
        \begin{tabular}{c|cccccccccccccccccccc}
            $c_D$ & 50 & 100 & 150 & 200 & 250 & 300 & 350 & 400 & 450 & 500 & 550 & 600 & 650 & 700 & 750 & 800 & 850 & 900 & 950 & 1000\\
            \hline
            Avg & 8 & 4 & 2 & 1 & 1 & 1 & 1 & 0 & 0 & 0 & 0 & 0 & 0 & 0 & 0 & 0 & 0 & 0 & 0 & 0\\
            \hline
            95\% & 15 & 8 & 5 & 4 & 3 & 3 & 2 & 2 & 2 & 2 & 1 & 1 & 1 & 1 & 1 & 1 & 1 & 1 & 1 & 1\\
        \end{tabular}%
        }
    }
    \\
    \subfloat[\label{tab:appjumpspercritn} Neptune]{
        \resizebox{\textwidth}{!}{%
        \begin{tabular}{c|cccccccccccccccccccc}
            $c_D$ & 50 & 100 & 150 & 200 & 250 & 300 & 350 & 400 & 450 & 500 & 550 & 600 & 650 & 700 & 750 & 800 & 850 & 900 & 950 & 1000\\
            \hline
            Avg & 10 & 5 & 3 & 2 & 1 & 1 & 1 & 1 & 1 & 1 & 0 & 0 & 0 & 0 & 0 & 0 & 0 & 0 & 0 & 0\\
            \hline
            95\% & 20 & 11 & 7 & 5 & 4 & 4 & 3 & 3 & 2 & 2 & 2 & 2 & 2 & 1 & 1 & 1 & 1 & 1 & 1 & 1\\
        \end{tabular}%
        }
    }
    \tablefoot{
    Table of the number of discontinuities detected per successful density profile over some selection criteria $c_D$ (in units of $\Delta \ell^{-1}$ kg\,m$^{-3}$), average (rounded to the nearest integer) and upper 95th percentile.
    }
    \label{tab:appjumpspercrit}
\end{table*}

%%%%%%%%%%%%%%%%%%%%%%%%%%%%%%%%%%%%%%%%%%%%%%%%%%%%%%%%%%%%%%%%%%%%%%%%%%%%%%%%%%%%%%

\subsection{Discontinuity detection}

%%%%%%%%%%%%%%%%%%%%%%%%%%%%%%%%%%%%%%%%%%%%%%%%%%%%%%%%%%%%%%%%%%%%%%%%%%%%%%%%%%%%%%

The following is a description of how density discontinuities are detected and categorised given a discontinuity criterion $c_D$. 
Note that the base working environment here is an array of discrete entries listing the increase from one level surface to the next.

%%%%%%%%%%%%%%%%%%%%%%%%%%%%%%%%%%%%%%%%%%%%%%%%%%%%%%%%%%%%%%%%%%%%%%%%%%%%%%%%%%%%%%

Our task is to find every discontinuity of any arbitrary width. For this, we must first define what a discontinuity is. For that, we use the following definitions.
\begin{definition}[Jump]
	A jump is defined as an interval of size $k$ such that the total increase over that interval is more than $k\cdot c_D$, that is, the total average increase over this interval is more than the discontinuity criterion.
\end{definition}
\begin{definition}[Maximally steep jump]
	A jump is considered maximally steep if shrinking it by one on the left or right would lower its steepness or the removed entry exceeds the discontinuity criterion.
\end{definition}
\begin{definition}[Jump cluster]
	A jump cluster is defined as the recursive union of a jump with all other jumps that intersect it.
\end{definition}
Note the jump cluster need not necessarily be a jump.
\begin{definition}[Discontinuity]
	A discontinuity is defined as the union of all maximally steep jumps in a jump cluster.
\end{definition}
A discontinuity therefore consists entirely of maximally steep jumps and is `minimally maximal'. 
It cannot be extended, since then at least one jump within the discontinuity would no longer be maximally steep. 
Neither can it be shortened, since then either its steepness would be lowered or sections above the discontinuity criterion would be removed.

%%%%%%%%%%%%%%%%%%%%%%%%%%%%%%%%%%%%%%%%%%%%%%%%%%%%%%%%%%%%%%%%%%%%%%%%%%%%%%%%%%%%%%

We provide some illustrative examples for $c_D=100$. 
We list the increases, not the values directly, and the first listed value has index 0.
The brackets denote a jump and the numbers within the brackets denote the indices. 
For example [1,1] denotes a jump of size one with a height of the value with index 1.
[1,3] denotes a jump of size three with a height equal to the sum of the values with indices 1, 2, and 3.

%%%%%%%%%%%%%%%%%%%%%%%%%%%%%%%%%%%%%%%%%%%%%%%%%%%%%%%%%%%%%%%%%%%%%%%%%%%%%%%%%%%%%%

\begin{itemize}

\item ..., 0, 101, 99, 101, 0, ... corresponds to one discontinuity at at [1,3] because [1, 1], [3, 3], and [1, 3] are maximally steep jumps.

\item ..., 0, 100, 0, 100, 0, ... corresponds to two discontinuities because [1, 3] is not a jump.

\item ..., 0, 100, 200, 100, 0, ... corresponds to one discontinuity at [1,3].

\item ..., 0, 99, 200, 99, 0, ... corresponds to one discontinuity at [2,2] because the jumps [1, 2], [2,3], and [1,3] are not maximally steep.

\item ..., 0, 100, 0, 200, 0, 100, 0, ... corresponds to one discontinuity at [1,5] because [1,3] and [3,5] are maximally steep jumps, resulting in a cluster spanning [1,5].

\item ..., 0, 100, 0, 100, 0, 100, 0, ... corresponds to three discontinuities.

\end{itemize}

%%%%%%%%%%%%%%%%%%%%%%%%%%%%%%%%%%%%%%%%%%%%%%%%%%%%%%%%%%%%%%%%%%%%%%%%%%%%%%%%%%%%%%

Regarding the numerical detection of a discontinuity, one could iterate over all interval sizes and find all corresponding jumps and then find their largest intersection of maximally steep jumps for each cluster. 
However, this method is very slow. 
We thus use an approximate solution using a convolution (sliding window). 
An averaging sliding window (for example [0.25, 0.25, 0.25, 0.25]), would only guarantee to find discontinuities of the size of the window. 
This is because smaller discontinuities would get squashed (in our example by the factor 0.25) and larger discontinuities could erroneously be split.
For example ..., 125, 125, 0, 125, 125, ... should be one discontinuity but would get identified as two with [0.5, 0.5] being the sliding window.
A constant sliding window (for example [1, 1, 1, 1]) is also not usable, since it would misidentify increases consistently below the density criterion as discontinuities, especially at large window sizes. 
Therefore, we used a sliding window with varying values: 
We chose a window size of $n=9$ and want to guarantee that we will find all discontinuities of any size up to $n=9$. 
The window hence must contain one entry $\geq 1$ to find discontinuities of length one, two entries $\geq 0.5$ to find discontinuities of length two, and so on. 
$n=9$ requires nine entries $\geq 1/9$ which ultimately yields [1/8, 1/6, 1/4, 1/2, 1, 1/3, 1/5, 1/7, 1/9] as a sliding window. 
Since it is highly likely that discontinuities larger than this window will have subsections within this window that are captured, this essentially guarantees all discontinuities are found.
In summary, this sliding window ensures that a discontinuity even of only size 1 will certainly at some point be detected by exceeding the discontinuity criterion.
The same holds true for a discontinuity spread over nine steps.
However, this method is still too generous for the same reason as the above problem with the constant sliding window of [1, 1, 1, 1]. 
Therefore, some post-processing has to be done, where we iteratively cut potential `discontinuities' on both sides as long as the smaller `discontinuity' has better steepness and as long as the cut itself is not steeper than the discontinuity criterion. 
If the final resulting `discontinuity' has a total height $\geq k\cdot c_D$, we have successfully detected a true discontinuity.

%%%%%%%%%%%%%%%%%%%%%%%%%%%%%%%%%%%%%%%%%%%%%%%%%%%%%%%%%%%%%%%%%%%%%%%%%%%%%%%%%%%%%%

\subsection{Further results}

%%%%%%%%%%%%%%%%%%%%%%%%%%%%%%%%%%%%%%%%%%%%%%%%%%%%%%%%%%%%%%%%%%%%%%%%%%%%%%%%%%%%%%

\begin{figure*}
	\centering
    \includegraphics[width=0.49\textwidth]{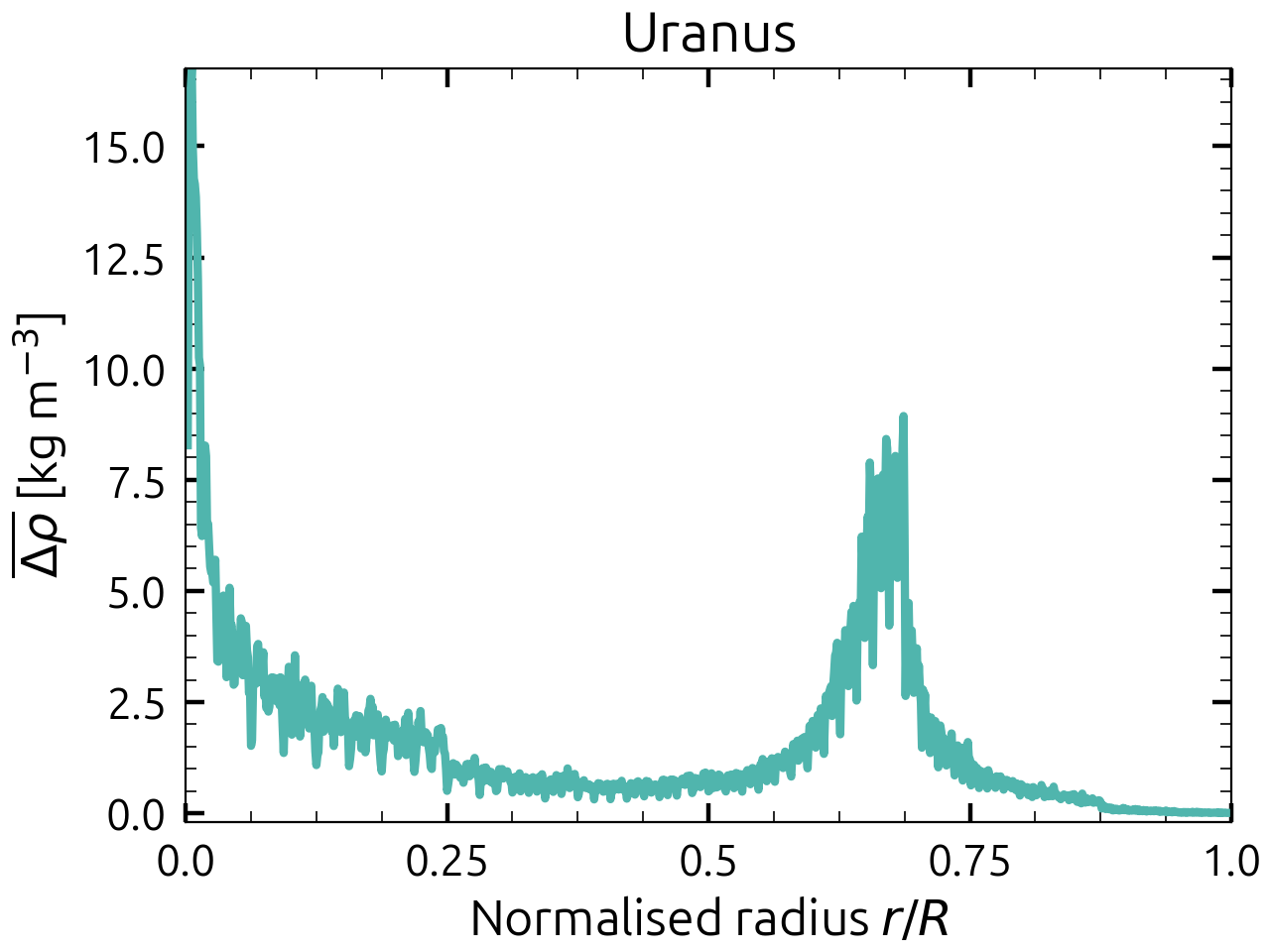}
	\includegraphics[width=0.49\textwidth]{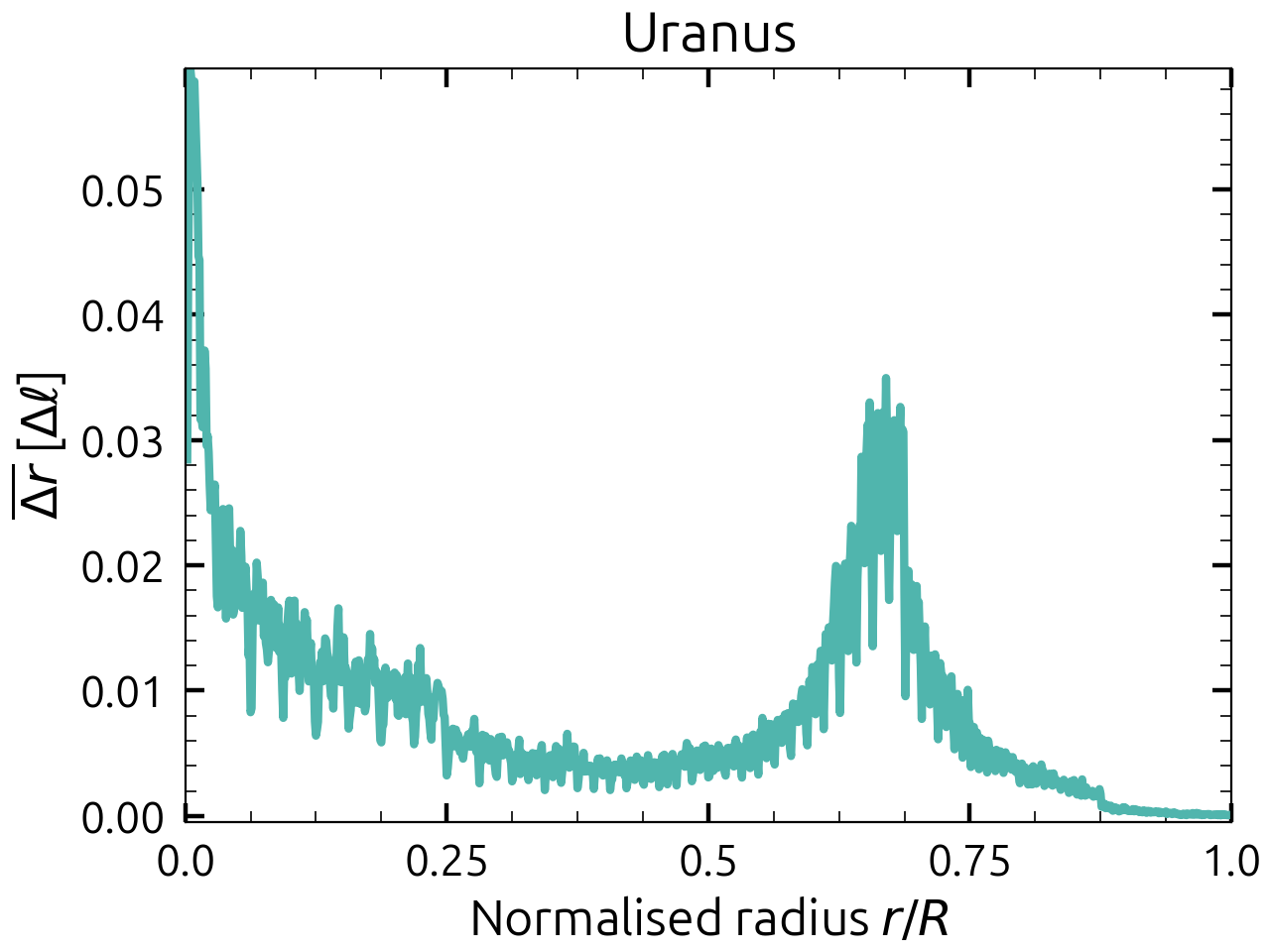}
	\caption{
    Average discontinuity heights (left) and widths (right) of successful density profiles.
    For example, discontinuities with an average location of $r/R=0.5$ possess an average width of $\overline{\Delta r}\sim0.005\Delta\ell$. 
    Profiles with no discontinuities also contribute to the averages with value zero.
    }
    \label{fig:jumpsdrhodru}
\end{figure*}

\begin{figure*}
	\centering
    \includegraphics[width=0.49\textwidth]{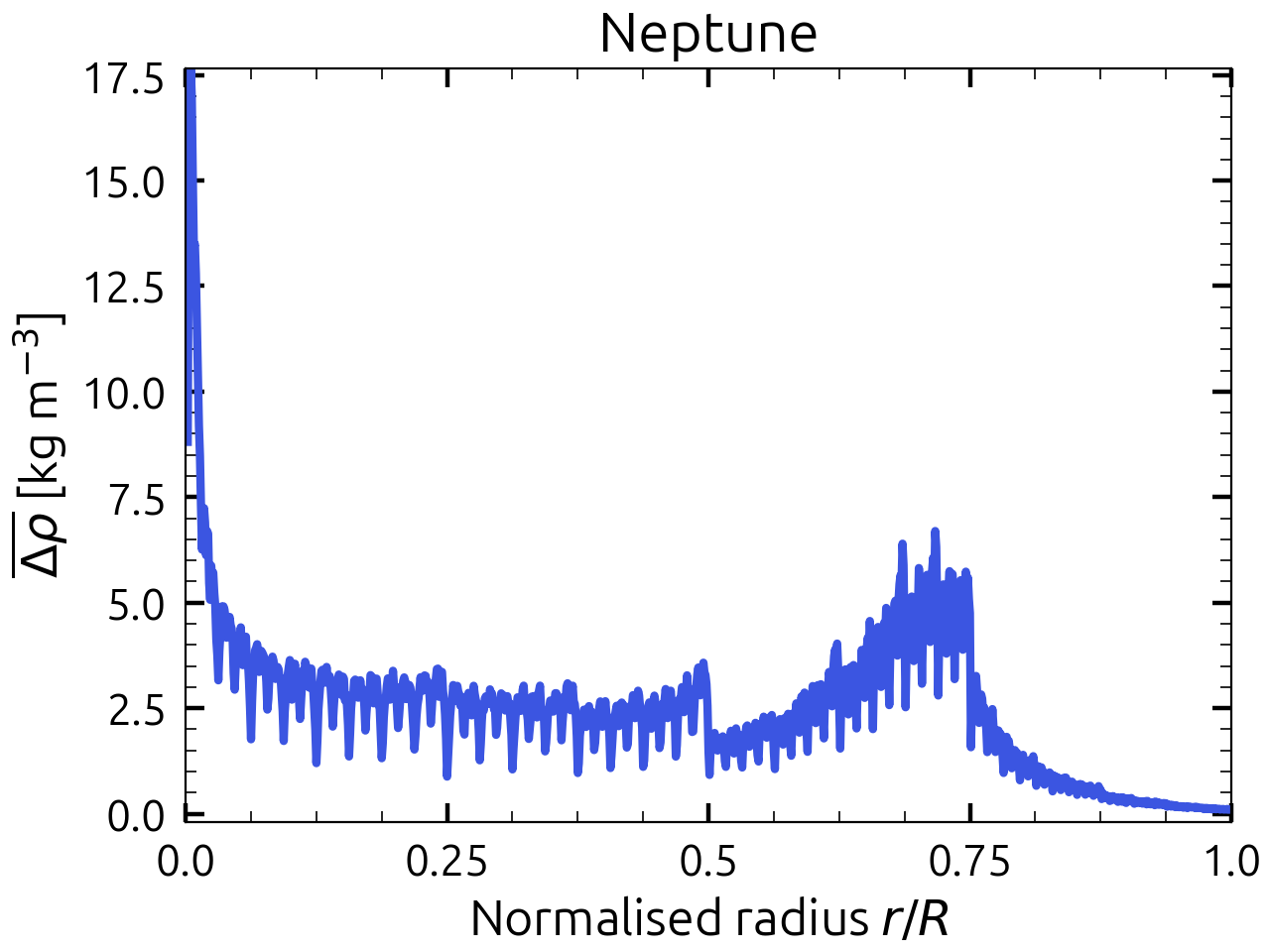}
	\includegraphics[width=0.49\textwidth]{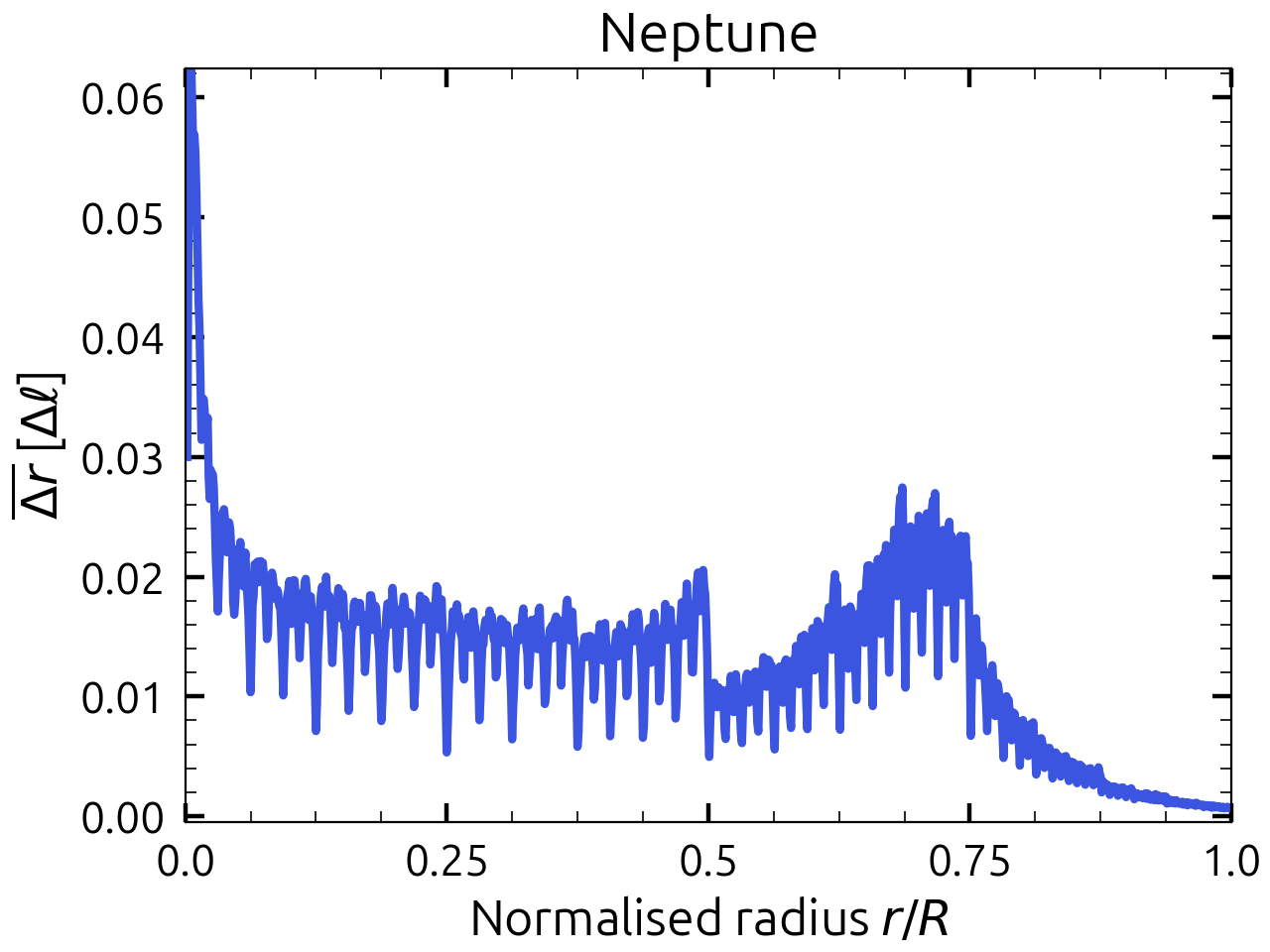}
	\caption{Same as Figure \ref{fig:jumpsdrhodru}, but for Neptune.}
    \label{fig:jumpsdrhodrn}
\end{figure*}

\begin{figure*}
	\centering
	\includegraphics[width=0.49\textwidth]{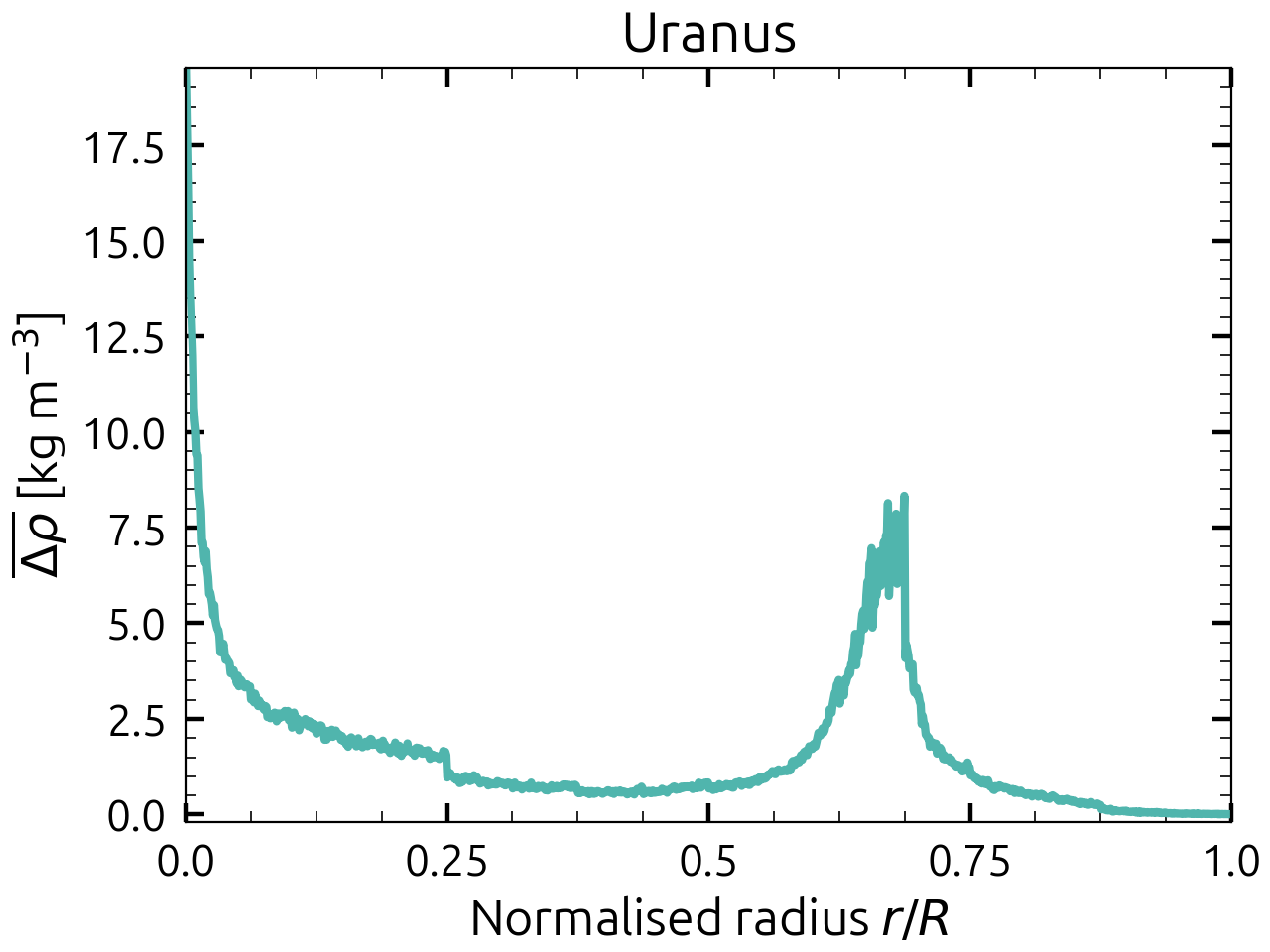}
    \includegraphics[width=0.49\textwidth]{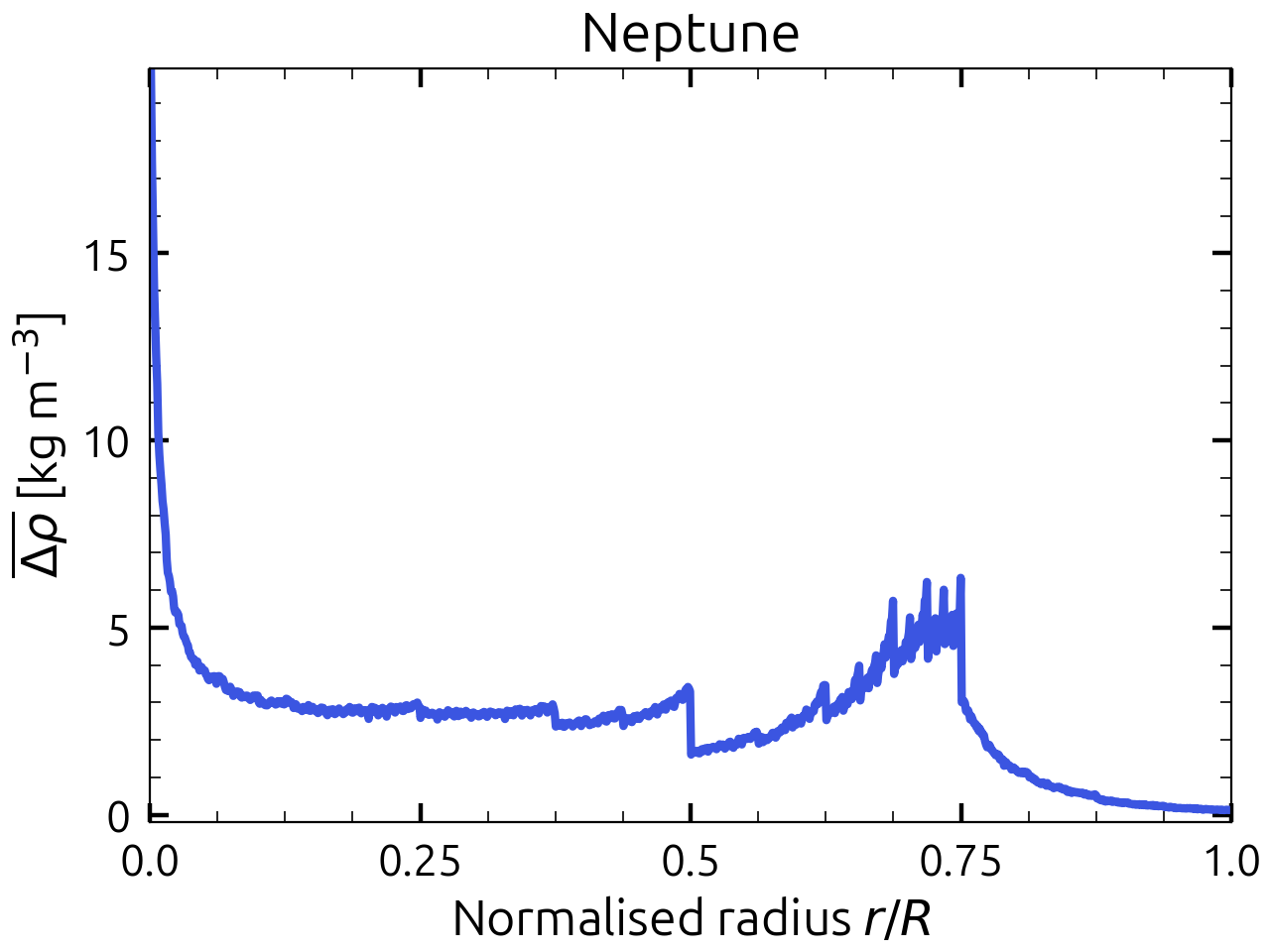}
	\caption{Average density increase as a function normalised radius for all successful density profiles. Increases below the discontinuity criterion contribute with value zero to the average.}
    \label{fig:jumpsraw}
\end{figure*}

We show the values of Figure \ref{fig:jumpspercrit} in Table \ref{tab:appjumpspercrit}.
Figures \ref{fig:jumpsdrhodru} (Uranus) and \ref{fig:jumpsdrhodrn} (Neptune) separately show the average discontinuity height $\overline{\Delta\rho}$ and width $\overline{\Delta r}$ combined in Figure \ref{fig:jumps} at their respective average normalised radii.
Note that profiles that have no discontinuity at any given location also contribute to the averages with value zero.
Artefacts from the starting model generation (see Section \ref{sec:discussion}) are clearly visible in Figure \ref{fig:jumpsdrhodrn}.
Finally, we show the `raw' average density increase for all successful profiles in Figure \ref{fig:jumpsraw}.
This Figure differs from the left panels of Figures \ref{fig:jumpsdrhodru} and \ref{fig:jumpsdrhodrn} because we do not show the average discontinuity height, but rather the general average density increase as a function of normalised radius for all successful density profiles.
However, increases below the discontinuity criterion still only contribute with value zero to the average.

%%%%%%%%%%%%%%%%%%%%%%%%%%%%%%%%%%%%%%%%%%%%%%%%%%%%%%%%%%%%%%%%%%%%%%%%%%%%%%%%%%%%%%

%%%%%%%%%%%%%%%%%%%%%%%%%%%%%%%%%%%%%%%%%%%%%%%%%%%%%%%%%%%%%%%%%%%%%%%%%%%%%%%%%%%%%%

%%%%%%%%%%%%%%%%%%%%%%%%%%%%%%%%%%%%%%%%%%%%%%%%%%%%%%%%%%%%%%%%%%%%%%%%%%%%%%%%%%%%%%

%%%%%%%%%%%%%%%%%%%%%%%%%%%%%%%%%%%%%%%%%%%%%%%%%%%%%%%%%%%%%%%%%%%%%%%%%%%%%%%%%%%%%%

\end{appendix}

\end{document}